\documentclass[aps,prb,amsmath,superscriptaddress,twocolumn,showkeys,english]{revtex4}
\usepackage{natbib}
\usepackage{graphicx}
\usepackage{amsmath, amsthm}
\usepackage{babel}
\usepackage{pstricks}
\usepackage{pst-node}
\usepackage{multido}
\usepackage{pstricks-add}
\usepackage{epsfig}
\usepackage{multirow}

\def\naive{na\"{\i}ve }

\makeatother

\begin{document}
\title{Grid Alignment in Entorhinal Cortex}

\author{Bailu Si}
\affiliation{Sector of Cognitive Neuroscience, International School for Advanced Studies,  via Bonomea 265, 34136 Trieste, Italy}
\email{{bailusi,ale}@sissa.it}
\author{Emilio Kropff}
\affiliation{Kavli Institute for Systems Neuroscience and Center for the Biology of Memory, Norwegian University of Science and Technology, 7489 Trondheim, Norway}
\email{emilio.kropff@ntnu.no}
\author{Alessandro Treves}
\affiliation{Sector of Cognitive Neuroscience, International School for Advanced Studies,  via Bonomea 265, 34136 Trieste, Italy}
\affiliation{Kavli Institute for Systems Neuroscience and Center for the Biology of Memory, Norwegian University of Science and Technology, 7489 Trondheim, Norway}

\begin{abstract}
The spatial responses of many of the cells recorded in all layers of rodent medial entorhinal cortex (mEC) show a triangular grid pattern, which appears to provide an accurate population code for position, and once established might be based in part on path-integration mechanisms. Competing models, each partially contradicted by experimental observations, try to explain how the grid-like pattern emerges in terms of network interactions, or of interactions with theta oscillations or, the one we have proposed, of mere single-unit mechanisms. 

Grid axes are tightly aligned across simultaneously recorded units. Recent experimental findings have shown that grids can often be better described as elliptical rather than purely circular and that, beyond the mutual alignment of their grid axes, ellipses tend to also orient their long axis along preferred directions. Are grid alignment and ellipse orientation the same phenomenon? Does the grid alignment result from single-unit mechanisms or does it require network interactions? 

We address these issues by refining our model, to describe specifically the spontaneous emergence of conjunctive grid-by-head-direction cells in layers III, V and VI of mEC. We find that tight alignment can be produced by recurrent collateral interactions, but this requires head-direction modulation. Through a competitive learning process driven by spatial inputs, grid fields then form already aligned, and with randomly distributed spatial phases. In addition, we find that the self-organization process is influenced by the behavior of the simulated rat. The common grid alignment often orients along preferred running directions. The shape of individual grids is distorted towards an ellipsoid arrangement when some speed anisotropy is present in exploration behavior. Speed anisotropy on its own also tends to align grids, even without collaterals, but the alignment is seen to be loose. Finally, the alignment of spatial grid fields in multiple environments shows that the network expresses the same set of grid fields across environments, modulo a coherent rotation and translation. Thus, an efficient metric encoding of space may emerge through spontaneous pattern formation at the single-unit level, but it is coherent, hence context-invariant, if aided by collateral interactions.
\end{abstract}

\maketitle

\section{Introduction}
Internal representations of space appear necessary for any agent, such as a rat or a robot, to move around in a spatial context and to distinguish between different contexts to which food or danger, for example, may be associated. Spatial cognition and memory have been long investigated in rodents, and an impressive body of results point at the major role being played by the hippocampus and related cortices, a region that in humans has been associated with the neural basis of episodic memory formation, since~\citep{Scoville1957}. Forty years ago, place cells were discovered in the rat hippocampus, showing specific firing activity whenever the rat enters a specific portion of the environment, the place field~\citep{OKeefe1971}. Head direction (HD) cells were first found in the rat postsubiculum, firing steadily when the animal points its head towards a specific direction in the environment~\citep{Taube1990}. These two distinct systems provide simple, distributed population coding of the location and heading direction of the animal. 

In recent years, cells with more complex spatial codes have been discovered in the rat medial entorhinal cortex (mEC), which sends strong projections to the hippocampus. Grid cells, found to be particularly abundant in layer II of mEC (perhaps about half of stellate cells there) have multiple firing fields positioned on the vertices of remarkably regular triangular grids, spanning the environment which the animal explores~\citep{Fyh+04,Hafting2005}. Conjunctive grid-by-head-direction cells, found along a smaller proportion of pure grid cells in the deeper layers of mEC, show firing selectivity to HD in addition to (perhaps slightly less precise) spatial tuning as grid cells~\citep{Sargolini2006}. The precise geometric tessellation provided by the activity of grid cells has stimulated a series of experimental and theoretical studies on the mechanisms underlying the emergence and the function of grid cells~\citep{Giocomo2011}.

Theoretical models of grid cell formation may be grouped in three main categories. The first type of models shows how grid fields may emerge from the attractor states induced, in continuous attractor networks, from structured recurrent connectivity~\citep{McN+06,Fuh+06,Guanella2007,burak2009,Navratilova2011}. The spatial layout of recurrent collateral connections of the network, assumed to be permanent or at least present during the developmental stage of grid cell formation, ensures that triangular spatial firing patterns are stable states of recurrent dynamics. Grid units in a continuous attractor network model are able to perform path integration by propagating the activity in the network in correspondence with the movement of the animal, in a way similar to previous models for place cells or HD cells~\citep{Zhang1996,Samsonovich1997,Walters2010}. 

The second class of models relates the periodic firing of grid cells to sub-threshold membrane potential oscillations, and proposes that grid fields may result from interference between a theta-related baseline oscillator and other velocity-controlled oscillations which originate either in different dendrites of a neuron or in different neurons~\citep{Bur+07,Giocomo2007,Burgess2008,Hasselmo2008,Zilli2010}. The frequency difference between these velocity-controlled oscillators is small, so that the low frequency ``envelope'' of the interference pattern corresponds to the spatial periodicity of grid cells. This hypothesis is in part supported by recent findings that the spatial periodicity of grid cells is susceptible to the suppression of theta oscillations by pharmacological silencing of the medial septum~\citep{Koenig2011,Brandon2011}. 

The third class of models on grid cell formation argues that grid fields may not require detailed ad hoc mechanisms like structured connectivity or theta oscillations, rather they may emerge spontaneously from a general feature of cortical cell activity, like firing rate adaptation~\citep{Kro+08} or, equivalently, other types of temporal modulation~\citep{Garden2008}.   Such temporal modulation is shown in computer simulations to sculpt the spatial modulation of grid cells through a self-organization process that, averaged over a long developmental time of one or two weeks~\citep{Langston2010,Wills2010}, leaves as a footprint on each unit the regular periodicity found in real grid cells. A simple analytical model ``explains'' this spontaneous pattern formation as an unsupervised optimization process at the single-unit level~\citep{Kro+08}.

An interesting aspect of this model, which motivates the present study, is that the emergence of perfect grid symmetry requires a perfectly isotropic distribution of rat trajectories and speed, once averaged over the long (but not infinite) learning period. Any deviation from perfect isotropy, for example because the animal spends the relevant developmental period (for a rat, somewhere between P15-P35, say) mainly in a rectangular cage and tends to move along the walls, or because it runs a bit faster along some particular directions, would be expected to induce distortions in all grids units. Excitingly, such deviations in the geometry of grid maps have been observed recently and described as an {\em ellipticity} effect~\citep{Stensland2010}. This phenomenon cannot be explained as small random deviations from the perfect symmetry, because of its extent and remarkable consistency across the population. Not only, as observed earlier, do grid axes show nearly the same alignment across all simultaneously recorded grid units, but also the long axes of the corresponding ellipses appear to loosely orient with either one of two major ellipse clusters, depending on their spacings~\citep{Stensland2010}. 

Can the mutual alignment result from the same single-unit mechanisms that produce the individual grid patterns? With perfect isotropic grids, the answer is obviously negative, and in fact Kropff and Treves (2008) pointed at a mechanism that can align grid cells at a population level, but based on collateral interactions between them. With the observed anisotropies, however, the situation might be different, as both alignment and common orientation might develop along the (common) anisotropy axes. 

In the computer simulation study reported here we find, again, that developing a tight alignment requires network interactions, through recurrent collateral connections with a specific structure, modulated by HD. As a first focus of this paper, we then describe how collateral connections may align grid cells, and then link this to the emergence of a coherent population {\em ellipticity} when the behavior of the animal is biased by running direction or speed. We then argue that anisotropy at the single-unit level is in principle sufficient to also produce some alignment, but fails in practice to make different units as tightly aligned as experimentally observed.    

Finally, we also investigate how well the same model describes {\em global remapping}~\citep{Colgin2008}. When a rat is taken into a new environment, or when the environment is manipulated in a way that it looks new, place maps in the hippocampus undergo an apparently random shuffling, or switch from active to inactive and vice versa. Grid maps, however, behave coherently at the population level, undergoing a common rotation and spatial shift, in such a way that spatial overlaps between maps are preserved across rooms~\citep{Fyh+07}. We thus address the question of how, within our model, coherent grid fields may emerge in the same network for multiple environments. 

The rest of the article is organized as follows. The network model is first introduced in Section~\ref{sec-model}. In Section~\ref{sec-sylind} grid alignment in a cylinder environment is studied, and it is compared to what occurs in a square environment in Section~\ref{sec-box}. We investigate in Section~\ref{sec-speed} the effects of exploration with speed anisotropy. Grid realignment in multiple environments is discussed in Section~\ref{sec-remap}. Finally, the results of the study are discussed in Section~\ref{sec-discuss}. 

\section{Network model}\label{sec-model}

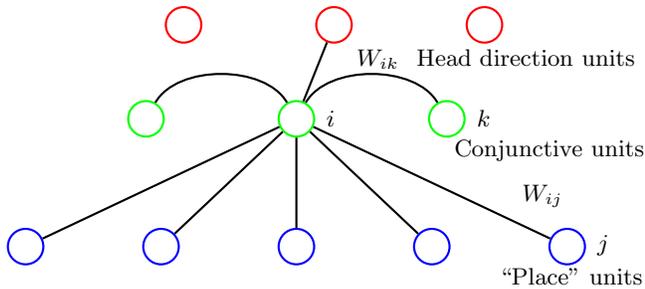
\begin{figure}
\centering
\begin{minipage}{8.4cm}
\begin{pspicture}(8.4,4.2)
  \multido{\iP=1+1,\nX=0.4+1.8}{5}{\cnode[linecolor=blue](\nX, 0.8){0.25}{P\iP}}
  \multido{\iG=1+1,\nX=2.0+2.0}{3}{\cnode[linecolor=green](\nX, 2.5){0.25}{G\iG}}
  \multido{\iH=1+1,\nX=2.5+2.0}{3}{\cnode[linecolor=red](\nX, 3.75){0.25}{H\iH}}
  \multido{\iP=1+1}{5}{%
    \ncline{-}{P\iP}{G2}%
  }
 \ncline{-}{H2}{G2}
  \nccurve[angleA=60,angleB=120]{-}{G1}{G2}
  \nccurve[angleA=120,angleB=60]{-}{G3}{G2}
  \put(4.8,3.2){$W_{ik}$}
  \put(6.1,2){Conjunctive units}
  \put(4.4,2.4){$i$}
  \put(6.4,2.4){$k$}
  \put(7,1.4){$W_{ij}$}
  \put(6.7,0.3){``Place'' units}
  \put(8,0.75){$j$}
  \put(5.6,3.2){Head direction units}
\end{pspicture}
\end{minipage}
\caption{A sketch of the network model for conjunctive cells in deep layers of mEC. Connections are shown only for unit $i$. Place units are fully connected to conjunctive units. Conjunctive units are connected to all other conjunctive units without self-connections. Each conjunctive unit is assumed to be modulated by one HD unit, representing the overall effect of angular modulation from the local network.}
\label{fig-simp-net}
\end{figure}

In Fig.~\ref{fig-simp-net} we present a diagram of the network that we use for simulations. It is intended to model conjunctive grid-by-head-direction cells in layers III, V and VI of mEC. In layer V, which receives strong projections from the subiculum and the CA1 region of the hippocampus, about 20\% of the putative pyramidal cells are estimated to be conjunctive grid-by-head-direction cells, along with a very small proportion of pure grid cells and more than 60\% of HD cells~\citep{Boccara2010}. In layer III, which projects to CA1, the proportion of conjunctive cells is similar to that of layer V, but the proportion of HD cells is much lower (20\%) and there is an extra 20\% of pure grid cells. These layers present a prominent recurrent connectivity, denser in layer V (around 12\%) but also important in layer III (around 9\%)~\citep{Dhillon2000}.  Layer II of mEC, where the highest proportion and the best quality of pure grid cells are found (around 50\%), lacks two critical elements that in our model go hand by hand: recurrent collateral connections and head direction information. In this paper, we assume that layer II maps can self-organize using inputs which include conjunctive maps, an assumption that is discussed in a separate study currently in progress. Here, we focus therefore on the learning process in layers III to VI, and neglect the possible influence there of the feedback from layer II.

The critical difference with previous versions of the model \citep{Kro+08} is that now we assign a central role to head direction information, which, as we show, is a key element for grid alignment. A preferred head direction, $\theta_i$, is introduced arbitrarily for each conjunctive unit $i$, to modulate its inputs. $\theta_i$ is uniformly sampled from all angles. This assumption is based on our own analysis of the HD selectivity of the conjunctive cells recorded by~\citep{ Sargolini2006}, in which we did not find significant clustering of HDs with respect to either the reference frame or grid axes. Each conjunctive unit receives afferent spatial inputs, which as discussed in~\citep{Kro+08} we take for simplicity to arise from regularly arranged ``place'' units, and collateral inputs from other conjunctive units (Fig.~\ref{fig-simp-net}). The overall input to conjunctive unit $i$ at time $t$ is then given by $h_{i}^t$
\begin{equation}
 h_{i}^t = f_{\theta_i}(\omega_t)(\sum_{j} W_{ij}^{t-1}r_{j}^t + \rho \sum_k W_{ik} \Psi_k^{t-\tau} )\label{eq-conjunc-affrent},
\end{equation}
with $\rho=0.2$ a factor weighing, relative to feed-forward inputs,  collateral inputs relayed with a delay $\tau = 25$ steps. In the model, each time step corresponds to $10$ msec in real time, so the collateral interaction describe temporally diffuse and delayed processes, rather than straightforward AMPA-mediated excitation. $f_{\theta_i}(\omega_t)$ is a tuning function that has maximal value when the current HD $\omega^t$ of the simulated rat is along the preferred direction $\theta_i$ of the unit, as in~\citep{Zhang1996}
\begin{equation}
f_{\theta}(\omega) = c+ (1-c)\exp[\gamma(\cos (\theta - \omega)-1)],\label{eq-hd-tuning}
\end{equation}
where $c=0.2$ and $\gamma=0.8$ are parameters determining the minimal value and the width of HD tuning.

The weight $W_{ij}^t$ connects place unit $j$ to conjunctive unit $i$, while $W_{ik}$ connects conjunctive unit $k$ to unit $i$. For simplicity, throughout the paper we only consider the self-organization of the feed-forward weights, keeping the collateral weights fixed at convenient values, which we will discuss in Sect.~\ref{collwei}. A model in which collateral weights are also the result of a perhaps slower self-organizing learning process will be discussed elsewhere. 

It is possible that conjunctive cells receive place-cell-like inputs from the hippocampus already when they develop their firing maps. Studies on the development of the spatial representation system in the rat may be taken to show that place cells and head-direction cells develop adult-like spatial and directional codes somewhat earlier than grid cells do~\citep{Langston2010,Wills2010}. The firing rate of a place unit $j$ is modeled by an exponential function centered in its preferred firing location $\vec{\bf x}_{j0}$
\begin{equation}
r_j^t = \exp(-\frac{\vert\vec{\bf x}^t- \vec{\bf x}_{j0} \vert^2}{2\sigma_p^2}) ,
\end{equation}
where $\vec{\bf x}^t$ is the current location of the simulated rat. $\sigma_p = 5 cm$ is the width of the firing field. The place field of a place unit is arranged such that the distances to neighboring fields of other place units are about $5 cm$. Note that the precise and regularly arranged location-specific inputs of place units are used only to speed up the learning process. The network model works in qualitatively the same way with broad spatial inputs, but at the cost of longer learning time~\citep{Kro+08}. 

The firing rate $\Psi_i^t$ of conjunctive unit $i$ is determined by
\begin{equation}
\Psi_i^{t} = \Psi_{sat}\arctan [g^{t} (\alpha_i^t - \mu^{t})]\Theta(\alpha_i^t - \mu^{t}),\label{eq-output}
\end{equation}
where $\Psi_{sat}=\pi/2$, so that the maximal firing rate is $1$ (in arbitrary units).  $\Theta(\cdot)$ is the Heaviside function. The variable $\alpha_i^t$ represents a time-integration of the input $h_i$, adapted by the dynamical threshold $\beta_i$
\begin{align}
\alpha_i^{t}&=  \alpha_i^{t-1} + b_1 (h_i^{t-1} - \beta_i^{t-1} -\alpha_i^{t-1}),\nonumber\\
\beta_i^{t} &=  \beta_i^{t-1} + b_2 (h_i^{t-1} - \beta_i^{t-1}),
\end{align}
where $\beta_i$ has slower dynamics than $\alpha_i$ ($b_2$ is set to $b_2 = b_1/3$, and in our time steps corresponds to a time scale of roughly 300 $(ms)^{-1}$, with $b_1 = 0.1\simeq 100 (ms)^{-1}$). Due to the adaptation dynamics, a unit that fires strongly in the recent past tends not to fire in the near future, because of a high threshold. 

In Eq.~\ref{eq-output}, the gain $g^t$ and threshold $\mu^t$ are two parameters chosen at each time step to keep the mean activity $a = \sum_{i}\Psi_i^t/N_{mEC}$ of the conjunctive units and the sparsity $s = (\sum_{i} \Psi_i^t)^2/(N_{mEC}\sum_{i} {\Psi_i^t}^2)$ within a 10\% relative error bound from pre-specified values, set at $a_0=0.1$ and $s_0=0.3$ respectively. The values of $g^t$ and $\mu^t$ are determined through iterations
\begin{align}
\mu^{t,l+1}&=\mu^{t,l}+ b_3(a^{l}-a_0),\nonumber\\
g^{t,l+1}&= g^{t,l} + b_4 g^{t,l}(s^{l}-s_0).\label{eq-gain-thresh}
\end{align}
Here $l$ is the index of the iteration within simulation step $t$. $b_3=0.01$ and $b_4=0.1$ are positive step sizes in iteration. $a^l$  and $s^l$ are the mean and the sparsity of the conjunctive units when the gain $g^{t,l}$ and threshold $\mu^{t,l}$ are applied. The gain and threshold at the end of the iteration are used in Eq.~\ref{eq-output} to determine the activity of conjunctive units.

With constant running speed, neural fatigue is invariant with respect to all directions. Conjunctive units self-organize firing fields into hexagonal/triangular grids in two-dimensional space, as the virtual rat explores the environment. This configuration of fields corresponds to the minimum of an energy function~\citep{Kro+08}, consistent with an hexagonal tiling being the most compact one to arrange circles in two-dimensional space~\citep{ Petkovic2009}.

\subsection{Learning feed-forward weights}
The feed-forward weights are adaptively modified according to a Hebbian rule
\begin{equation}
\tilde W_{ij}^{t} = W_{ij}^{t-1} + \epsilon (\Psi_i^{t} r_j^t -  \bar \Psi_i^{t-1} \bar r_j^{t-1}).
\end{equation}
Here $ \bar \Psi_i^{t}$ and $\bar r_j^{t}$ are estimated mean firing rates of conjunctive unit $i$ and place unit $j$
\begin{align} 
\bar \Psi_i^{t} &= \bar \Psi_i^{t-1} + \eta \left(\Psi_i^{t} - \bar \Psi_i^{t-1}\right),\nonumber\\
 \bar r_j^{t} &= \bar r_j^{t-1} + \eta \left(r_j^{t} - \bar r_j^{t-1}\right) \label{eq-ave-act},
\end{align}
while $\epsilon=0.005$ is a moderate learning rate, intended to produce gradual weight change, and $\eta = 0.05$ is a time averaging factor.

After each learning step, the new weights are normalized into unitary norm
\begin{equation} 
\sum_j {W_{ij}^t}^2 = 1\label{eq-norm}.
\end{equation}
Through competitive learning, conjunctive units that win the competition are associated to the input units that provide strong inputs. As a result, firing fields of conjunctive units are anchored to places where inputs are strong simultaneously with recovery from adaptation, and are stabilized as the learning proceeds. There is experimental evidence of the dependence of grid fields on sensory information~\citep{Bar+07,Sav+08}. For example, grids can expand and contract in response to expansions and contractions of a familiar environment. Some still unpublished observations indicate that the learning of stable grid maps in a novel environment for non-over-trained adult rats could take several days of training~\citep{Bar+09}.

\subsection{Collateral weights}\label{collwei}

Two characteristic features of grid cells are that they align their axes and maintain fixed spatial phases relative to each other. Such alignment and relative phases are invariant to environmental changes, including those under which place cells in the hippocampus remap completely. 

In our network model, we include these two constraints on grid cell population activity with the help of collateral weights. Head direction information is used to gate the interaction between two units that fire in sequence, so that the second unit will develop fields along its preferred direction close to those of the first unit, which fires earlier. If the interaction were not gated by HD, the second unit would tend to form a field that is a ring surrounding the field of the first unit. 

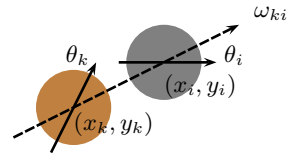
\begin{figure}
\centering
\begin{pspicture}(3.8,2.2)
\cnode[fillstyle=solid,fillcolor=brown,linecolor=brown](0.8,0.6){.5}{A}
\cnode[fillstyle=solid,fillcolor=gray,linecolor=gray](2,1.2){.5}{B}
\psline[linewidth=1pt,linestyle=dashed,dash=4pt 1pt]{->}(0,0.2)(3,1.7)
\psline[linewidth=1pt]{->}(0.5,0)(1.1,1.2)
\psline[linewidth=1pt]{->}(1.4,1.2)(2.7,1.2)
 \put(0.7,1.2){$\theta_k$}
 \put(2.8,1.2){$\theta_i$}
 \put(3.2,1.8){$\omega_{ki}$}
\put(0.8,0.3){$(x_k, y_k)$}
\put(2,0.8){$(x_i, y_i)$}
\end{pspicture}
\caption{Assignment of the fixed collateral weight from unit $k$ to unit $i$.}
\label{fig-fixed-collat}
\end{figure}

The collateral weights are assigned before any learning takes place in the feed-forward weights. The weight $W_{ik}$ for the connection from unit $k$ to $i$ is calculated in the following way. Each conjunctive unit $i$ is temporarily assigned an auxiliary field, the location $(x_i,y_i)$ of which is randomly chosen among the place fields in the environment. Once the weights are fixed, these auxiliary fields have nothing to do with the conjunctive units any longer. The collateral weight between unit $k$ and $i$ is calculated as
\begin{equation}
W_{ik} =  [ f_{\theta_k}(\omega_{ki})f_{\theta_i}(\omega_{ki})\exp(-\frac{d_{ki}^2}{2\sigma_{f}^2}) -\kappa]^+,\label{eq-fix-weight}
\end{equation}
where $[\cdot]^+$ is a threshold function, with $[x]^+=0$ for $x<0$, and $[x]^+ = x$ otherwise. $\kappa=0.05$ is an inhibition parameter to favor sparse weights.  $f_{\theta}(\omega)$ is the HD tuning function defined in Eq.~\ref{eq-hd-tuning}. $\omega_{ki}$ is the direction of the line from field $k$ to $i$. $\sigma_{f}=10 cm$ is the width of spatial tuning. $d_{ki}=\sqrt{[x_{i} - (x_k+ \ell \cos{\omega_{ki}})]^2 + [y_{i} - (y_k+ \ell \sin{\omega_{ki}})]^2}$ is the distance between field $k$ and $i$ with offset $\ell=10 cm$. 

Normalization is performed on $ W_{ik}$ similarly as in Eq.\ref{eq-norm}. The resulting weight structure allows strong collateral interactions between units that have similar HDs, and meanwhile avoids co-activation, to produce grids with certain spatial shifts. 

In the following sections, we are going to show how, with the same set of fixed collateral weights and adaptable feed-forward weights, units in the network develop grid fields in single and multiple environments with various boundary shapes, and with different exploration behaviors of the virtual rat.

\section{Grid alignment in cylinder environments}\label{sec-sylind}
To test whether triangular grid fields can emerge and mutually align in our network model, we first simulate the network for a virtual rat randomly exploring a cylinder environment, where trajectories are completely isotropic. For simplicity, we assume that when the rat runs in a certain direction, it points its head toward the running direction (RD) most of the time. Therefore, in our simulations the head direction and the running direction are not distinguished.

Each step in the simulation is taken to correspond to $10 msec$  of real time. Each run of a simulation lasts for $8\times 10^6$ steps, or over $22 h$ real time. The standard speed of the rat is $v_s = 0.4 m/s$, towards the peak speed of real rats in active exploration, so that, considering the time a real rat spends not in exploration, a run of the simulation is intended to correspond to the developmental time scale for the emergence of grid cells \citep{Langston2010,Wills2010}. At each step, the running direction is chosen by using a pseudorandom Gaussian distribution with mean equal to the direction in the previous time step and an angular standard deviation $\sigma_{RD}=0.2$ radians. Very importantly, the new direction cannot lead the rat outside the limits of the environment. If so, the selection process is repeated until a valid direction is chosen. This has the effect of favoring trajectories along the walls of the environment. The overall appearance of the trajectory is comparable with those of actual rats, reflecting also the natural tendency of \naive animals to run along the walls of the enclosure.
The number of place units is $500$ and the number of conjunctive units is $250$. Place units are fully connected to conjunctive units. Each conjunctive unit receives connections from all other conjunctive units except itself. In the following simulations, we use the above mentioned default parameters, including the number of units in the network, the number of simulation steps, the speed and the noise in RD, unless stated differently.

\begin{figure*}
\begin{center}
\begin{minipage}{17.4cm}
\pspicture(0,0)(17.4,9)

\rput[bl](0,4.6){\epsfig{file=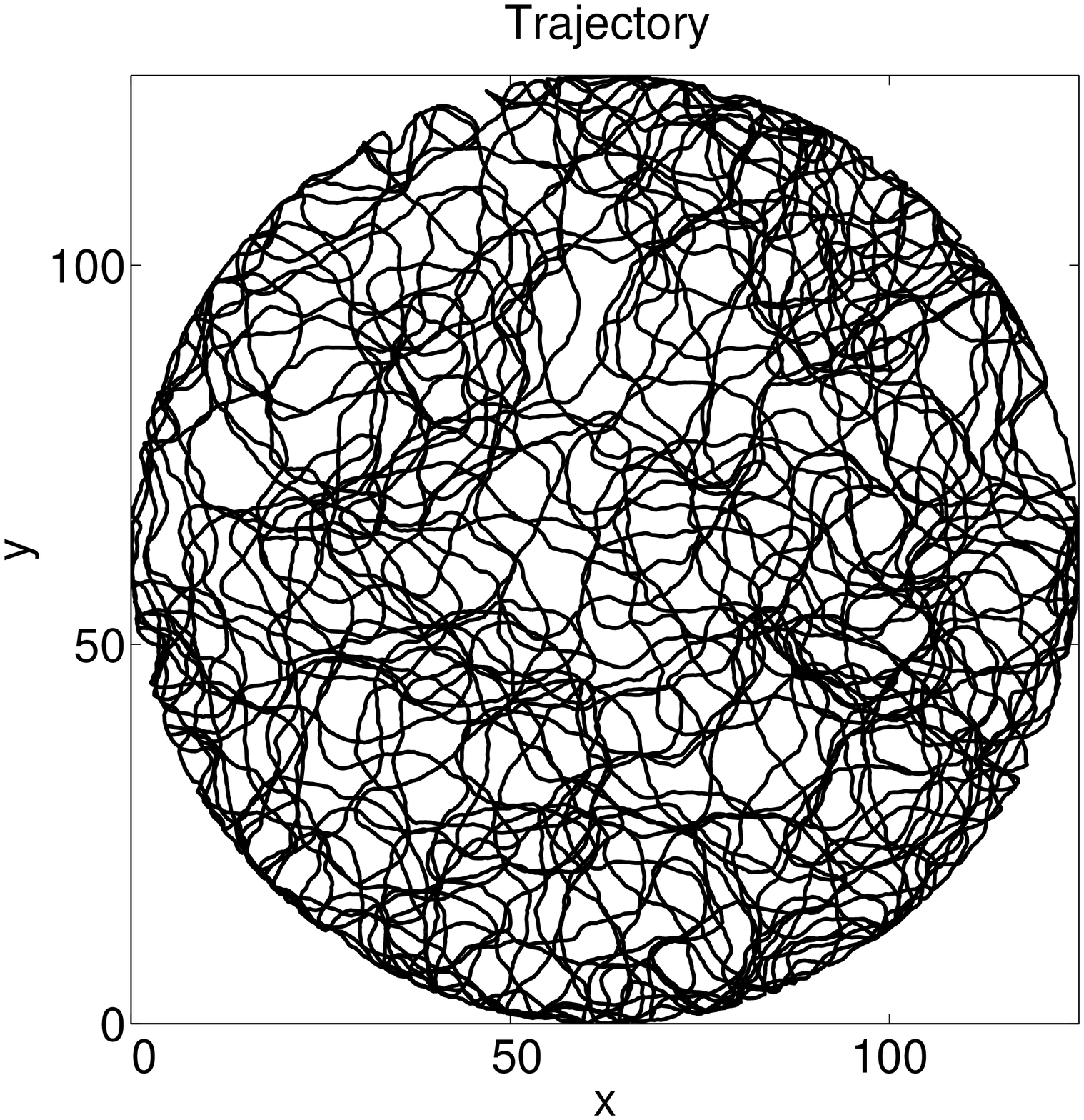,height=4.3cm}}
\rput[bl](5,4.6){\epsfig{file=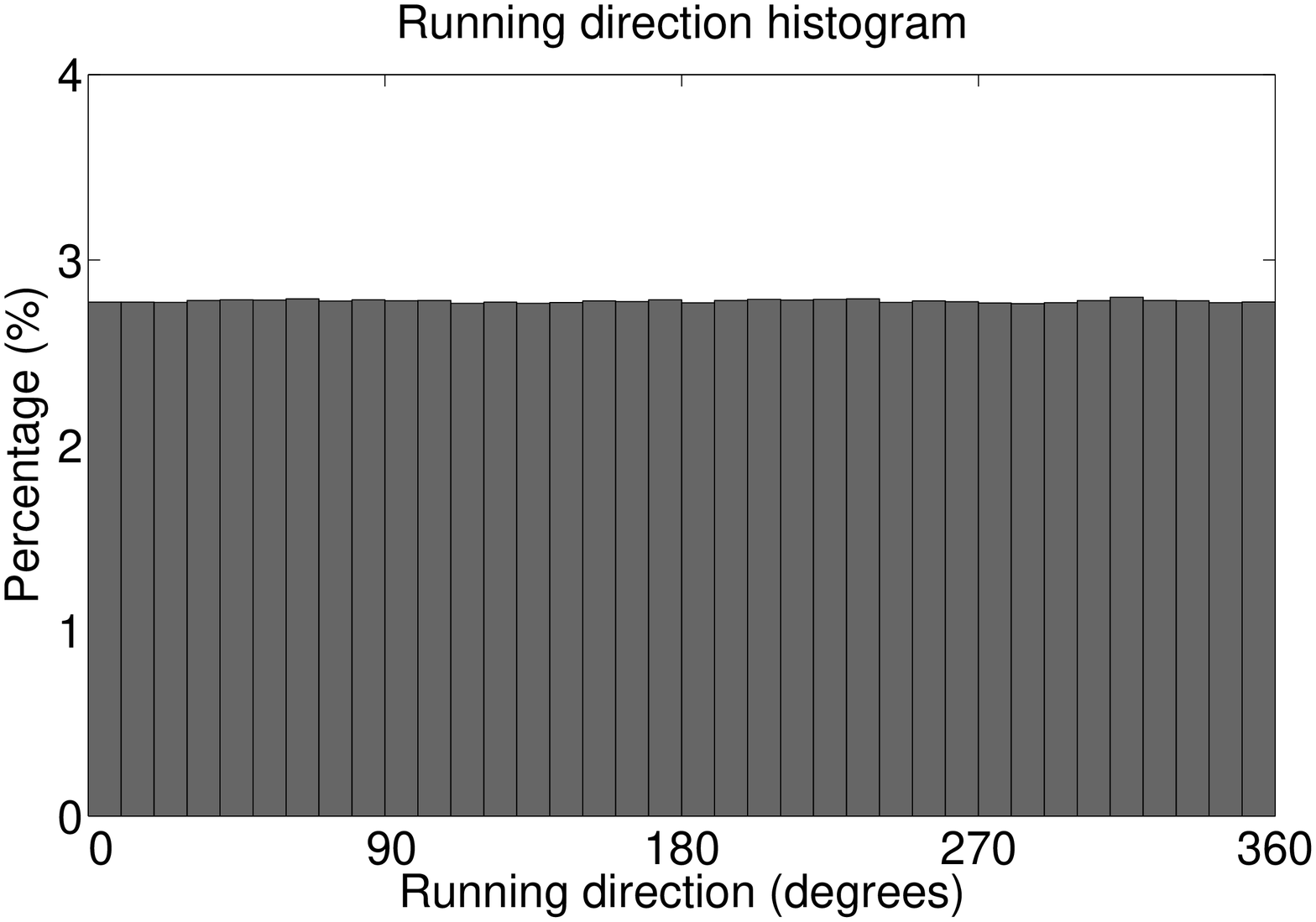,width=5.6cm}}
\rput[bl](11.2,4.6){\epsfig{file=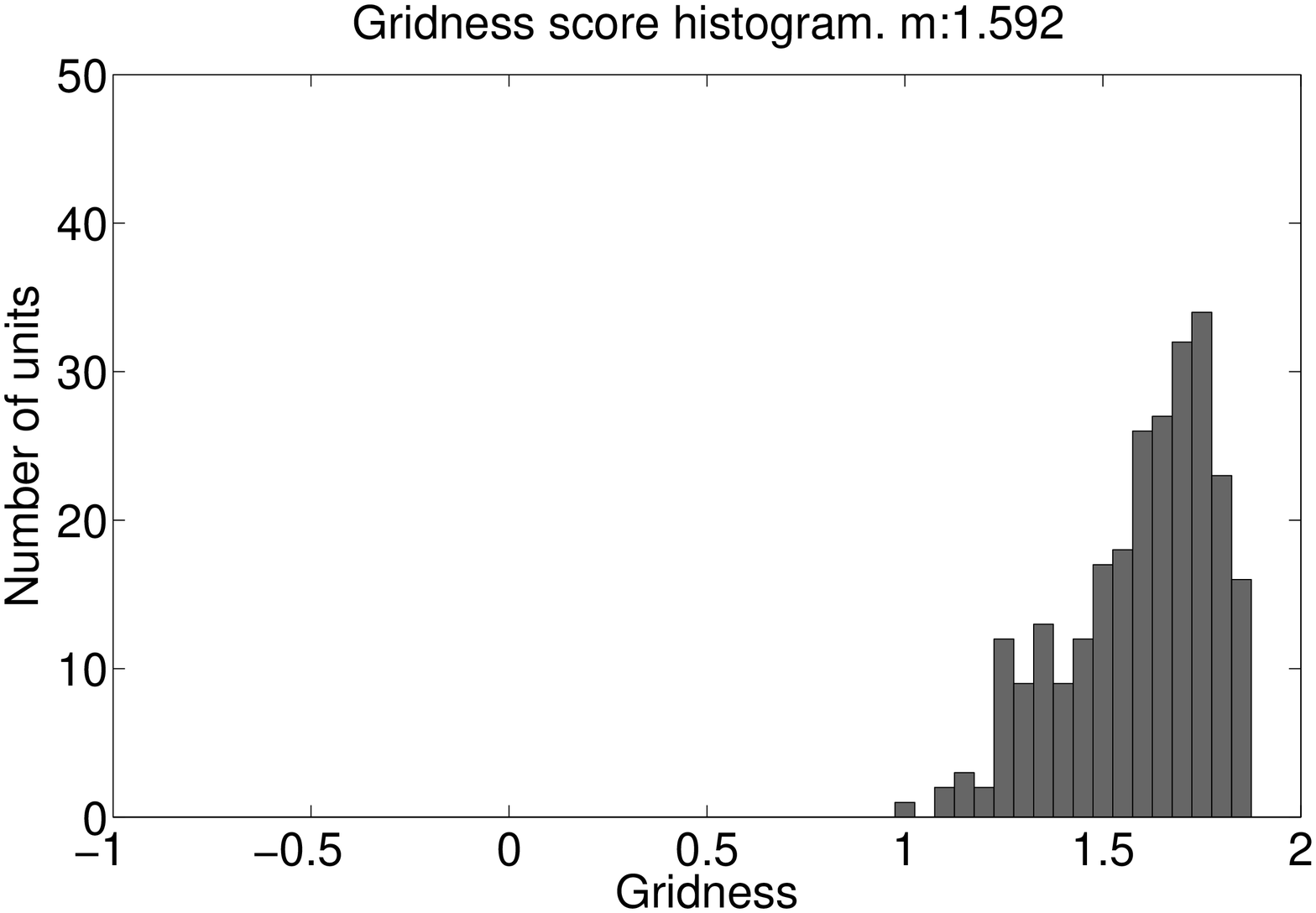,width=5.6cm}}

\rput[bl](0,0){\epsfig{file=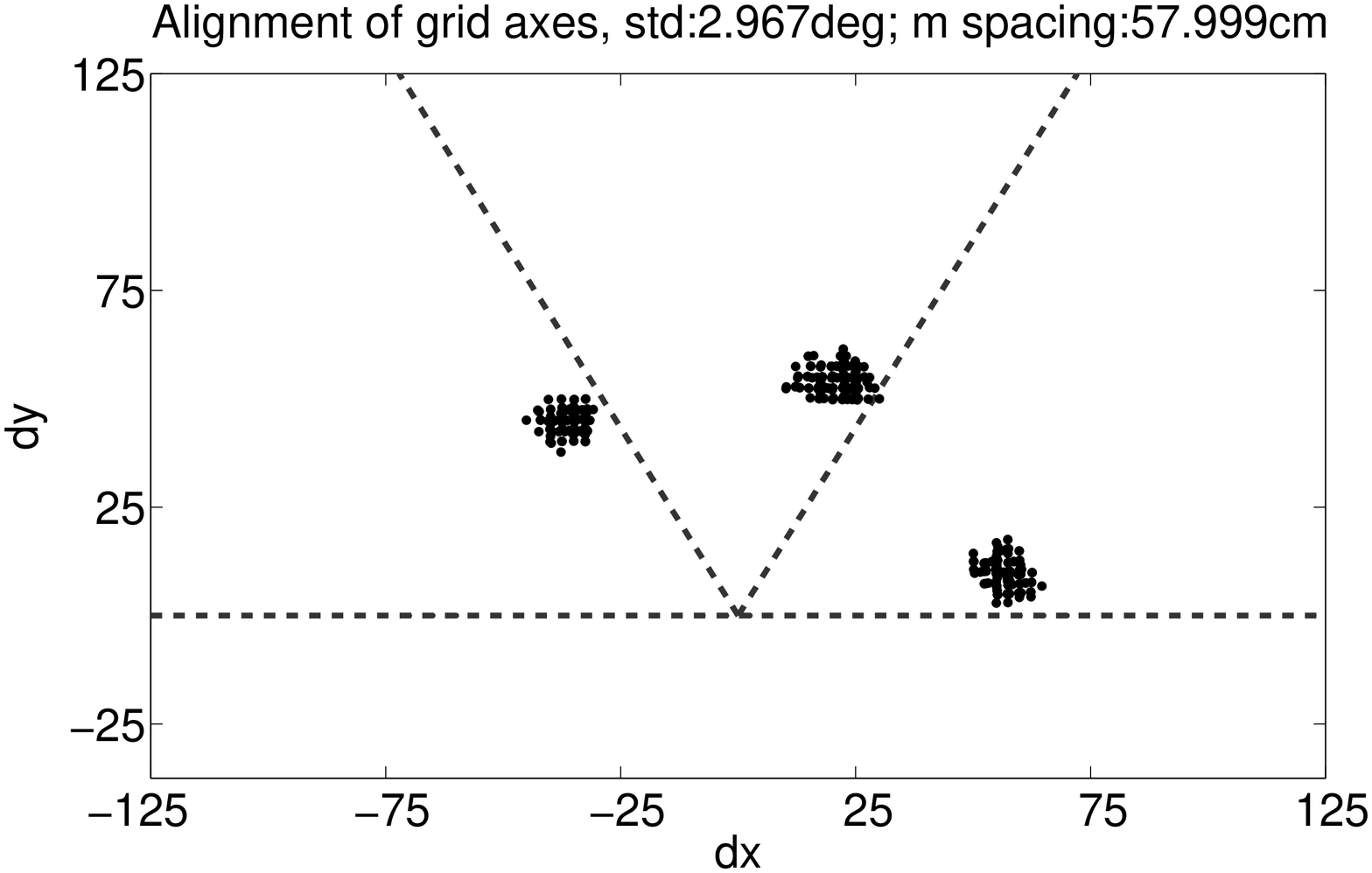,width=5.6cm}}
\rput[bl](5.8,0){\epsfig{file=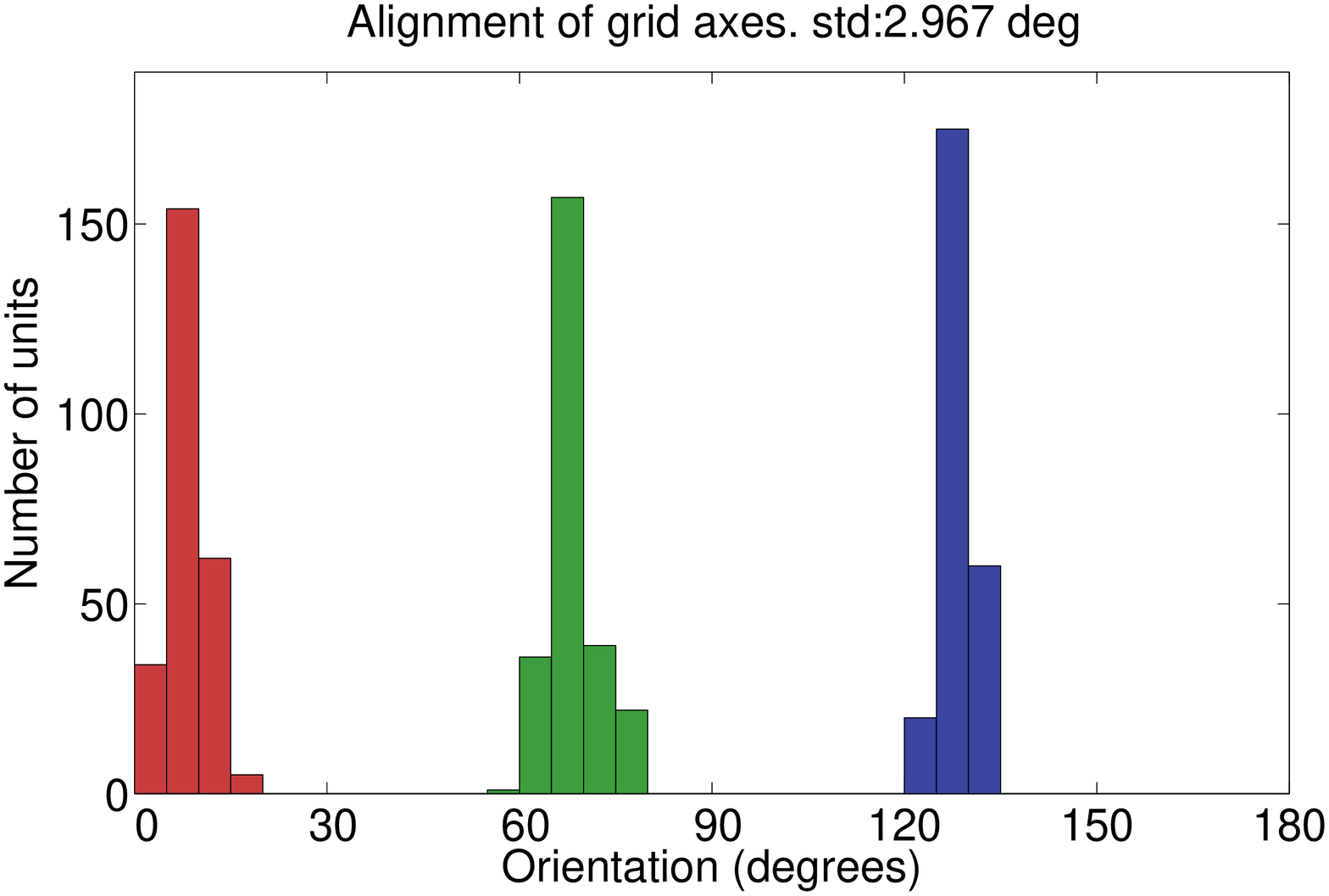,width=5.6cm}}
\rput[bl](12,0){\epsfig{file=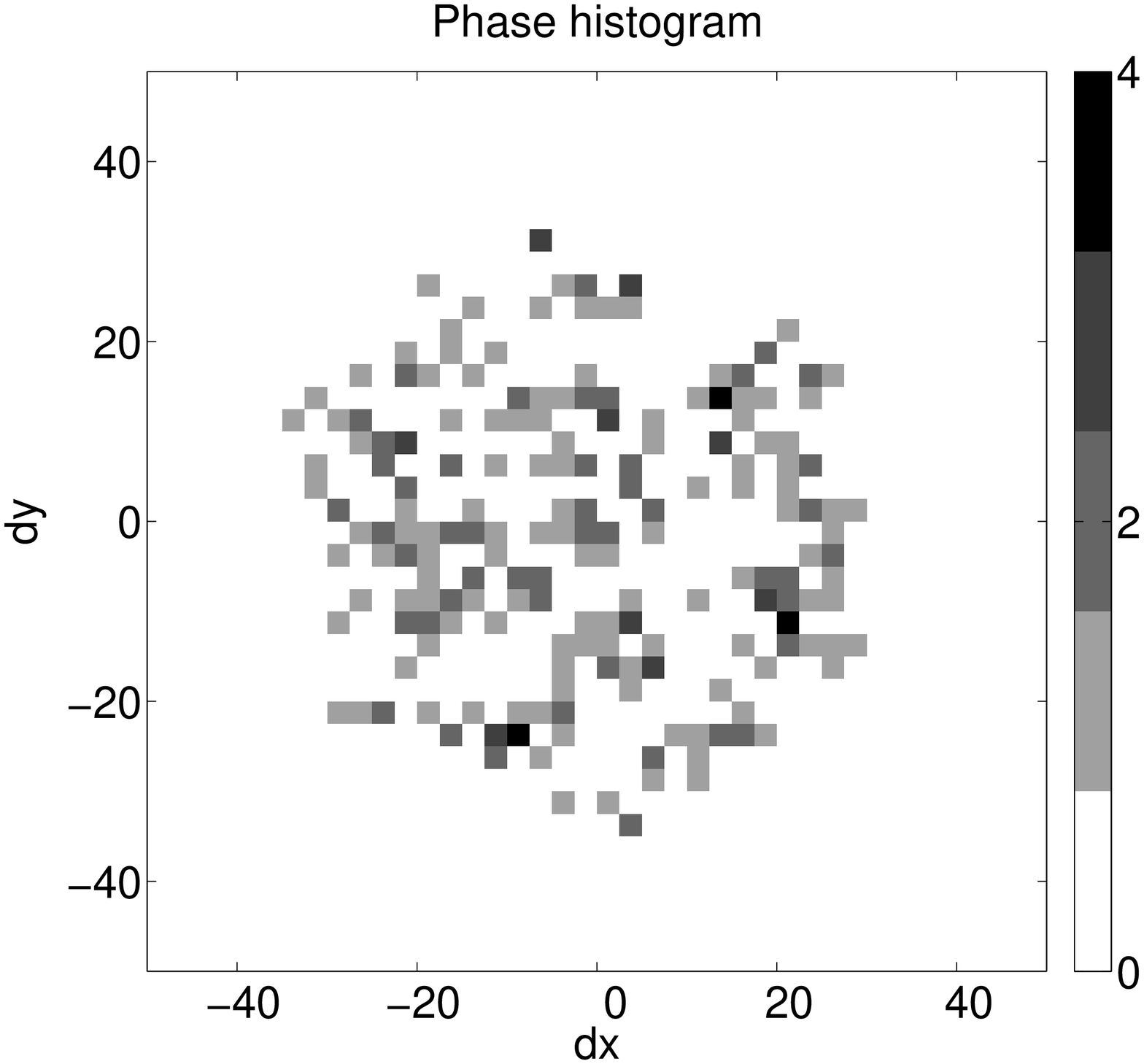,height=4cm}}

\rput[bl](0.4,8.6){\text{\small{\bf a}}}
\rput[bl](5,8.6){\text{\small{\bf b}}}
\rput[bl](11.2,8.6){\text{\small{\bf c}}}

\rput[bl](0.4,3.8){\text{\small{\bf d}}}
\rput[bl](6,3.8){\text{\small{\bf e}}}
\rput[bl](12.2,3.8){\text{\small{\bf f}}}
\endpspicture
\end{minipage}
\end{center}
\caption{Grid alignment in a cylinder environment. (a) A trajectory of the virtual rat with $3\times 10^4$ steps; (b) RD distribution of the trajectory in the simulation; (c) Histogram of the gridness scores of all conjunctive units in the simulation; (d) Scatter plot of the locations of the three peaks found in autocorrelograms (as shown by the white markers in Fig.~\ref{fig-field-circu}b); (e) Histograms of the angles of the three grid axes, plotted with different colors for each axis. The alignment coherence score, i.e. standard deviation (in degrees) of the orientation averaged over the three grid axes, is indicated at the top of the panel; (f) Two dimensional histogram of spatial phases relative to the best grid. The grayscale encodes the number of units, the phases of which fall into a $2.5 cm \times 2.5 cm$ spatial bin. White color represents zero, and darker colors represent larger numbers.}
\label{fig-circu}
\end{figure*}

In the first simulation covered in this section, the rat runs in a cylinder environment with a diameter of $125 cm$. Fig.~\ref{fig-circu}a plots part of a typical trajectory. The RD of the rat in this environment is uniformly distributed (Fig.~\ref{fig-circu}b), making the cylinder environment a suitable control condition to compare to anisotropy later. 

\begin{figure*}
\begin{center}
\pspicture(0,0)(10.5,14)

\rput[bl](0,0){\epsfig{file=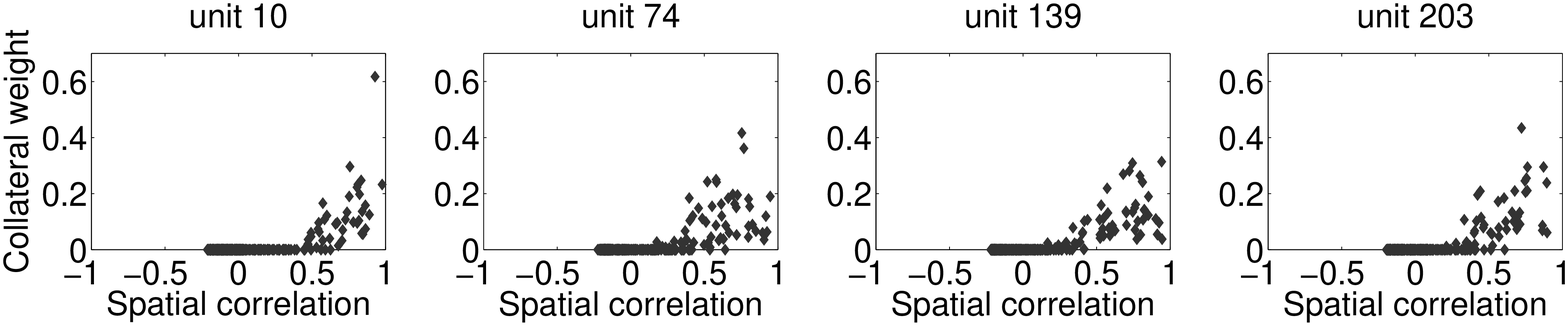,width=11cm}}
\rput[bl](0,2.5){\epsfig{file=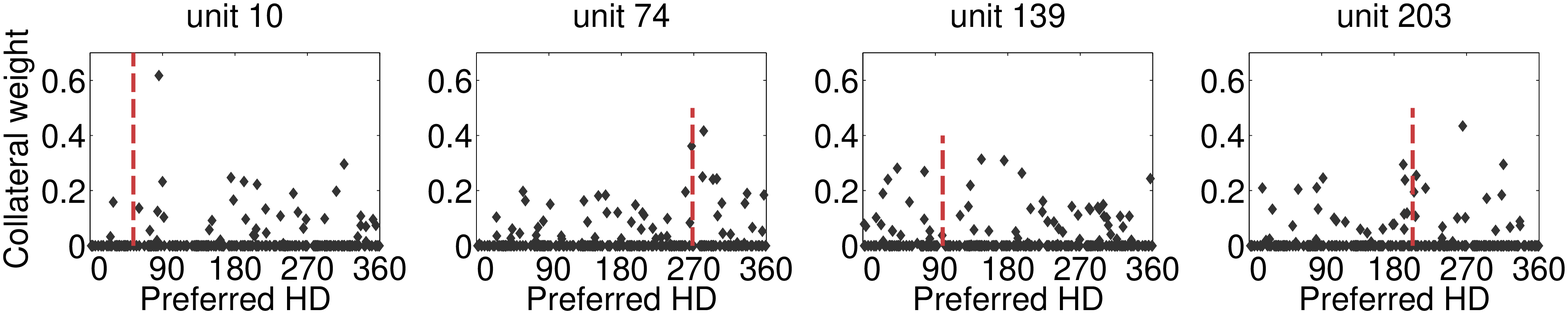,width=11cm}}
\rput[bl](0,5){\epsfig{file=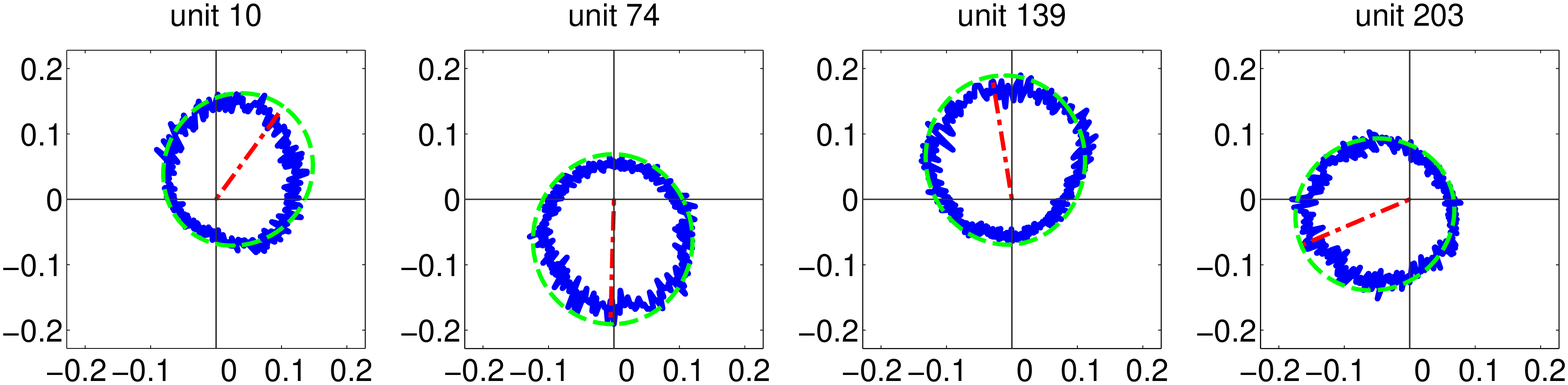,width=10.5cm}}
\rput[bl](0,8){\epsfig{file=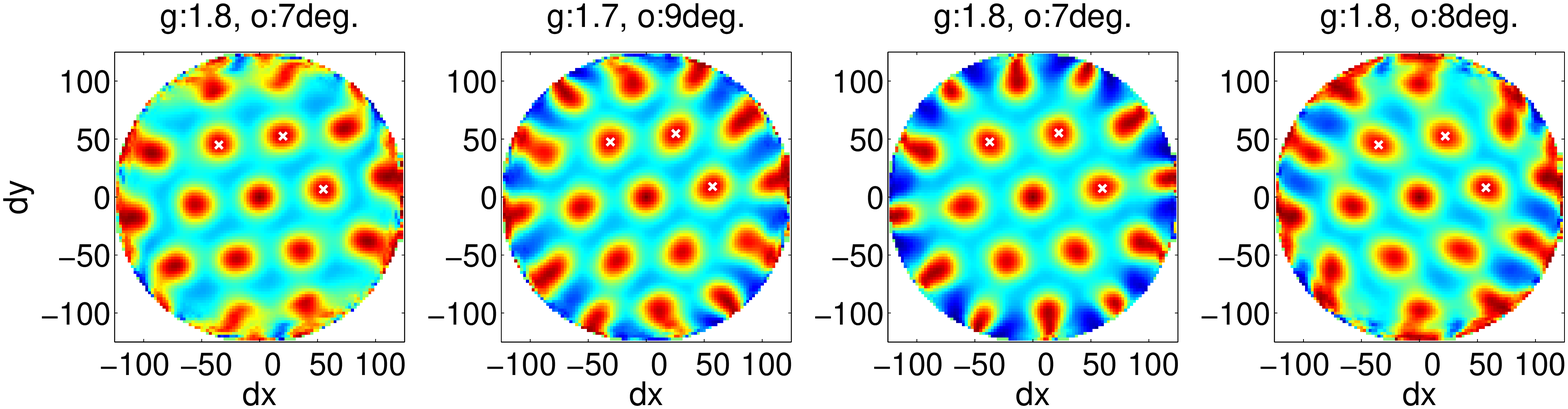,width=10.5cm}}
\rput[bl](0,11.1){\epsfig{file=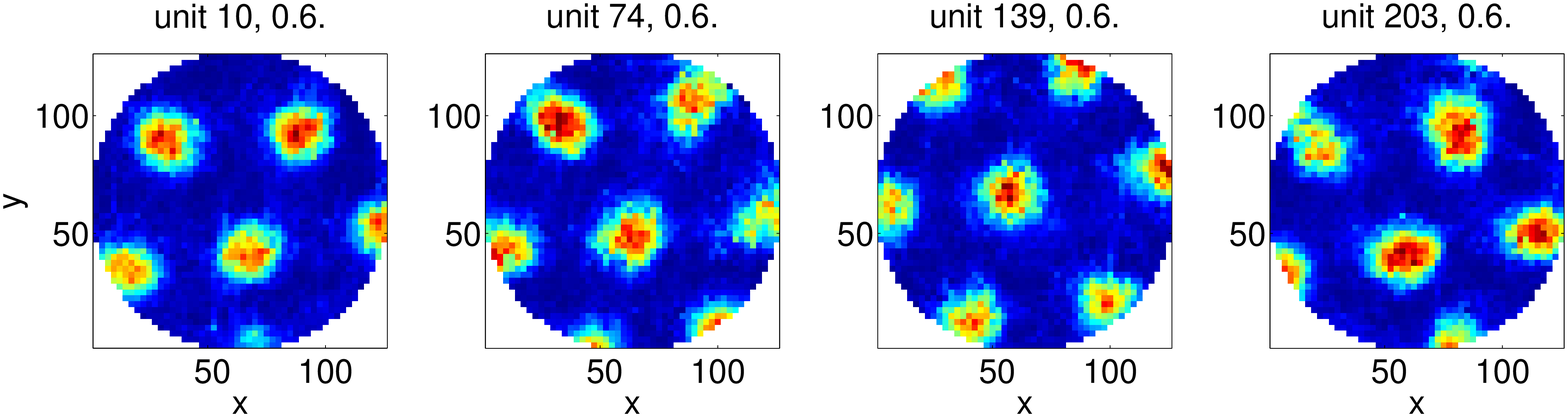,width=10.5cm}}

\rput[bl](-0.2,13.7){\text{\small{\bf a}}}
\rput[bl](-0.2,10.7){\text{\small{\bf b}}}
\rput[bl](-0.2,7.5){\text{\small{\bf c}}}
\rput[bl](-0.2,4.6){\text{\small{\bf d}}}
\rput[bl](-0.2,2){\text{\small{\bf e}}}
\endpspicture
\end{center}
\caption{Collateral connections align grid fields. (a) Spatial firing rate maps of example units in a cylinder environment. Unit number and maximal firing rate (in arbitrary units) are indicated above each rate map; (b) Autocorrelograms of the maps shown in a. Gridness score and grid orientation (in degrees) are indicated above each autocorrelogram; (c) Angular firing maps (blue solid lines) of the same units as in a with, in green, the tuning curve of the HD input to each unit shown. The red dash-dot lines indicate preferred HD of the units; (d) Collateral weight $W_{ik}$  as a function of $\theta_k$. The broken line is $\theta_i$; (e) Scatter plots between collateral weights and the spatial correlations of the fields between pre- and post-synaptic units.}
\label{fig-field-circu}
\end{figure*}

Fig.~\ref{fig-field-circu}a shows four examples of the spatial maps of the conjunctive units in the network. The multiple firing fields of the units locate on the vertices of triangular grids. To better reveal grid structure, the autocorrelograms of the spatial firing maps are calculated (Fig.~\ref{fig-field-circu}b). The \{x,y\} pixel in an autocorrelogram is the correlation between the original firing map and its shifted version  by the vector \{x,y\}. The peaks away from the center of the autocorrelogram indicate the directions and distances at which the shifted map has a high overlap with the unshifted one. The six maxima that are closer to the center of each autocorrelogram are then detected, and the three maxima with positive y coordinates (white markers in Fig.~\ref{fig-field-circu}b) define the orientations of the three grid axes (the autocorrelogram is redundant by definition, the lower part resulting from the rotation of the upper part by 180 degrees). The orientation of the first grid axis, i.e. the one with the lowest angle with respect to the x axis, is used as the orientation of a grid. The mean distance of the three maxima away from the center is defined as the spacing of a grid. The average spacing of the grids that self-organize in our network is about 58cm, comparable to the grid spacings observed in experiments~\citep{Hafting2005}. As noted previously, the spacing results from the parameters describing neuronal fatigue in the model~\citep{Kro+08}.

Averaged into angular bins, the angular map of the firing of each unit shows HD selectivity, matching the HD tuning curve of the unit (Fig.~\ref{fig-field-circu}c). This obviously results from the HD modulation on the inputs to conjunctive units (Eq.~\ref{eq-conjunc-affrent}-\ref{eq-hd-tuning}).

The spatial periodicity of conjunctive units is measured by the gridness score, as in~\citep{Sargolini2006}. We thus cut a ring area in each autocorrelogram with inner and outer radii chosen so as to contain the six maxima closest to the center and exclude the rest. We then correlate the original ring with its rotated versions, expecting to obtain for a perfect grid a very positive (negative) correlation coefficient for rotations at even (odd) multiples of 30 degrees. The gridness index is defined as the difference between the mean correlation for even rotations (60 and 120 degrees), and odd rotations (30, 90 and 150 degrees). Note that the gridness score is in the range [-2, 2].

The histogram of the gridness scores of all units is presented in Fig.~\ref{fig-circu}c. Most units have high gridness score. Fig.~\ref{fig-circu}d shows for all the 250 conjunctive units in the simulation the locations of the three maxima defining the grid axes. They clearly cluster into three clouds, indicating similar orientations and spacings for all units. The orientation histograms of the three grid axes are displayed in Fig.~\ref{fig-circu}e, one color for each. To quantify the coherence in aligning grids, we define a {\em cross-population alignment coherence score} by the standard deviation of the orientations averaged over the three grid axes, with low values indicating tight grid alignment. In the simulation shown in Fig.\ref{fig-circu}, the alignment coherence score is 2.967 degrees, indicating (very) tight grid alignment. 

While grids are aligned, the spatial phases of all grids (shown relative to the best grid, as a conventional reference) are distributed in an area with diameter similar to the grid spacing (Fig.~\ref{fig-circu}f). This means that grids cover the whole environment more or less randomly. 

The collateral weights are sparse and are stronger for units with similar HDs  (broken lines in Fig.~\ref{fig-field-circu}d), a relationship given by Eq.~\ref{eq-fix-weight}. When the connectivity of collateral connections is reduced to 50\%, the network is still able to align grids (data not shown). Therefore, the exact strength of the collateral connectivity is not critical for the model. Collateral weights are strongest between units that are active in locations a small distance apart and that are moderately correlated in firing, due to the delay parameter $\tau$ in their inputs (Fig.~\ref{fig-field-circu}e). The network is not too sensitive to $\tau$ as long as $\tau \geq 13$ steps, i.e. $130 msec$ in real time. This is the time it takes the virtual rat to travel the minimal distance between two input place fields. Smaller delays do not change the gridness scores or the alignment of grid maps, but cause the spatial phases of the grid fields to collapse, showing a non-random phase distribution (data not shown). The phase collapse due to small $\tau$ can be understood by the associative learning between conjunctive units and place units. Small $\tau$ would cause two strongly connected conjunctive units to fire consecutively in short intervals, during which the place units show similar population activity. Associative learning therefore would produce similar grid fields for both units. To avoid that, the delay between two strongly connected conjunctive units needs to be long enough so that grid fields are able to anchor to different locations, resulting in random spatial phases. The exact mechanism that generates this delay is something that our simple rate-based model does not attempt to describe, and it could go from several types of slow synaptic potentials to interactions with rhythms, such as phase precession.  

In summary, with fixed collateral weights, the grid fields of model units align to a common orientation while the relative spatial phases between them are randomly distributed in the environment. The firing of each unit is conjunctively modulated by a spatial triangular grid and by its HD input.

\subsection{Grid development with speed variation}\label{sec-acceleration}

Constant running speed is not a requirement of the model. To check this,  we have simulated a rat with variable speeds, and left unchanged the method of choosing running directions. The trajectory of the simulated rat is then composed of epochs with positive or negative accelerations. The lengths of epochs are Poisson random numbers with mean length 3, roughly matching  available behavioral data~\citep{Sargolini2006}.  The speed at the end of each epoch is drawn from a truncated Gaussian distribution. A two-sided truncation is applied to keep the speed both positive and symmetrically distributed around the mean. The mean of the truncated Gaussian distribution is the same as the constant speed used in most simulations in this paper, i.e. 0.4m/s. The speed within each epoch is a linear interpolation between the speeds at the start and end of the epoch.

\begin{figure*}%
\centering
\begin{minipage}{16.4cm}
\pspicture(0,0)(16.4,12.5)
\rput[bl](0,9.4){\epsfig{file=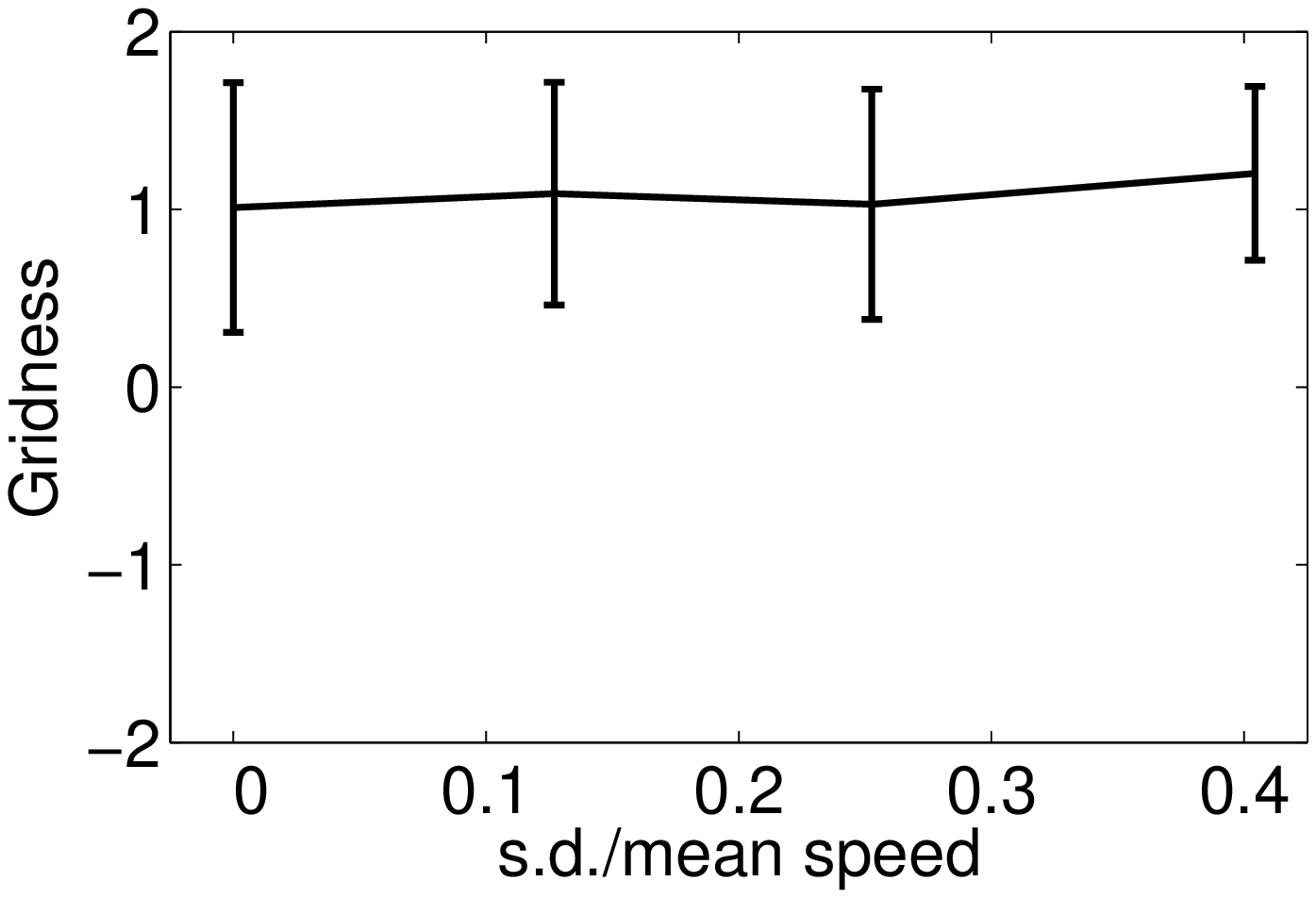,width=4.4cm}}
\rput[bl](0,6.3){\epsfig{file=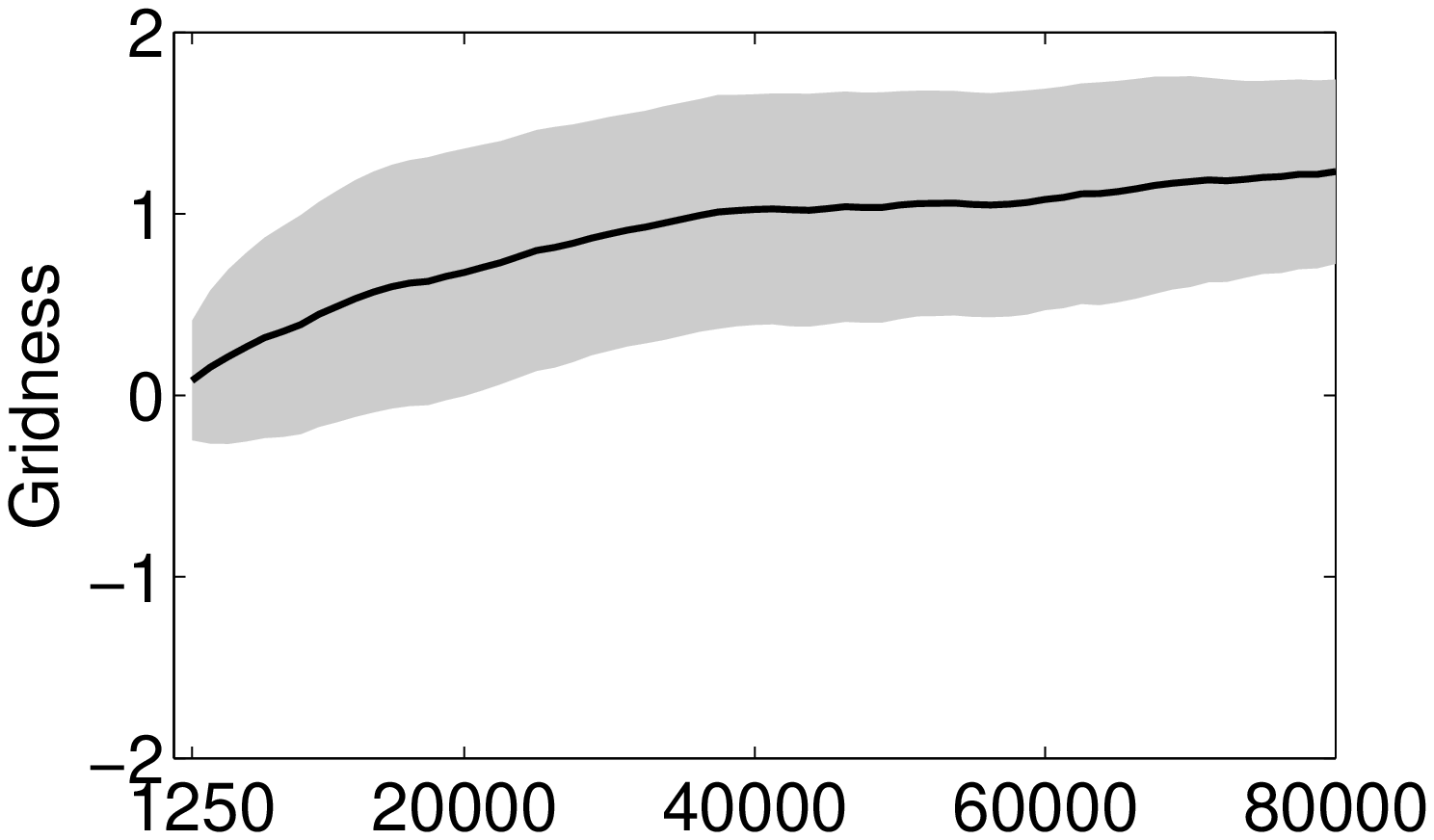,width=4.5cm}}
\rput[bl](0,3.2){\epsfig{file=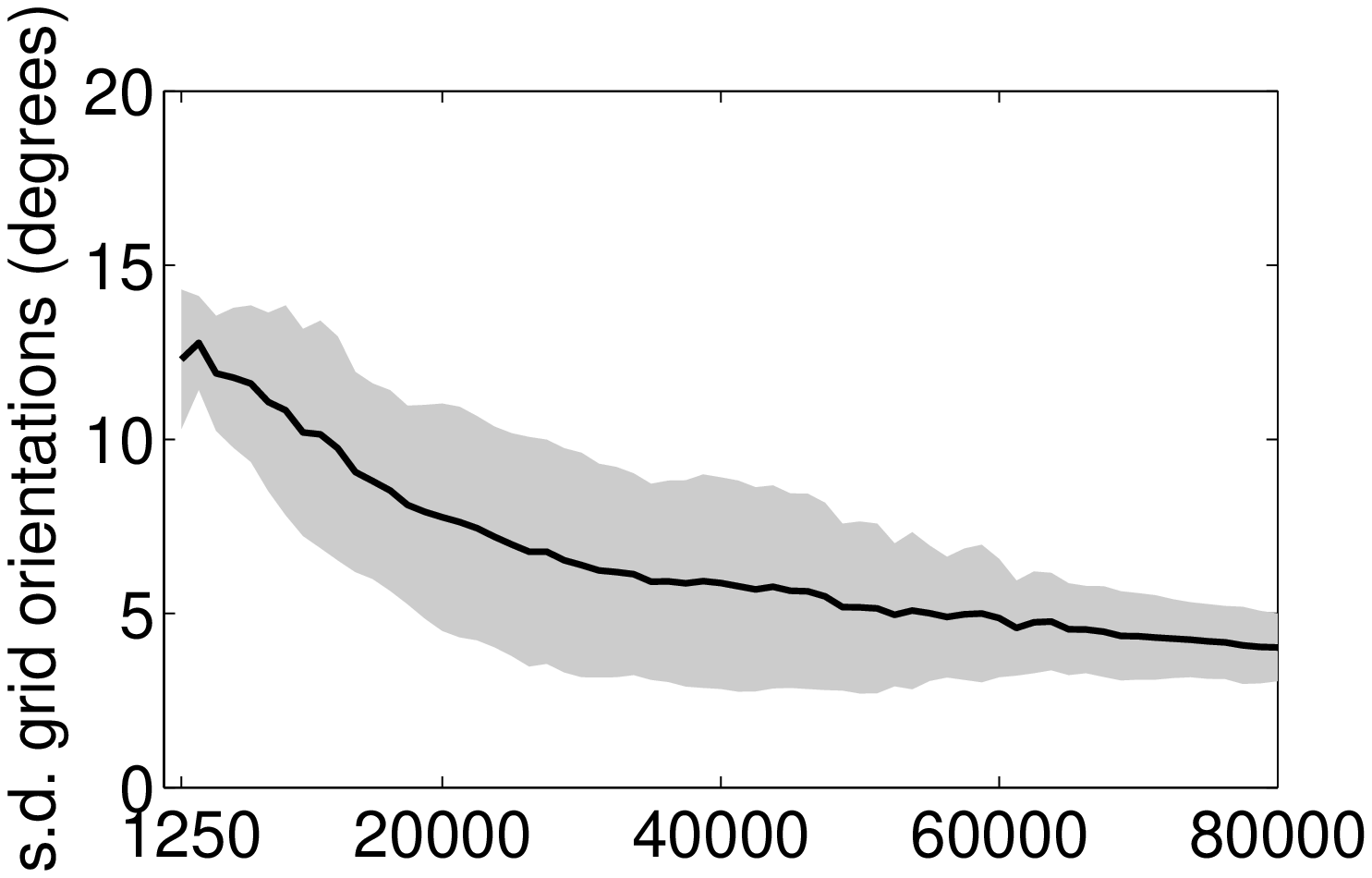,width=4.5cm}}
\rput[bl](0,0){\epsfig{file=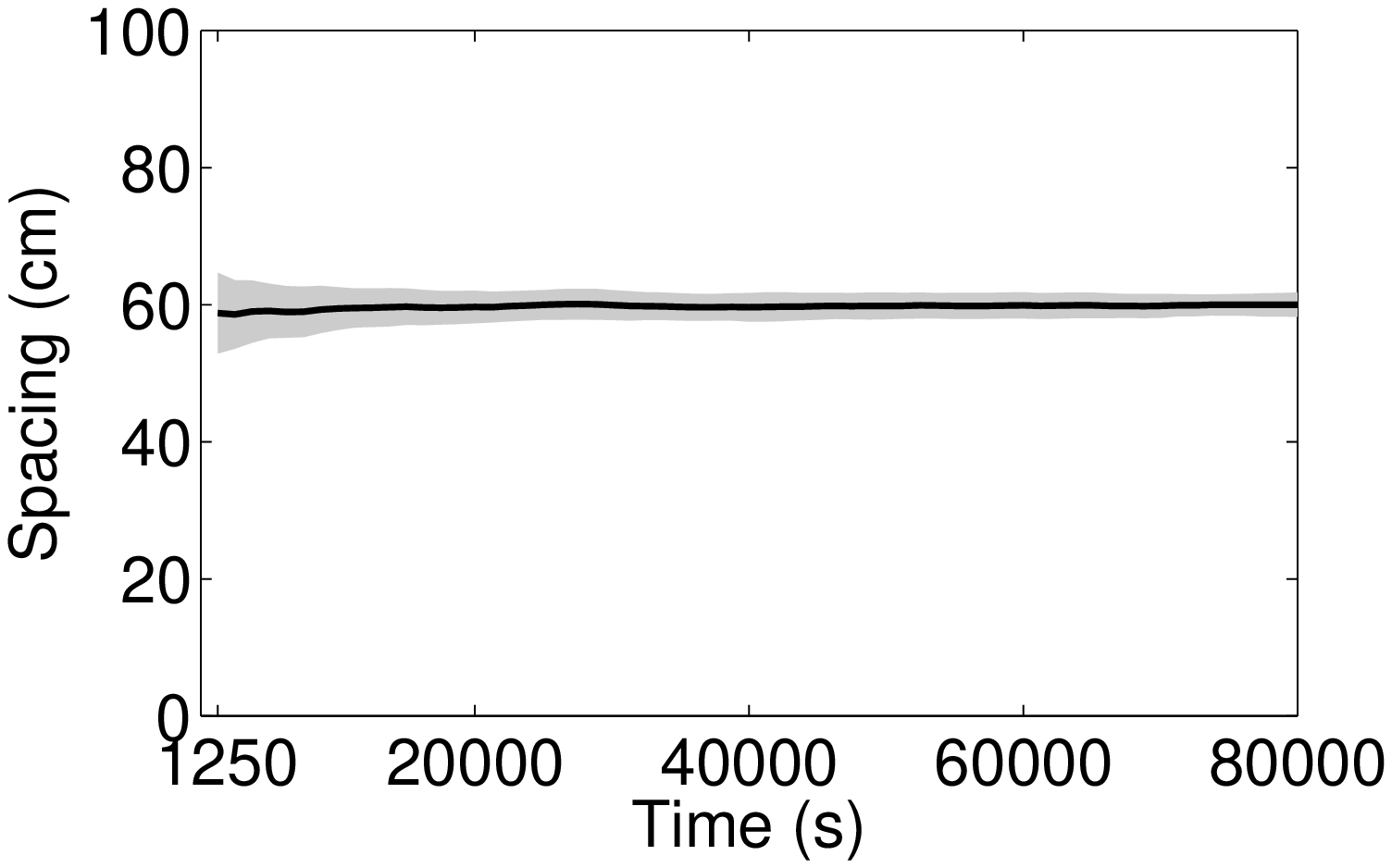,width=4.5cm}}
\rput[bl](4.6,0){\epsfig{file=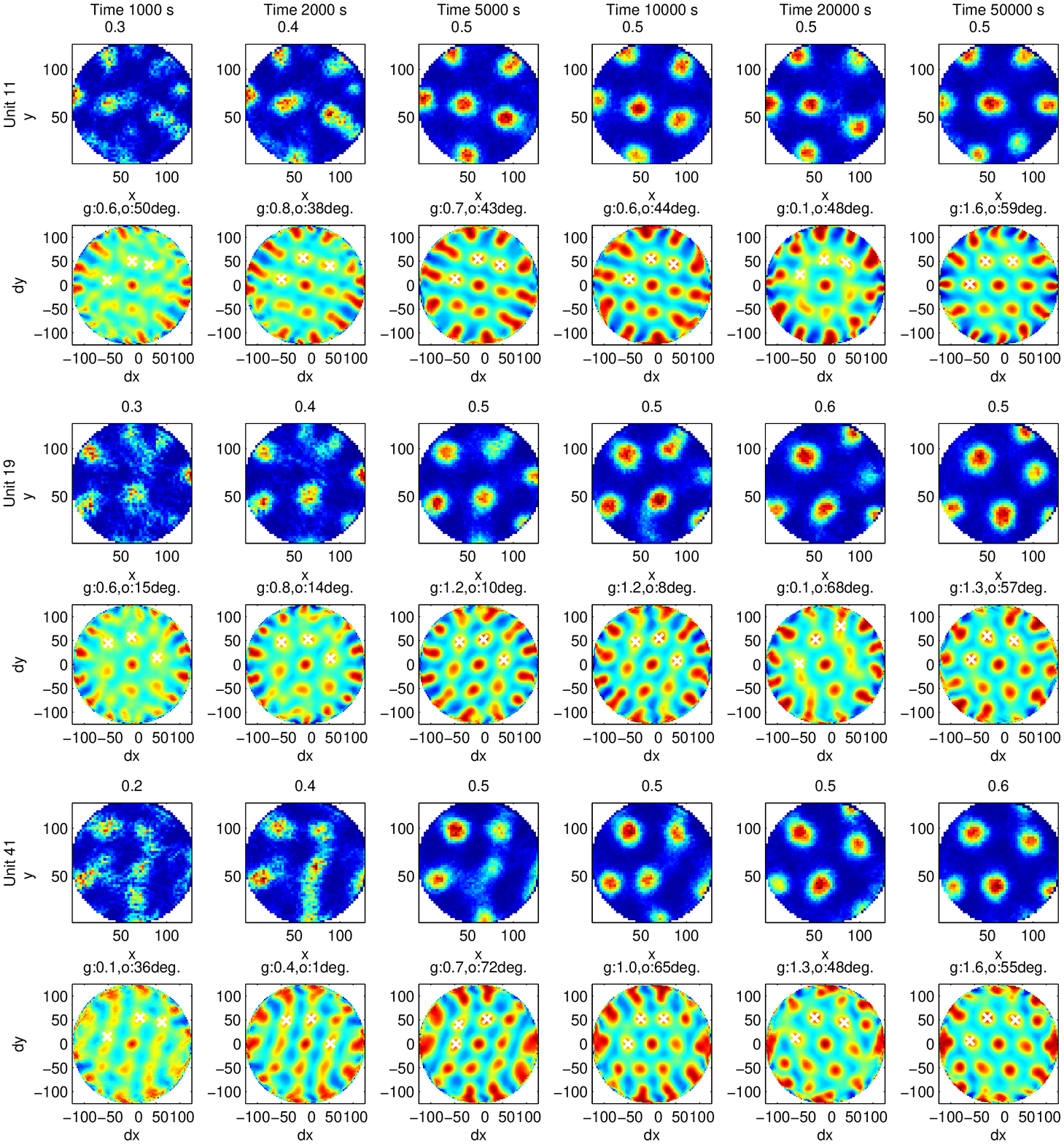,height=12.5cm}}

\rput[bl](0,12.4){\text{\small{\bf a}}}
\rput[bl](0,8.9){\text{\small{\bf c}}}
\rput[bl](0,6.1){\text{\small{\bf d}}}
\rput[bl](0,2.7){\text{\small{\bf e}}}
\rput[bl](4.6,12.4){\text{\small{\bf b}}}

\endpspicture
\end{minipage}
\caption{The variation in speed does not influence grid formation.  (a) Grids show similar mean gridness scores at the end of simulations performed with different speed standard deviation relative to mean speed. Error bars indicate $\pm$ standard deviation across 5 simulations;  (b) Grid fields appear in the early phases of the simulation and stabilize with more experience in the environment. Grid fields and the corresponding autocorrelograms are ordered in rows for three example units from the same network. Maps formed after the same amount of exploration are arranged in the same column;  (c) Mean gridness increases with respect to the amount of exploration (speed standard deviation is 40\% relative to the mean speed). Gray area is the $\pm$ standard deviation; (d) Standard deviation of grid orientations; (e) Mean spacing of the grids with gridness larger than 0.25. }
\label{fig-development}
\end{figure*}

Regardless of different levels of speed standard deviation, grid fields form with high gridness scores (Fig.~\ref{fig-development}a), tight grid alignment and similar spacings (data not shown). Even for large speed standard deviation, 16.1 cm/s, i.e. 40\% relative to the mean speed, the changes in mean gridness,  mean standard deviation of grid orientations and mean gridness are all within $\pm$ one standard deviation error bar when averaged across simulations.

\subsection{Gradual grid development}\label{sec-develop}

Although grid formation is a gradual process, grids are expressed at relatively early stages of development~\citep{Langston2010,Wills2010}. In adult rats, grids appear after a first exposure to a novel environment, but need several days of experience to become stable~\citep{Bar+09}. Consistent with these findings, conjunctive units in our network also show early expression of grids, within a gradual formation process. As shown in Fig.~\ref{fig-development}b, grids can be observed already 20 minutes after the simulated rat has started to explore the environment. With longer exploration, the grid fields become both more triangular and more coherently aligned to a common orientation, as quantified by the increasing mean gridness and the decreasing mean standard deviations of grid orientations (Fig.~\ref{fig-development}c,d). Grids stabilize after about 14 hours of continuous exploration. The mean spacing of the grids does not show big changes during development (Fig.~\ref{fig-development}e). Both the time scales of early grid appearance and of grid stabilization at the population level are comparable with experimental results, considering the time spent by a real rat in rest and sleep.

\section{The effect of the shape of the environment}\label{sec-box}

As shown in the previous section, grid fields align mutually to each other in cylinder environments. As cylinder has no preferred direction, and the common alignment emerging in one simulation bears no relation with the one emerging in another, leading to the expectation that different animals would show differently aligned grid units, if these were to form prevalently in cylinder environments. Rodent cages are usually rectangular, however, and a rectangle does have preferred directions. Is the orientation of grid units in different rats influenced by the shape of the training environment?  The current section is devoted to this issue, namely, to the coherence of {\it grid orientation} across rats, i.e., in our case, across simulations. In contrast to the standard simulations in the previous section, we simulate virtual rats in a 125 cm $\times$ 125 cm square environment, which leads to a (simulated) anisotropic exploration. We also vary the standard deviation $\sigma_{RD}$ in running directions, which affects the degree of behavioral anisotropy.

\begin{figure*}
\centering
\pspicture(0,0)(10.2,9.2)

\rput[bl](0,4.7){\epsfig{file=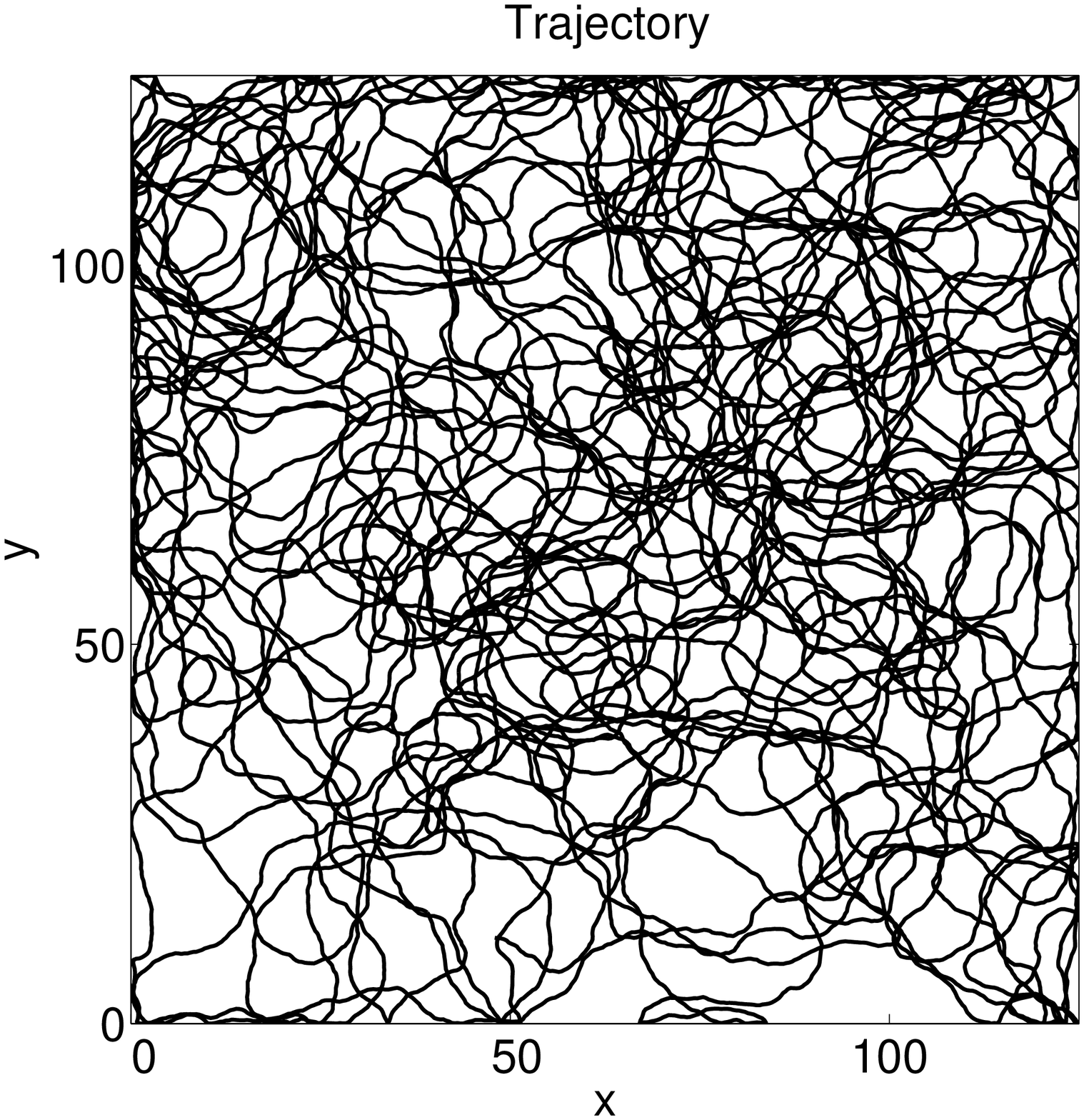,height=4.3cm}}
\rput[bl](4.5,4.7){\epsfig{file=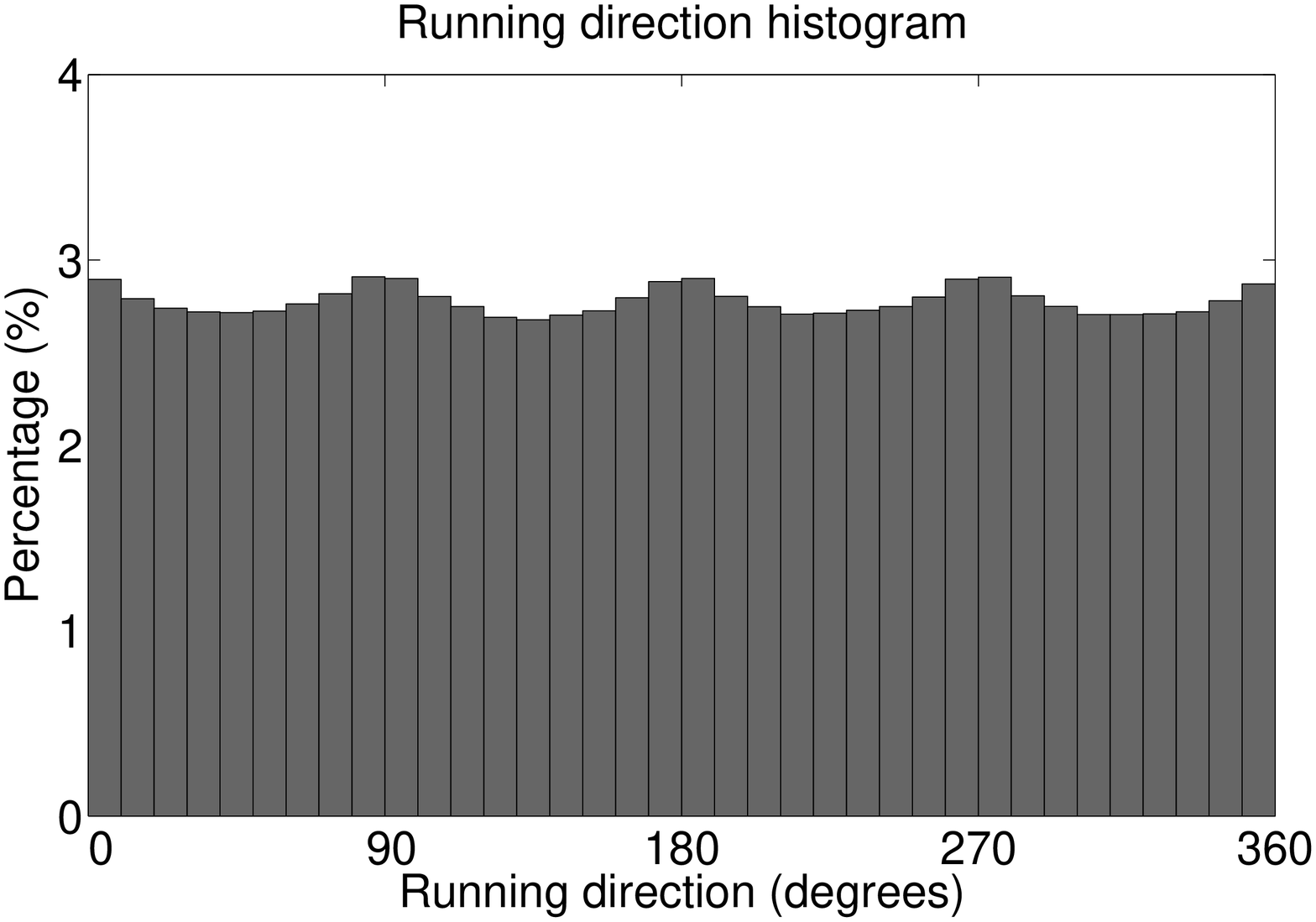,width=5.6cm}}
\rput[bl](0,0){\epsfig{file=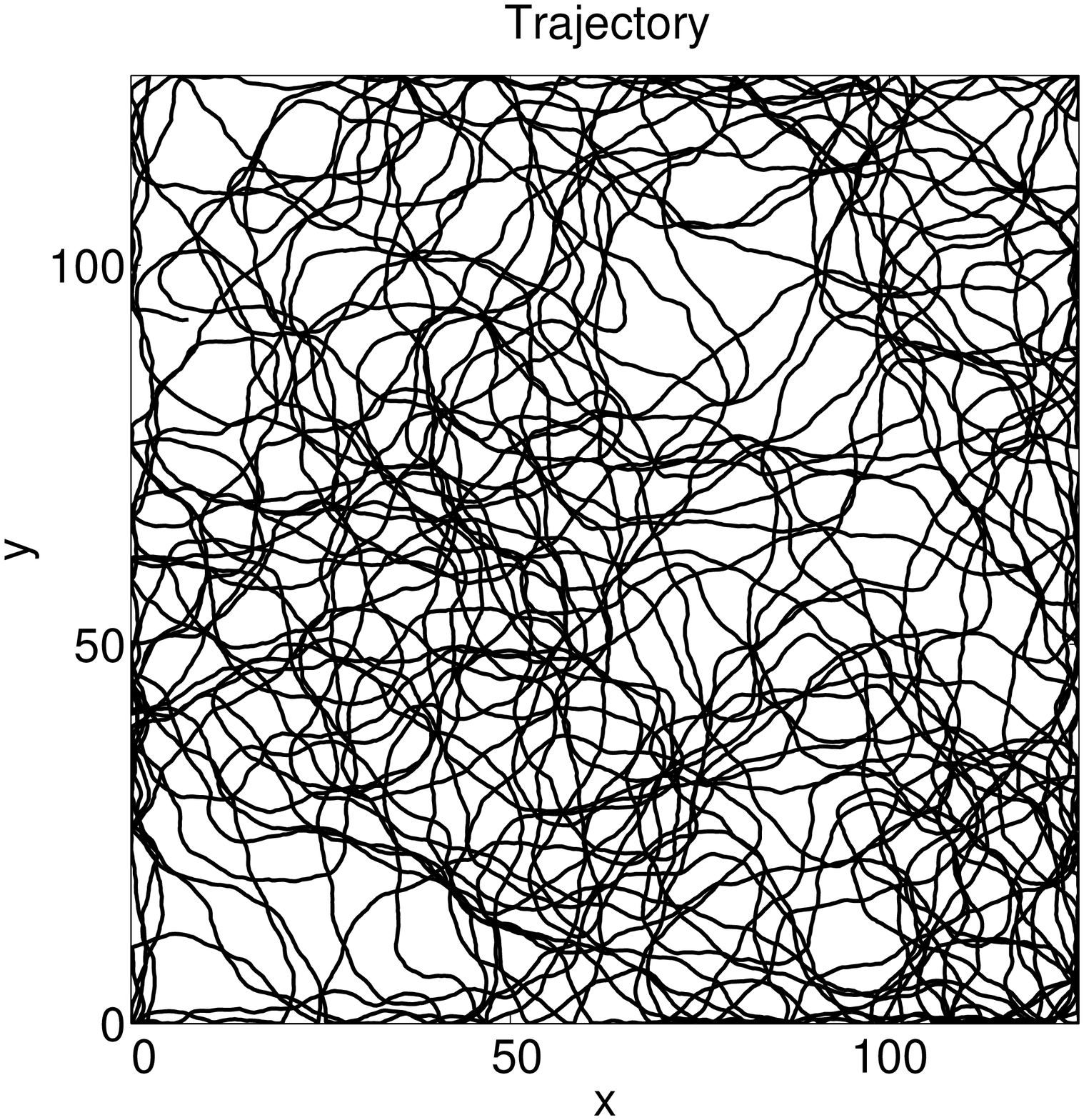,height=4.3cm}}
\rput[bl](4.55,0){\epsfig{file=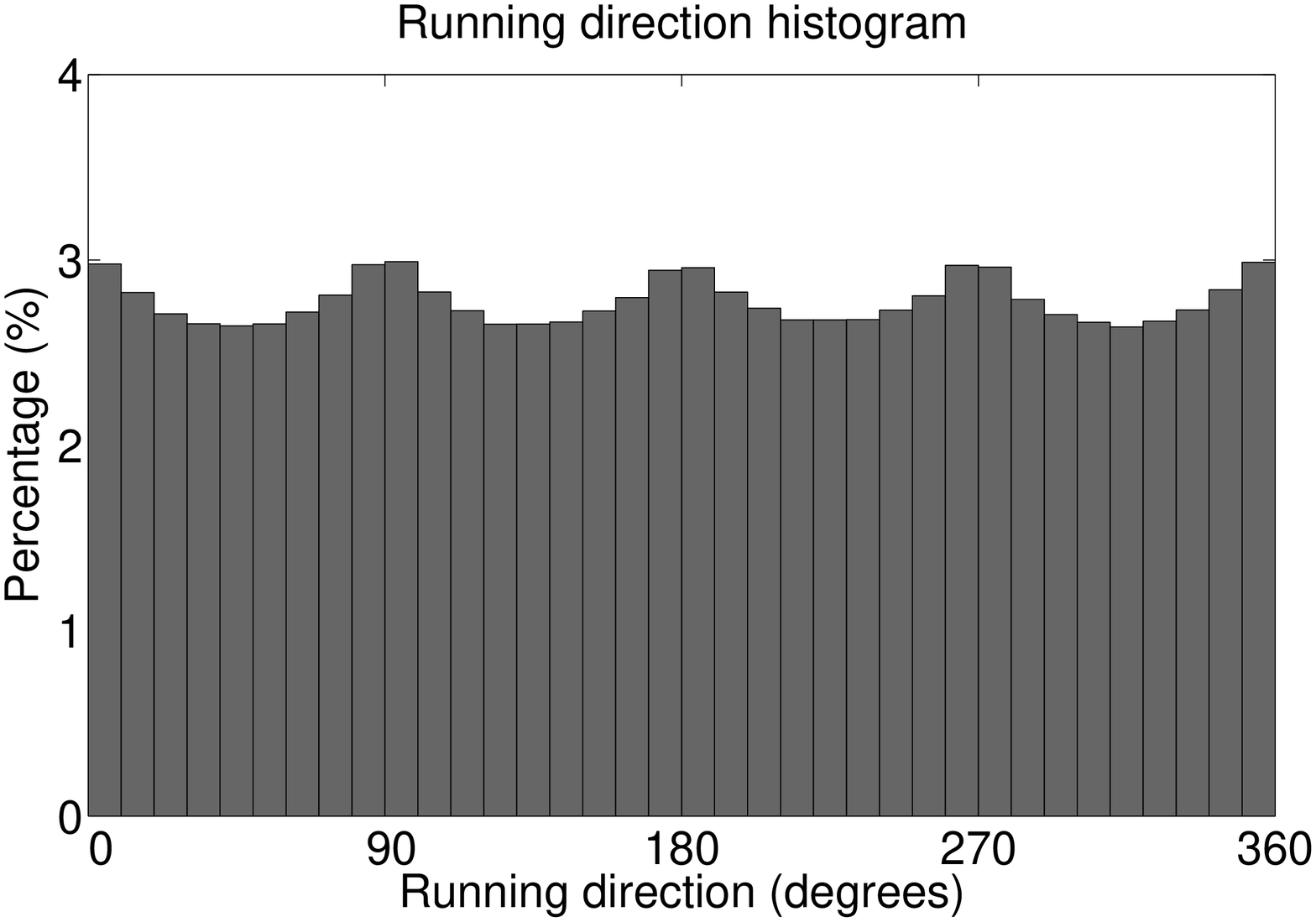,width=5.6cm}}

\psframe[linewidth=3pt,linecolor=cyan](-0.1,4.6)(10.2,9)
\psframe[linewidth=3pt,linecolor=blue](-0.1,-0.1)(10.2,4.3)

\rput[bl](0.2,8.6){\text{\small{\bf a}}}
\rput[bl](0.2,3.9){\text{\small{\bf b}}}
\endpspicture
\caption{Square environments induce more running along the sides. The left panels show parts of typical trajectories. The corresponding distributions of RDs in simulations are depicted in the right panels. (a) The default standard deviation in RD, $\sigma_{RD} =0.2$ radians; (b) A simulation with $\sigma_{RD}=0.15$ radians, demonstrating stronger anisotropy in the trajectories.}
\label{fig-box-path}
\end{figure*}

Rats, and especially \naive ones, have a natural tendency to run along the walls of the environment. Therefore their trajectories would reflect the anisotropy of the enclosing perimeter. Our virtual rat in a square box has a probability higher than chance of following the walls, simply because of the random walk mechanism described earlier. While in the cylinder environment the trajectory is necessarily isotropic, following the walls in a square box will result in the emergence of four preferred running directions (Fig.~\ref{fig-box-path} left column). The RD distributions for square environments exhibit four peaks centered around the directions parallel to the walls (Fig.~\ref{fig-box-path} right column). In contrast, in a cylinder environment, the RD distribution is uniform (Fig.~\ref{fig-circu}b). The non-uniformity of the RD distribution in square environments is stronger when the standard deviation in RDs is smaller (Fig.~\ref{fig-box-path}b).

In an environment with anisotropic boundary, if one running direction is preferred systematically, the network would associate the activity of conjunctive units more strongly to places following the preferred RD, and would effectively orient one of the grid axes along this direction, forcing the other two grid axes to follow. We would expect to see coherent grid orientation across rats in such conditions.

\begin{figure*}
\centering
\begin{minipage}{17.4cm}
\pspicture(0,0)(17.4,8.6)

\rput[bl](0.1,4.2){\epsfig{file=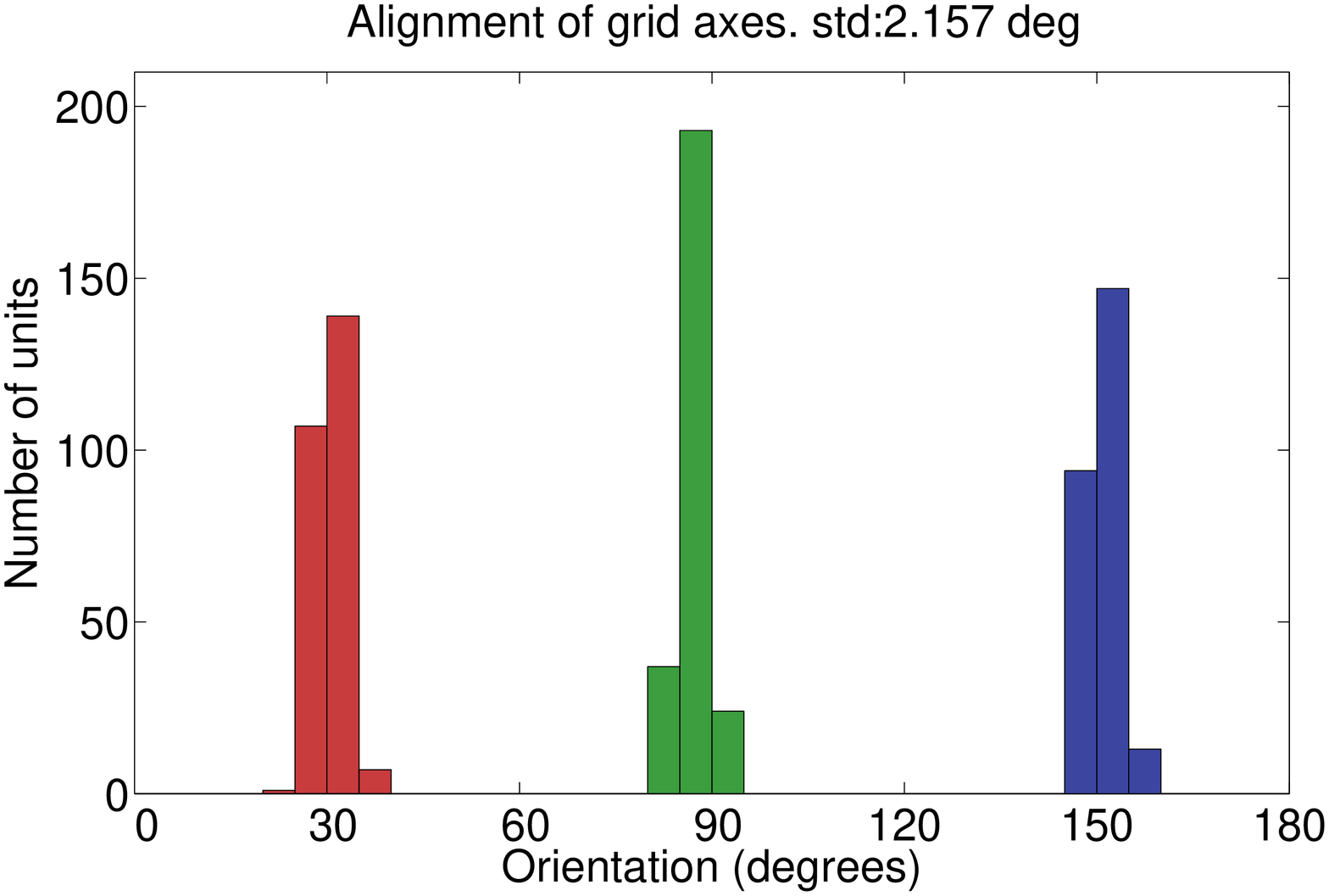,width=5.6cm}}
\rput[bl](6,4.2){\epsfig{file=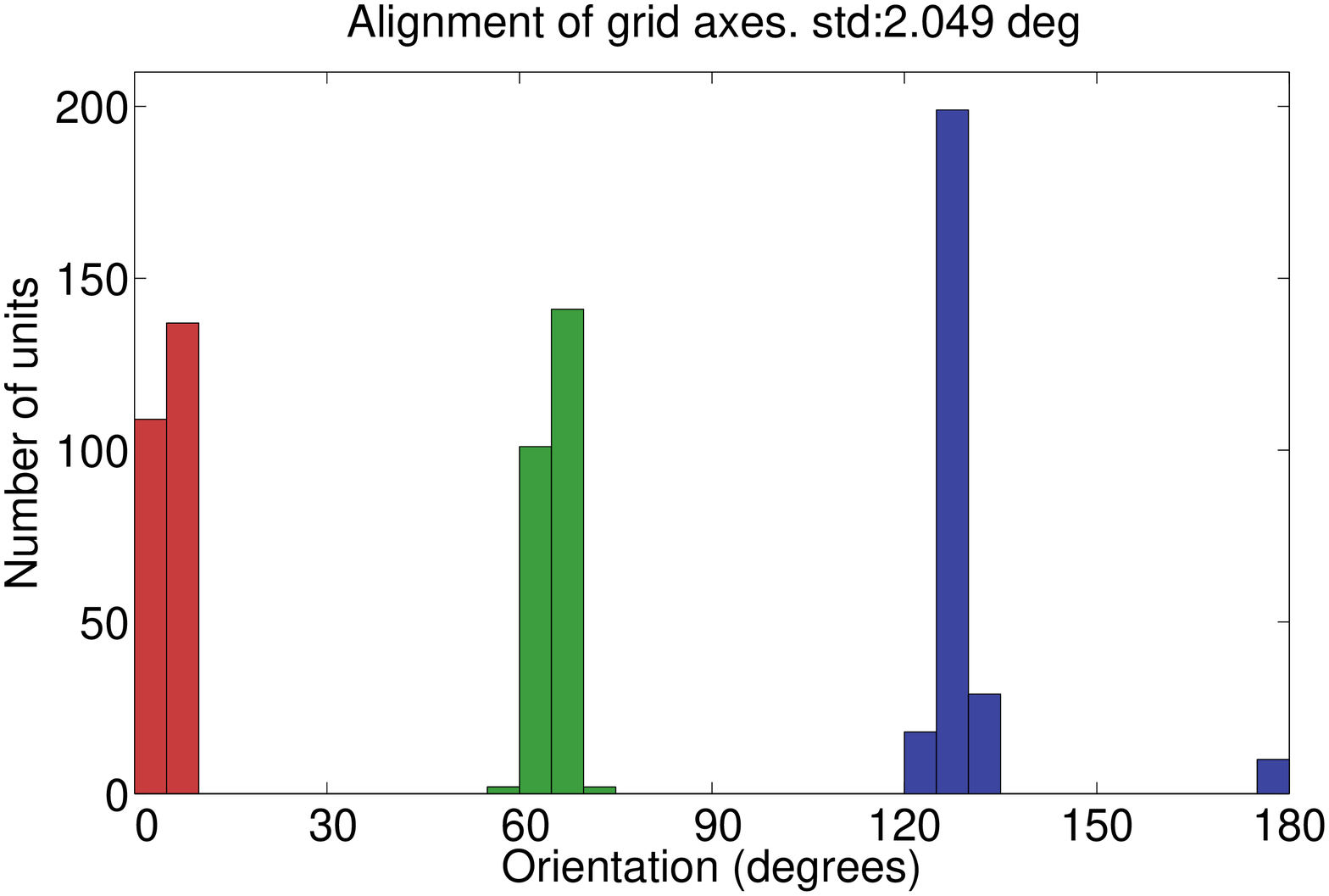,width=5.6cm}}
\rput[bl](11.8,4.2){\epsfig{file=figure/iso/run/iso-circu-mn20-distri-t8125000-orihist.eps,width=5.6cm}}
\rput[bl](0.1,0){\epsfig{file=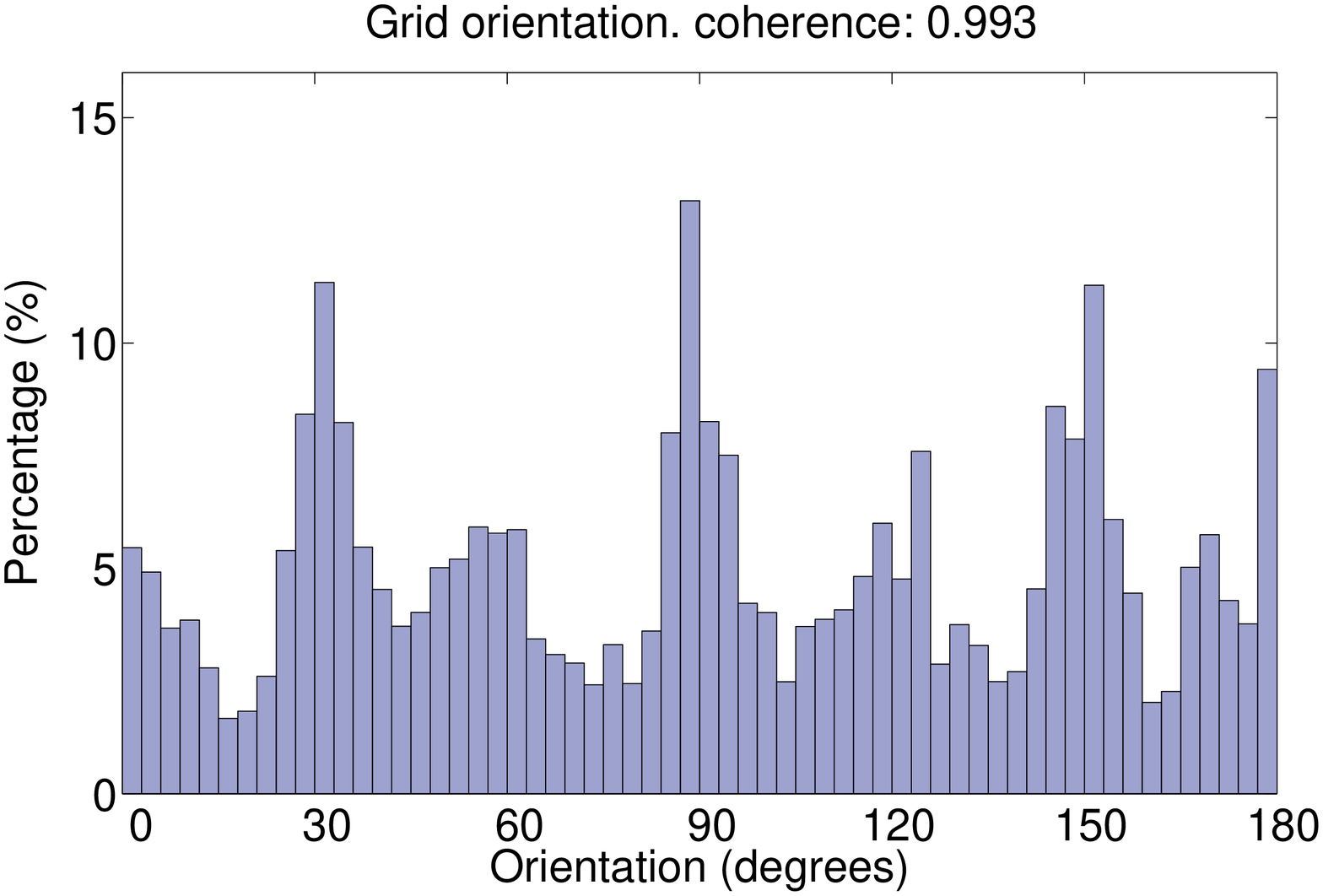,width=5.6cm}}
\rput[bl](6,0){\epsfig{file=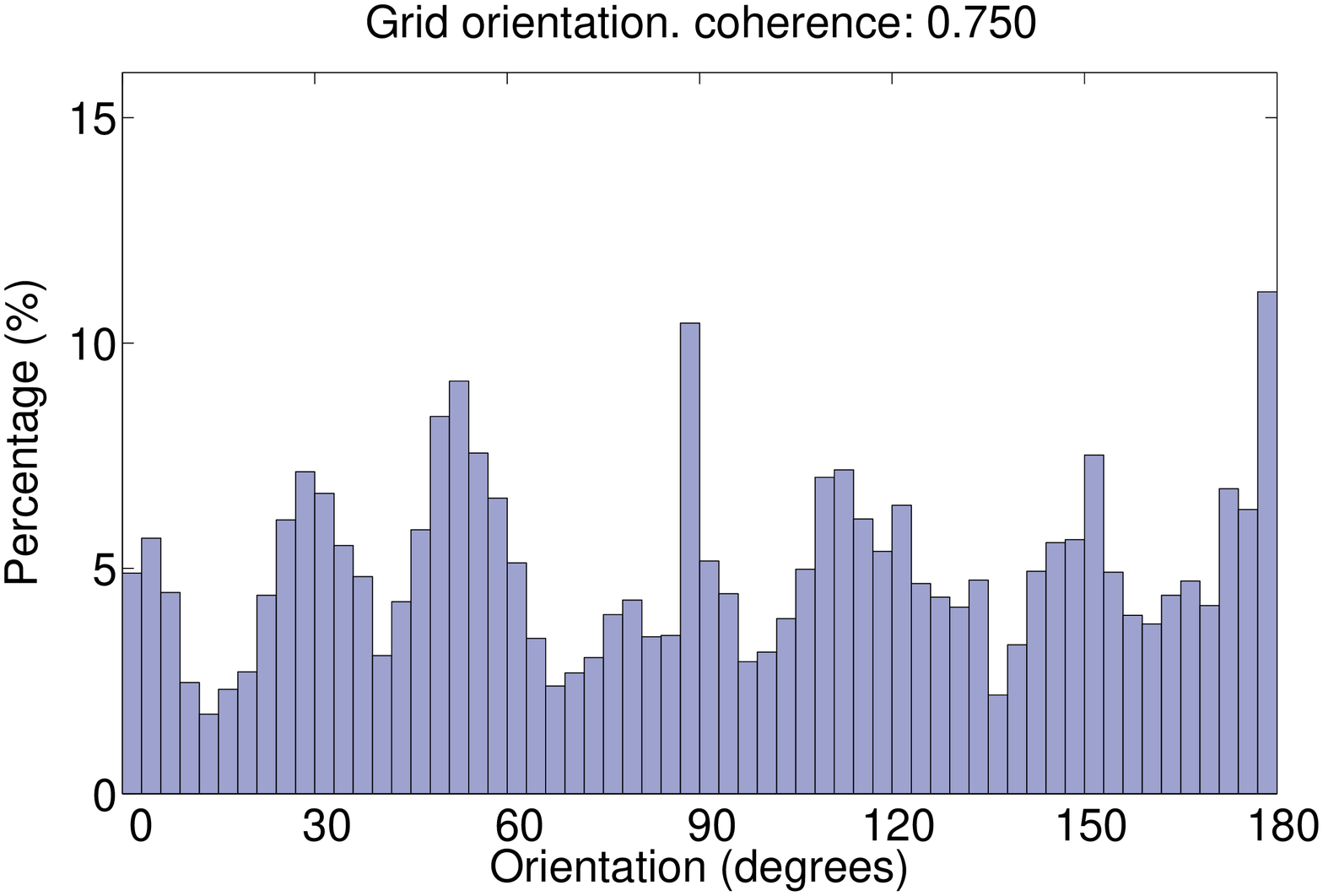,width=5.6cm}}
\rput[bl](11.8,0){\epsfig{file=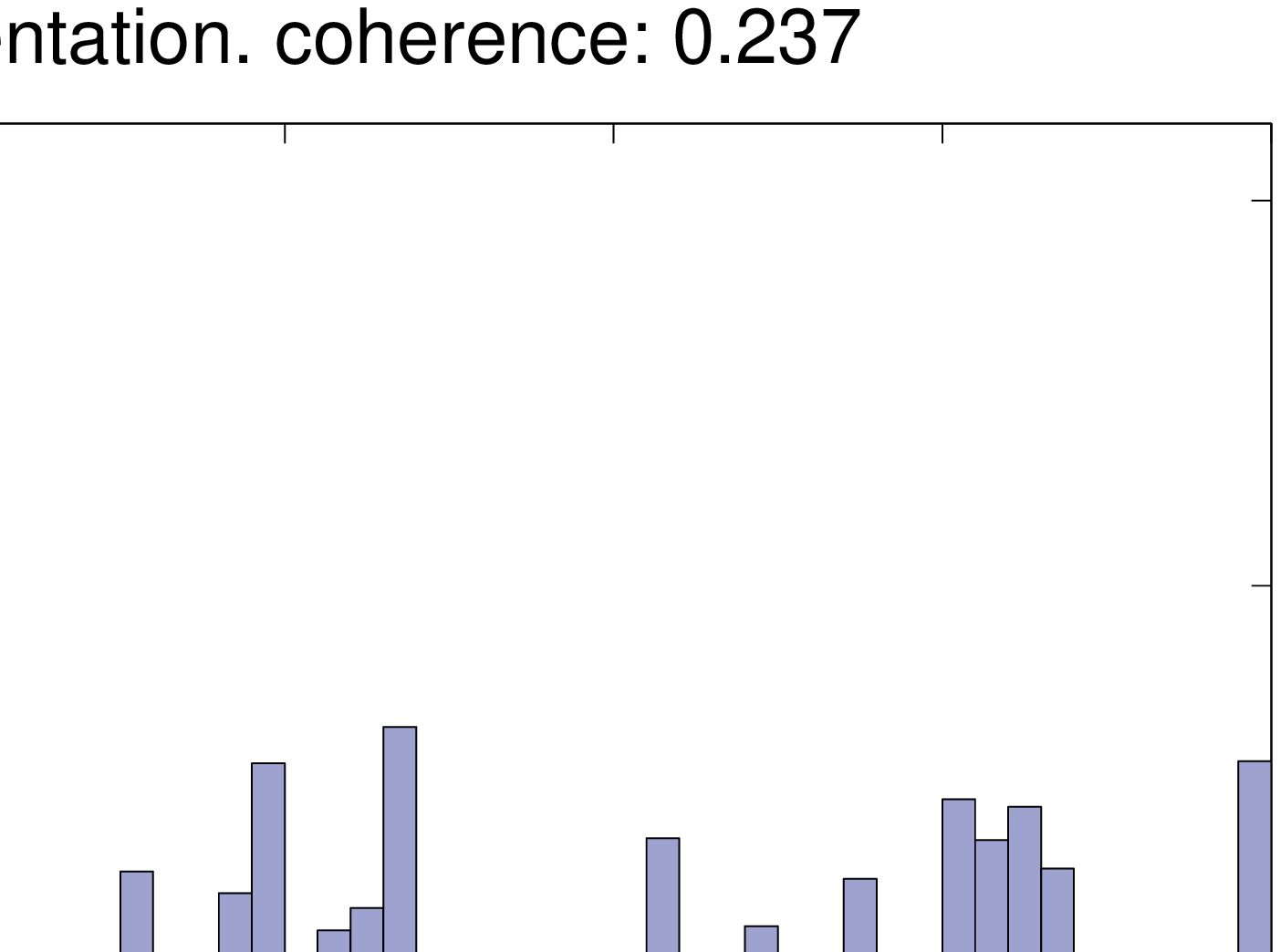,width=5.6cm}}

\psframe[linewidth=3pt,linecolor=cyan](0,-0.1)(5.75,8.3)
\psframe[linewidth=3pt,linecolor=blue](5.9,-0.1)(11.63,8.3)
\psframe[linewidth=3pt,linecolor=green](11.78,-0.1)(17.45,8.3)

\rput[bl](0.4,8){\text{\small{\bf a}}}
\rput[bl](6.2,8){\text{\small{\bf b}}}
\rput[bl](12,8){\text{\small{\bf c}}}

\rput[bl](0.4,3.8){\text{\small{\bf d}}}
\rput[bl](6.2,3.8){\text{\small{\bf e}}}
\rput[bl](12,3.8){\text{\small{\bf f}}}
\endpspicture
\end{minipage}
\caption{Square environments tend to orient the aligned grid fields along the directions of the sides, compared with cylinder environments. (a-c) Grid fields align to each other in square environments (left column: $\sigma_{RD}=0.2$ radians, middle column: $\sigma_{RD}=0.15$ radians) and in a cylinder environment (right column: $\sigma_{RD}=0.2$ radians) in three individual simulations. The alignment coherence score, i.e. the mean standard deviation (in degrees) of the orientations averaged over the three grid axes, is indicated at the top of each panel. The same data in Fig.~\ref{fig-circu}e is shown in c again; (d-f) The sum of the orientation distributions of the three grid axes averaged over 70 simulations. In square environments, the orientation distribution of the three grid axes shows periodic clusters, which are distinct from the random fluctuations of the average orientation distribution. The number at the top of each panel indicates the coherence of grid orientation.}
\label{fig-grid-ori-bound}
\end{figure*}
As in cylinder environments, gird fields show coherent alignment in individual simulations in square environments, irrespective of the RD standard deviation (Fig.~\ref{fig-grid-ori-bound}a-c). To see any common orientation across simulations, we performed multiple simulations in square and cylinder environments with RD standard deviation $\sigma_{RD}=0.2$ radians and in a square environment also with $\sigma_{RD}=0.15$ radians. In each of the three conditions, 70 independent simulations were conducted with different seeds for the random number generator. Averaged and normalized over the 70 trials, the sum of the mean orientation distributions of the three grid axes in square environments shows a significant concentration at multiples of 30 degrees (Fig.~\ref{fig-grid-ori-bound}d-e). That is, the common orientation of the grid fields is more likely to align along the walls of box environments. To quantify the coherence in orienting grids, a {\em cross-trial grid orientation coherence score} is defined for the one-dimensional mean orientation distribution of grid axes. The coherence score is computed in a similar way as the gridness score in Section~\ref{sec-sylind} but assuming 30-degree periodicity, thus evaluating even and odd multiples of 15 degrees. Note that the range of the cross-trial grid orientation coherence score is [-2,2]. The cross-trial coherence of grid orientation in square environments is much larger than that in cylinder environments (Fig.~\ref{fig-grid-ori-bound}d-e vs. f). 

Results in Fig.~\ref{fig-grid-ori-bound} indicate that although the running-direction anisotropy associated with training in square environments does not influence the coherence in grid alignment, it may strongly modulate the coherence in grid orientation across rats. That is, running-direction anisotropy produces a single orientation for the aligned grids in each simulation, but two clusters of prevailing common orientations across simulations, at 30 (or rather, 90) degrees of each other, i.e. with one grid axis aligned to a wall of the square environment.

In the real system, two main orientations are observed in the same experiment (across grid units with different characteristic grid spacings~\citep{Stensland2010}), not quite orthogonal to each other, but also not aligned to the walls of the testing environment. It could well be that these prevailing orientations reflect other environments, where rats were caged during development. A rat who develops in a square or rectangular cage, however, is expected to experience not only RD anisotropy, as tested in the model, but also speed anisotropy. To disentangle a potential contribution of the latter, in the following section we introduce speed anisotropy, and compare its effects with those of running-direction anisotropy.

\section{Speed anisotropy}\label{sec-speed}

\begin{figure}
\centering
\begin{minipage}{5cm}
\includegraphics[width=\linewidth]{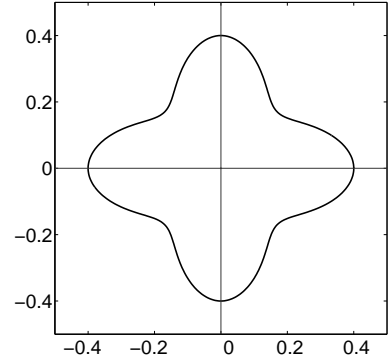}
\end{minipage}
\caption{Polar plot of the speed of the simulated rat with respect to running directions. The maximal speed is the default speed $0.4 m/s$, and the minimal speed is 60\% of $v_{s}$.}
\label{fig-quadrupole}
\end{figure}

Having observed that grid orientation is influenced by the non-uniformity of running directions, in this section we focus on another factor in the trajectories of the simulated rat, namely running speed, and investigate whether the shape of the grids can be influenced by an anisotropic speed distribution in exploration behavior, even with no anisotropy in the boundary conditions (i.e. in cylinder environments). 
In this section, the simulations are identical to those in Section~\ref{sec-sylind}, except that the virtual rat explores the cylinder environment running faster in four preferred directions, as visualized in Fig.~\ref{fig-quadrupole}
\begin{equation} 
v^t = v_{s} [q+ (1-q) \frac{|\sin(\omega^t)|^3 + |\cos(\omega^t)|^3 -1/\sqrt{2}}{1-1/\sqrt{2}}].
\end{equation}
Here $\omega^t$ is the current running direction of the rat. The speed now can go from a minimum of $v_{s}=24 cm/s$ to a maximum of $v_{s}=40 cm/s$ depending on the running direction. $q = 0.6$ is the ratio between the two speed extremes.

\begin{figure*}
\centering
\begin{minipage}{17.4cm}
\pspicture(0,0)(17.4,16)
\rput[bl](0,11.7){\epsfig{file=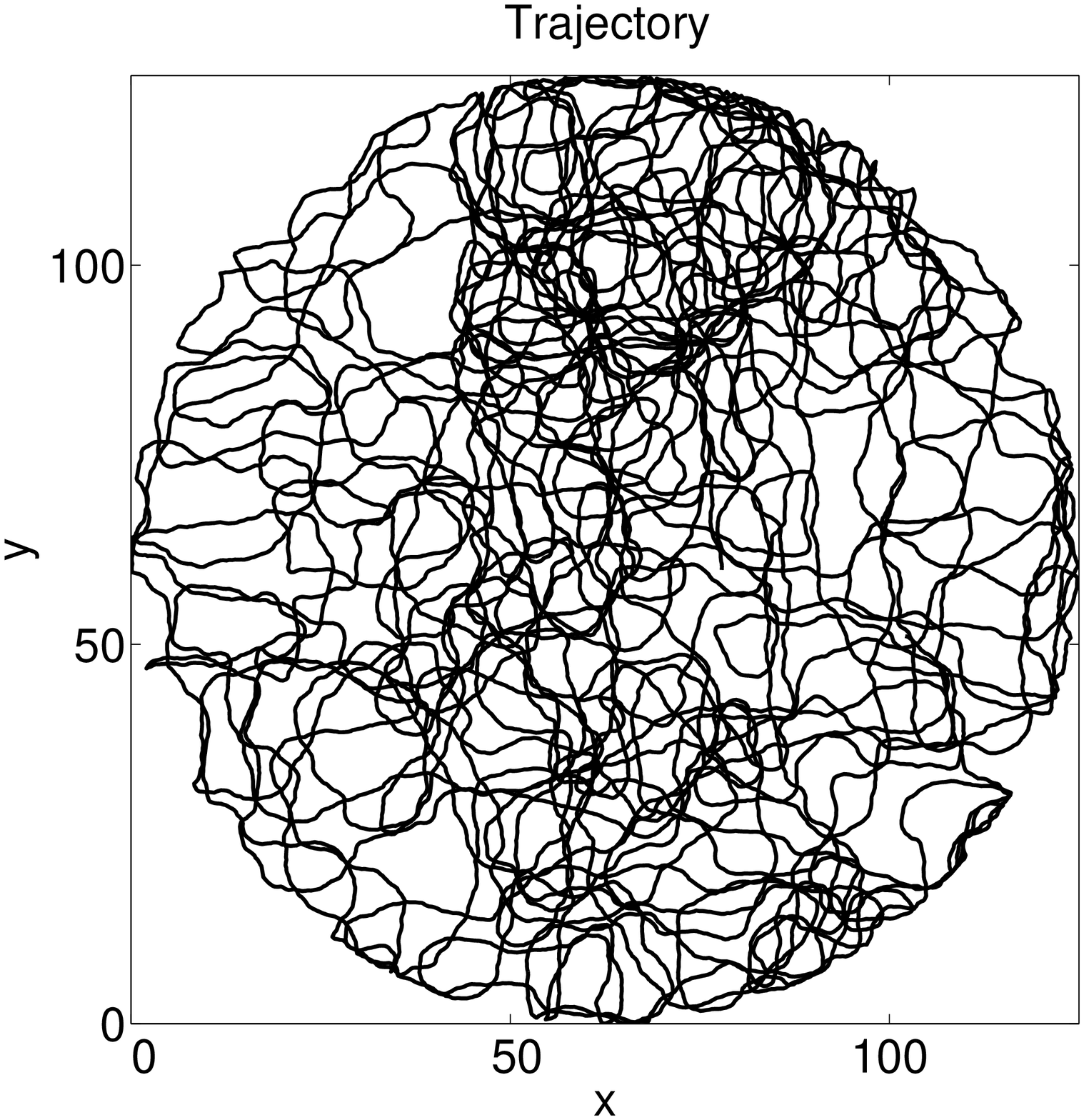,height=4.3cm}}
\rput[bl](5,11.7){\epsfig{file=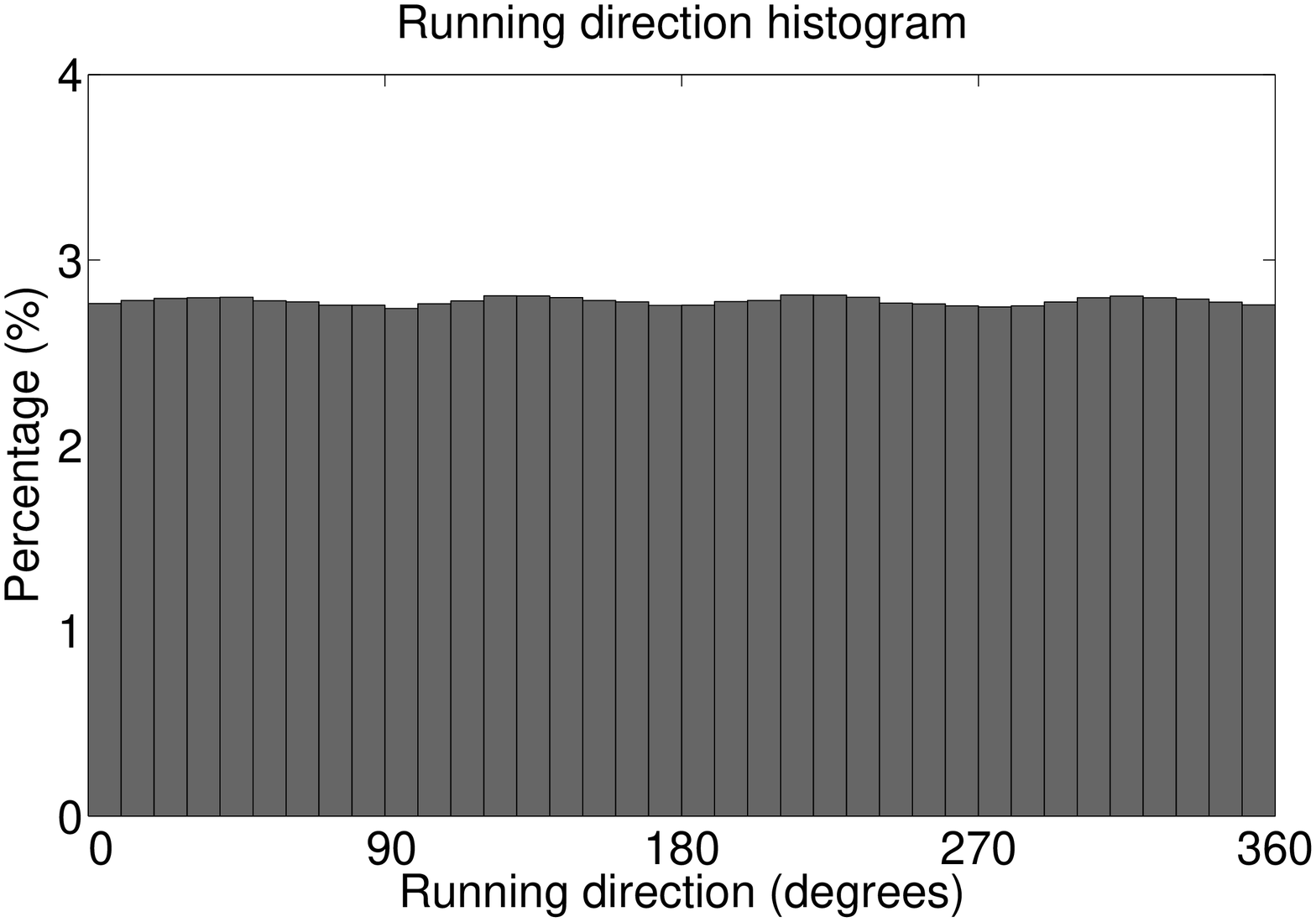,width=5.6cm}}
\rput[bl](11.2,11.7){\epsfig{file=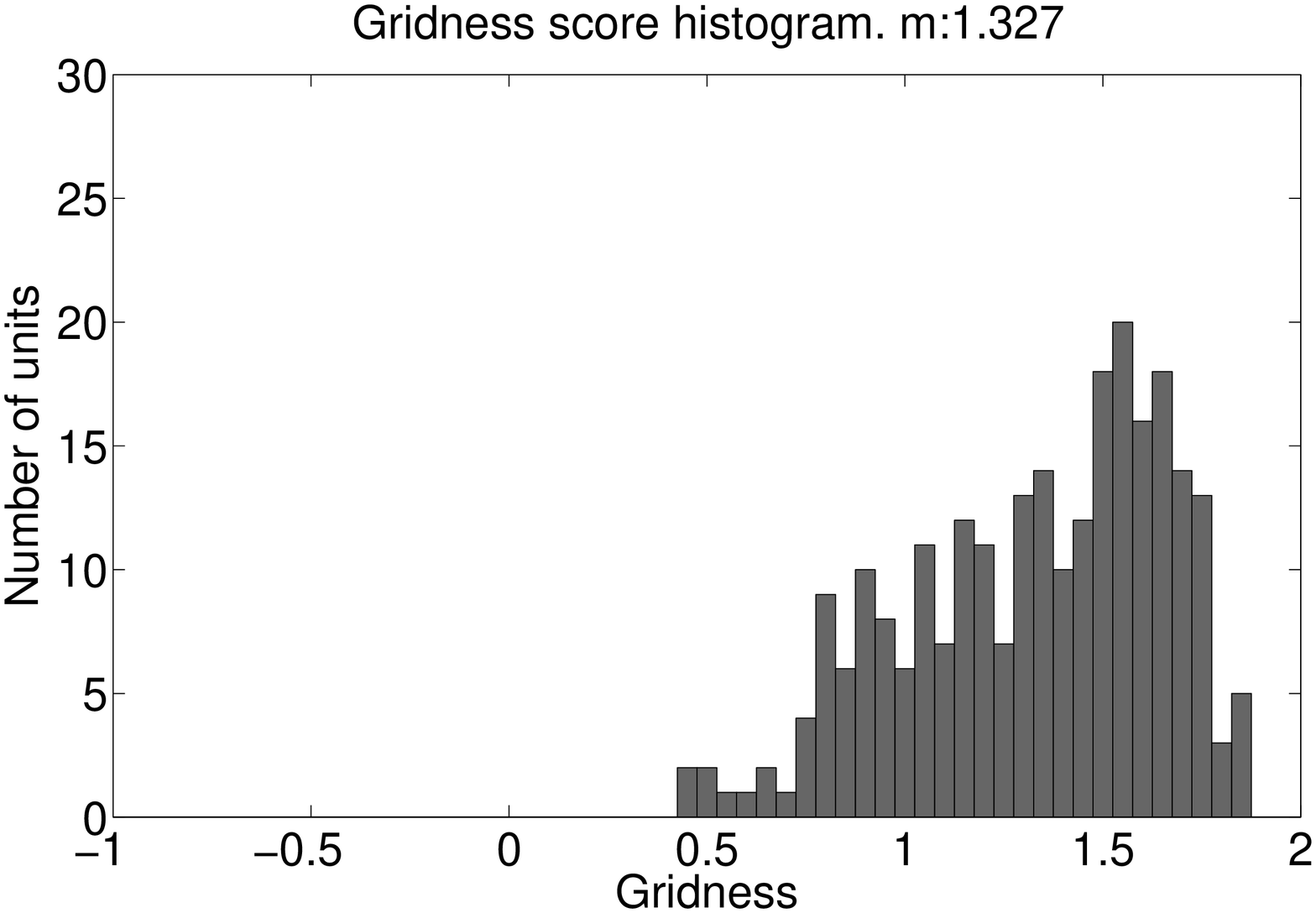,width=5.6cm}}

\rput[bl](0,7.5){\epsfig{file=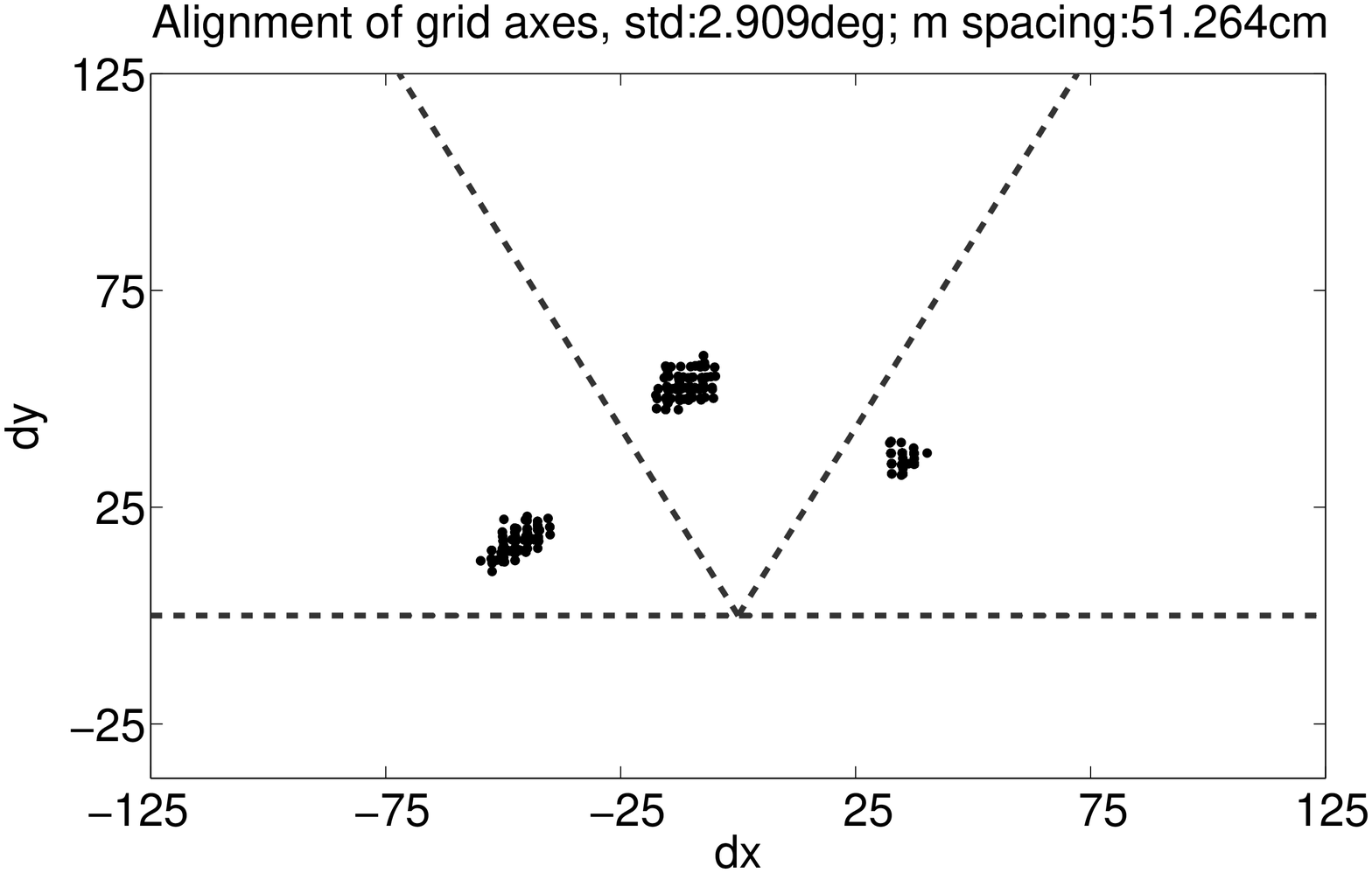,width=5.6cm}}
\rput[bl](5.8,7.5){\epsfig{file=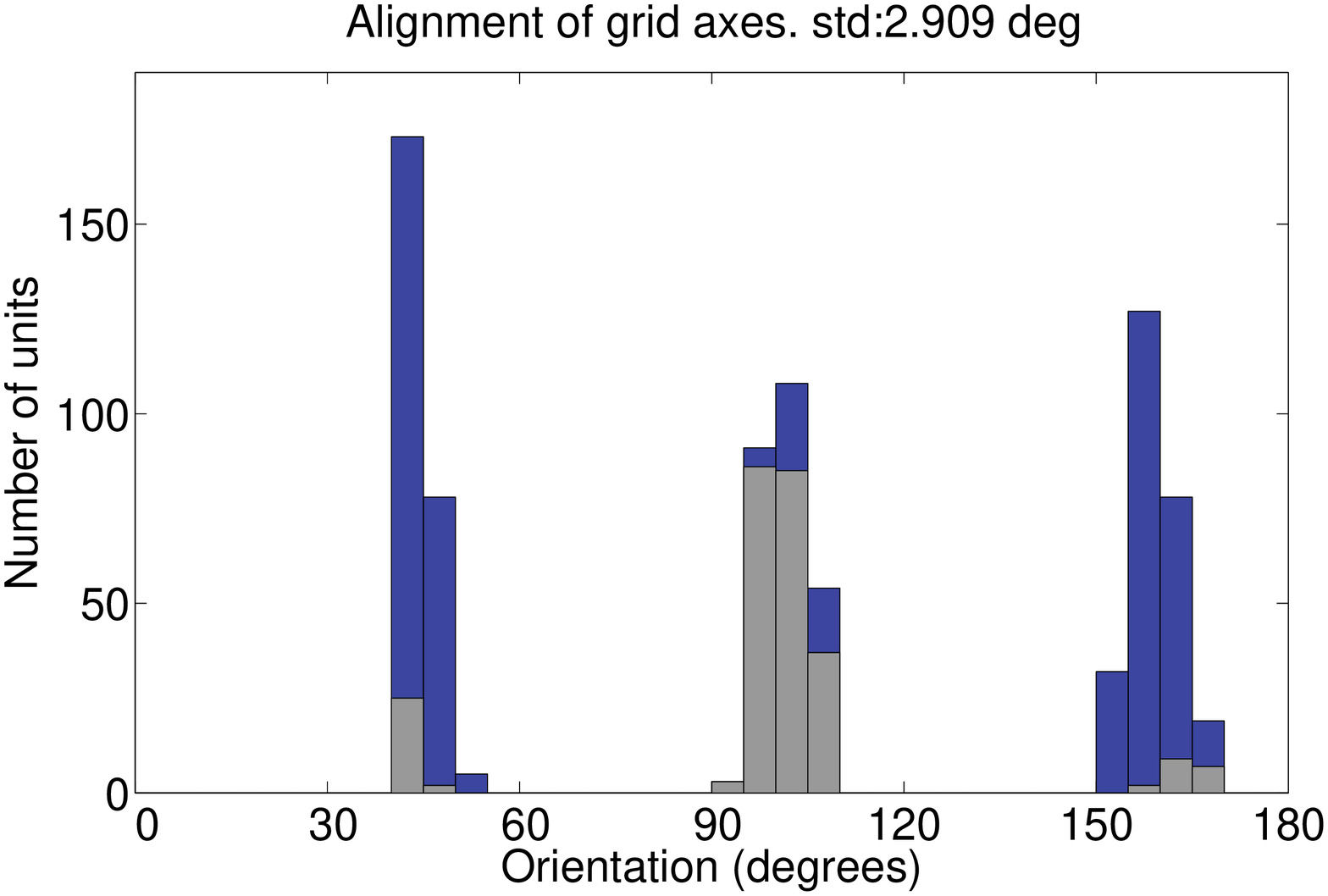,width=5.6cm}}
\rput[bl](12,7.5){\epsfig{file=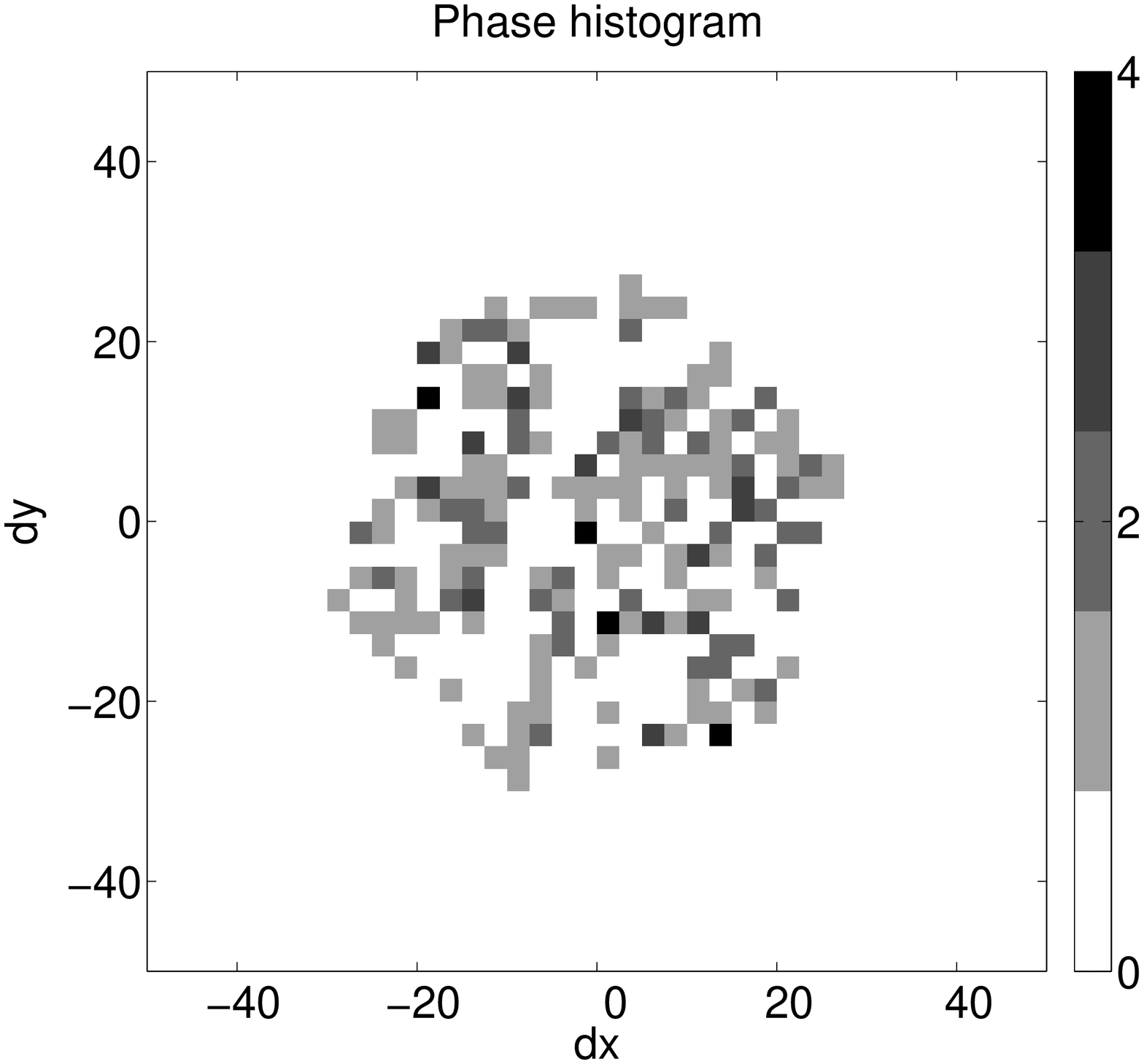,height=4cm}}

\rput[bl](0,2.8){\epsfig{file=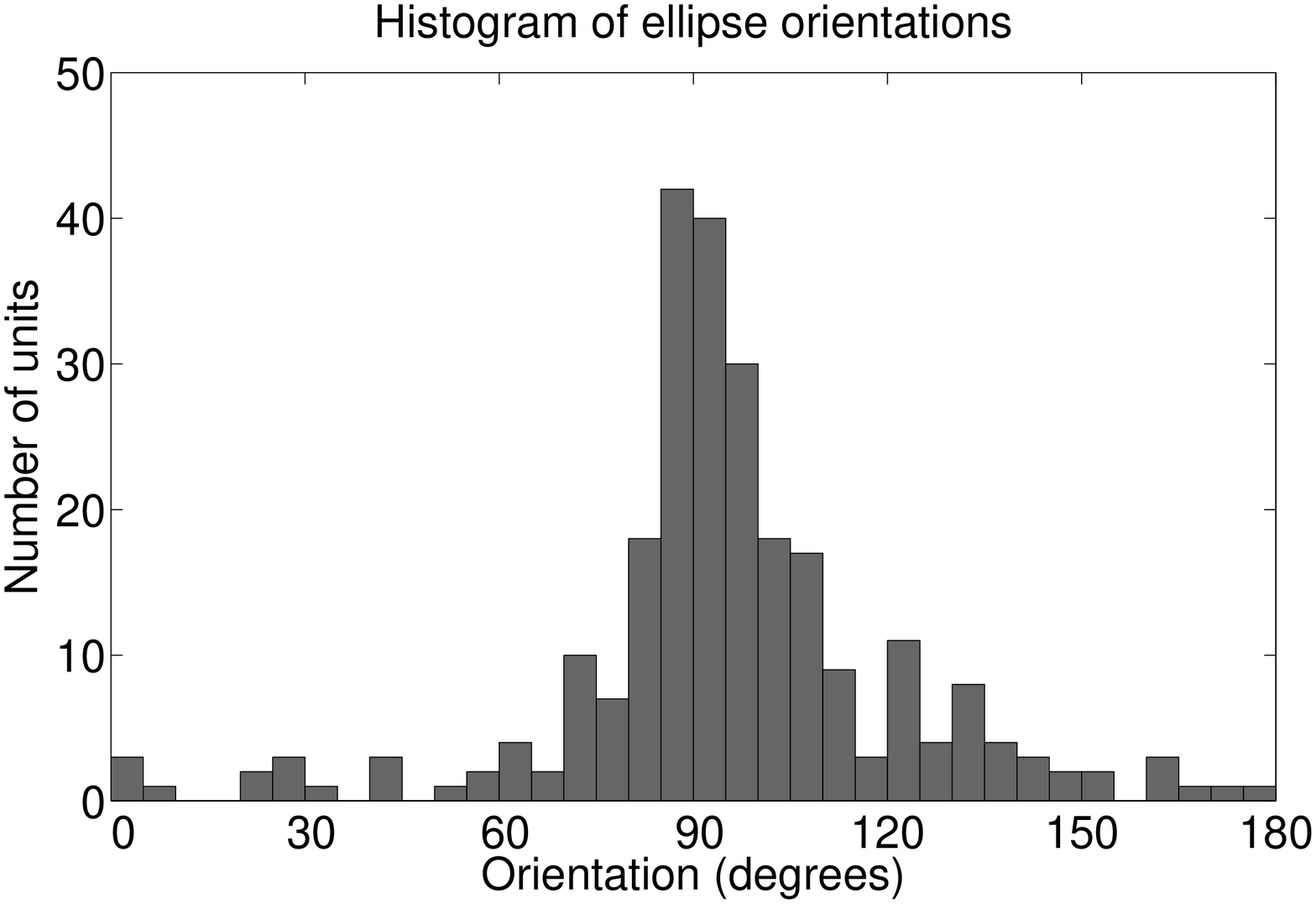,width=5.6cm}}
\rput[bl](5.6,2.8){\epsfig{file=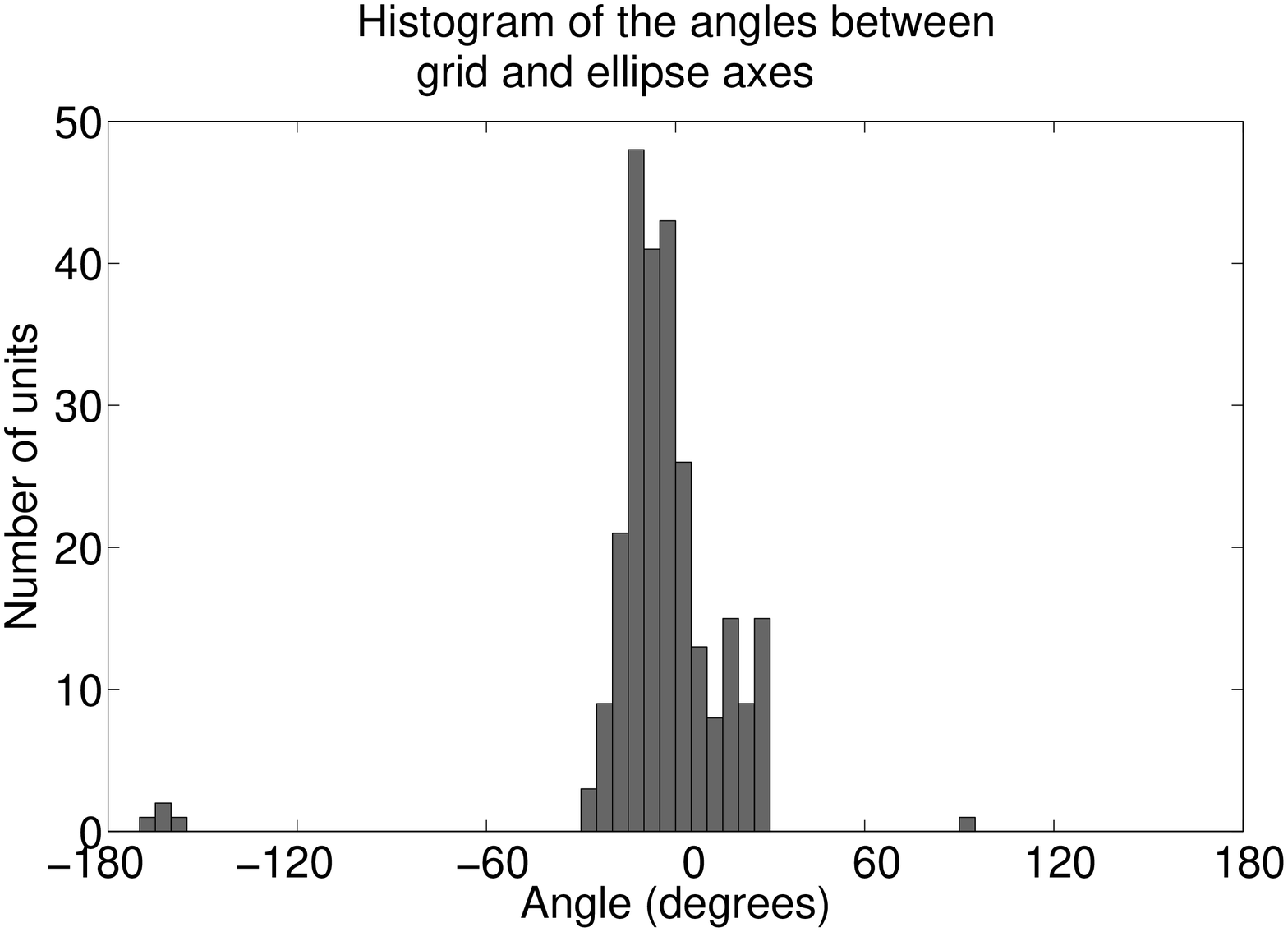,width=5.6cm}}
\rput[bl](11.2,2.8){\epsfig{file=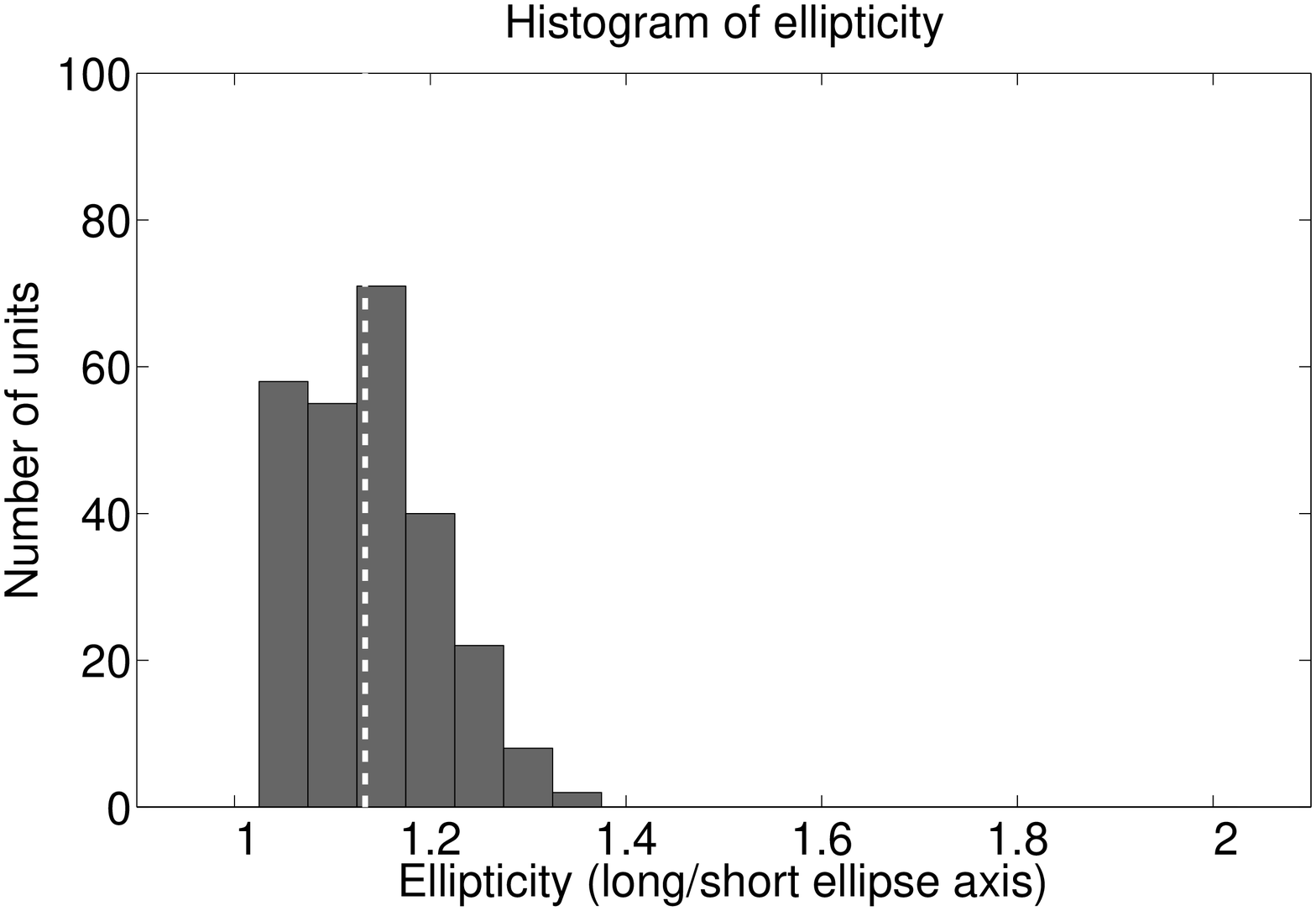,width=5.6cm}}
\rput[bl](0,0){\epsfig{file=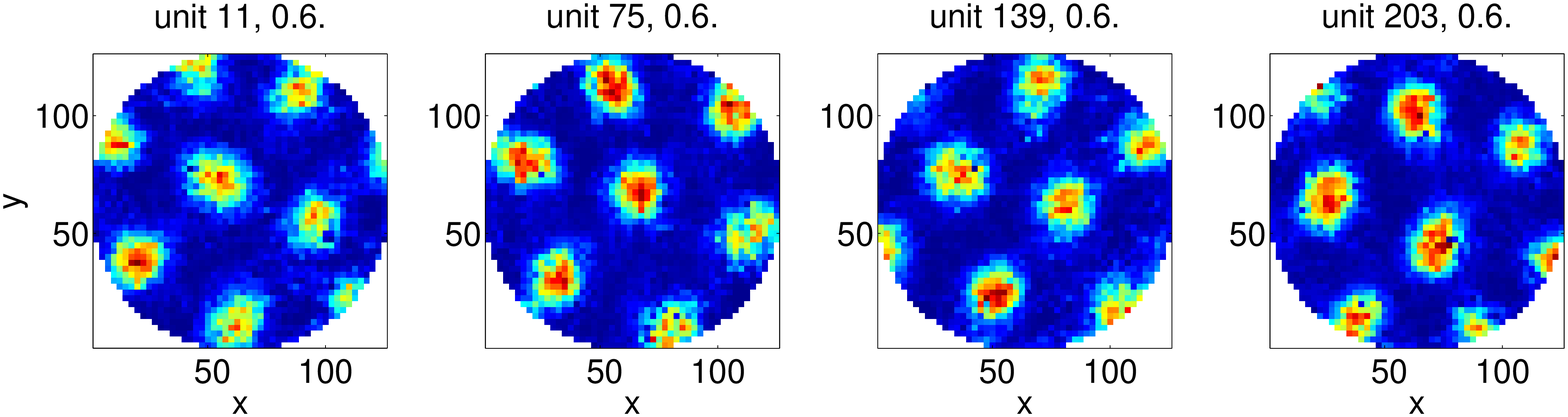,width=8.7cm}}
\rput[bl](8.7,0){\epsfig{file=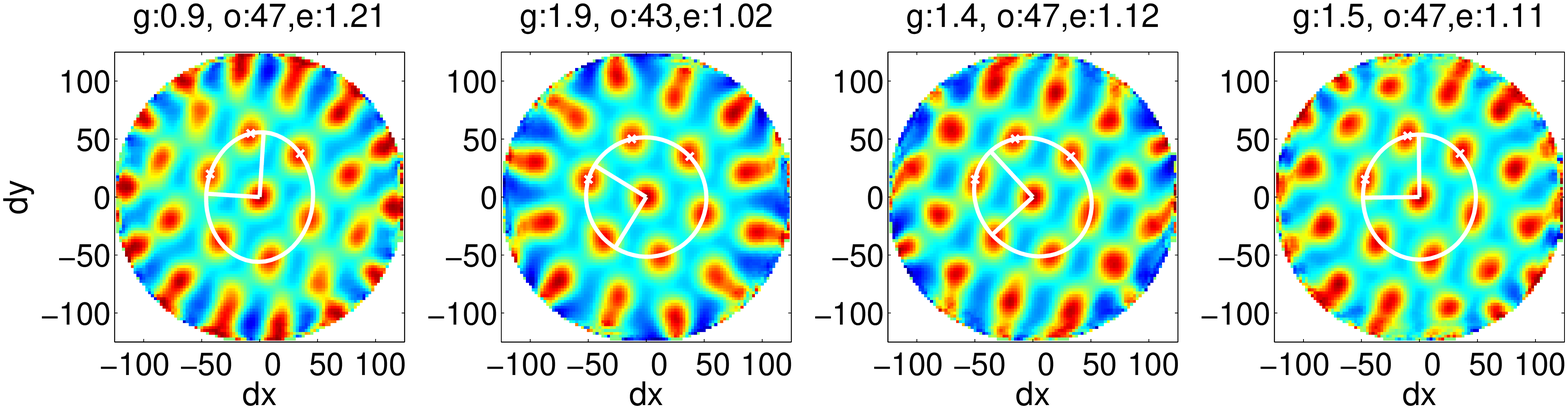,width=8.7cm}}

\rput[bl](0,15.8){\text{\small{\bf a}}}
\rput[bl](5,15.8){\text{\small{\bf b}}}
\rput[bl](11.2,15.8){\text{\small{\bf c}}}

\rput[bl](0.2,11.2){\text{\small{\bf d}}}
\rput[bl](5.8,11.2){\text{\small{\bf e}}}
\rput[bl](12,11.2){\text{\small{\bf f}}}

\rput[bl](0,7){\text{\small{\bf g}}}
\rput[bl](5.8,7){\text{\small{\bf h}}}
\rput[bl](11.3,7){\text{\small{\bf i}}}
\rput[bl](0,2.3){\text{\small{\bf j}}}
\endpspicture
\end{minipage}
\caption{Grid alignment during exploration in a cylinder environment with anisotropic speed. (a) A trajectory of the rat with $2\times 10^4$ steps in a cylinder environment; (b) RD distribution of an entire trajectory in a $8\times 10^6$-step  simulation; (c) Histogram of the gridness scores of all conjunctive units in the simulation; (d) The scatter plot of the locations of the three peaks found in autocorrelograms; (e) Histogram of the orientations of the three grid axes. Shown in front in gray is the orientation histogram of the long grid axes. Indicated at the top of the panel is the coherence score in grid alignment (mean standard deviation, in degrees, of the orientation averaged over the three grid axes), similar as the coherence in grid alignment in the cylinder environment with constant speed; (f) The histogram of spatial phases (again, relative to that of the best grid); (g) Histogram of the orientations of the major axes of the ellipses determined from each autocorrelogram; (h) Histogram of the angles between the long grid axis and the ellipse major axis; (i) Histogram of the ratios between the major and minor axes of ellipses. The white broken line indicates the median; (j) Examples of the fields of conjunctive units (left) and the corresponding autocorrelograms (right). The white curves show the ellipses determined from the three maxima found in autocorrelograms. Unit number, and maximal firing rate (in arbitrary units) are indicated above each rate map. Gridness score,orientation (in degrees) and ellipticity are indicated above each autocorrelogram.}
\label{fig-aniso}
\end{figure*}

As in the standard simulations in Section~\ref{sec-sylind}, the RD distribution of the trajectory is quite uniform (Fig.\ref{fig-aniso}a,b). In the RD distribution there is a tiny over representation of running directions parallel to the diagonals. This tiny distortion is originated by the fact that when the rat is running faster it takes a shorter time for it to reach the boundary, where it is forced to turn. However, the effect appears much smaller than that due to the square shape of the environment in Section~\ref{sec-box}, and its direct influence on grid alignment is negligible. 

Fig.~\ref{fig-aniso}j shows the spatial maps as well as the autocorrelograms of four example conjunctive units from the network. Similar to the case with constant speed, grids align with each other, and the phases are kept broadly distributed in the environment (Fig.~\ref{fig-aniso}d-f). However, the average gridness score is somewhat lower as compared to the case with constant speed (Fig.~\ref{fig-aniso}c). The reason is that, with speed anisotropy, the grids are distorted, showing grid axes with non-equal lengths, as can been seen carefully in Fig.~\ref{fig-aniso}j. Among the three axes of a grid, the axis that passes through the maximum farthest from the origin is identified as the long grid axis. In our simulations with speed anisotropy, conjunctive units tend to develop maps that share the same grid axis as the long one (in the simulation of Fig.~\ref{fig-aniso}e this corresponds to the axis with orientation around 100 degrees), indicating that grid distortion happens predominantly along one direction. Another way to quantify this distortion, introduced in~\citep{Stensland2010}, is to fit an ellipse that passes the six maxima close to the center of the grid autocorrelogram (white curves in Fig.~\ref{fig-aniso}j). These ellipses deviate from perfect circles, indicating modified grid structures. We find that such ellipses are roughly aligned in each map with the corresponding long grid axis, laying within 30 degrees of it (Fig.~\ref{fig-aniso}g-h). The flattening of an ellipse is measured by ellipticity, i.e. the ratio between the major and minor axes. Note that the ellipticity of a perfect circle is 1. In the simulation, the median of the ellipticity distribution of the grids is about 1.15 (Fig.~\ref{fig-aniso}i).

\begin{figure*}
\centering
\begin{minipage}{17.4cm}

\pspicture(0,0)(17.4,8.6)
\rput[bl](0,4.35){\epsfig{file=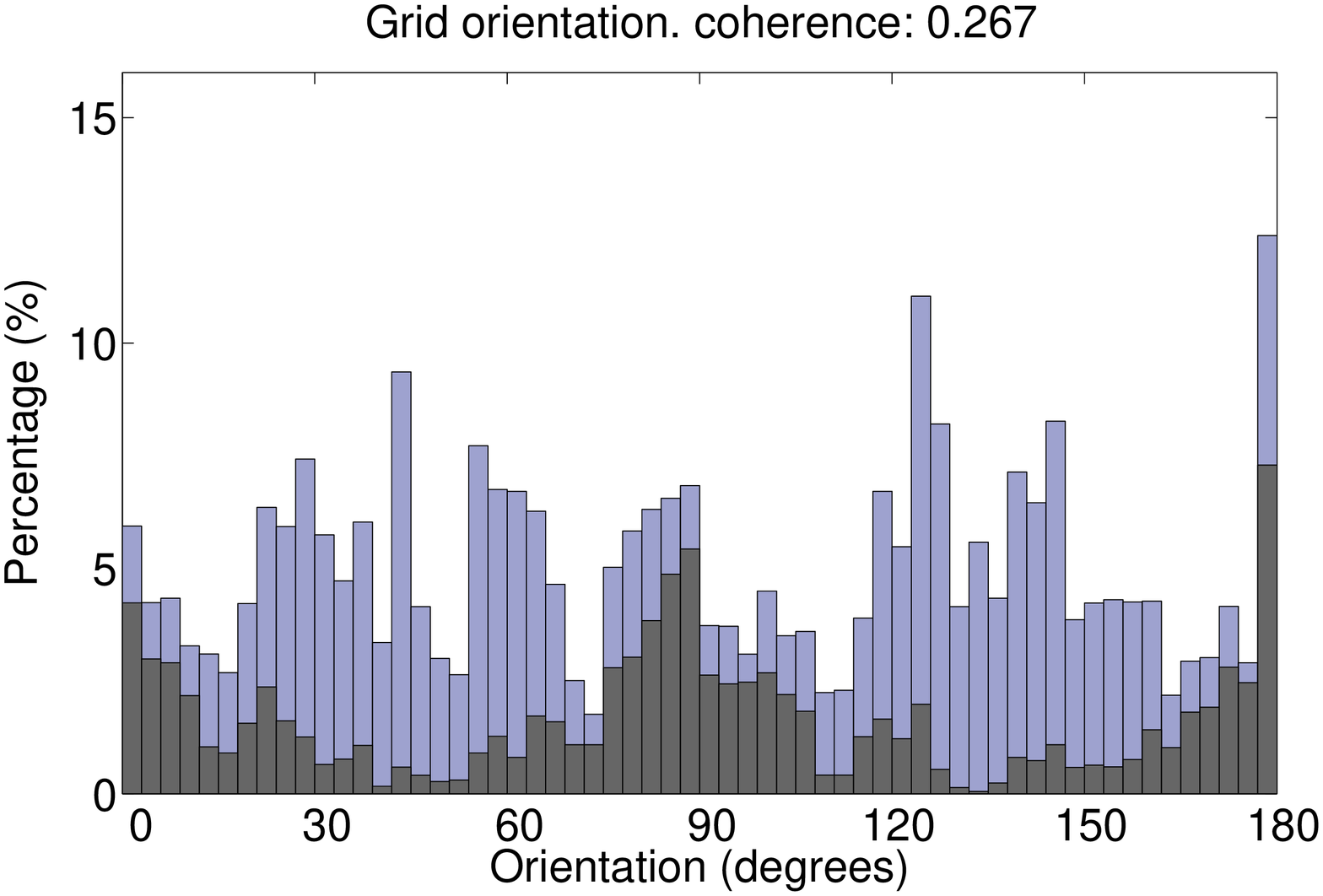,width=5.8cm}}
\rput[bl](5.8,4.35){\epsfig{file=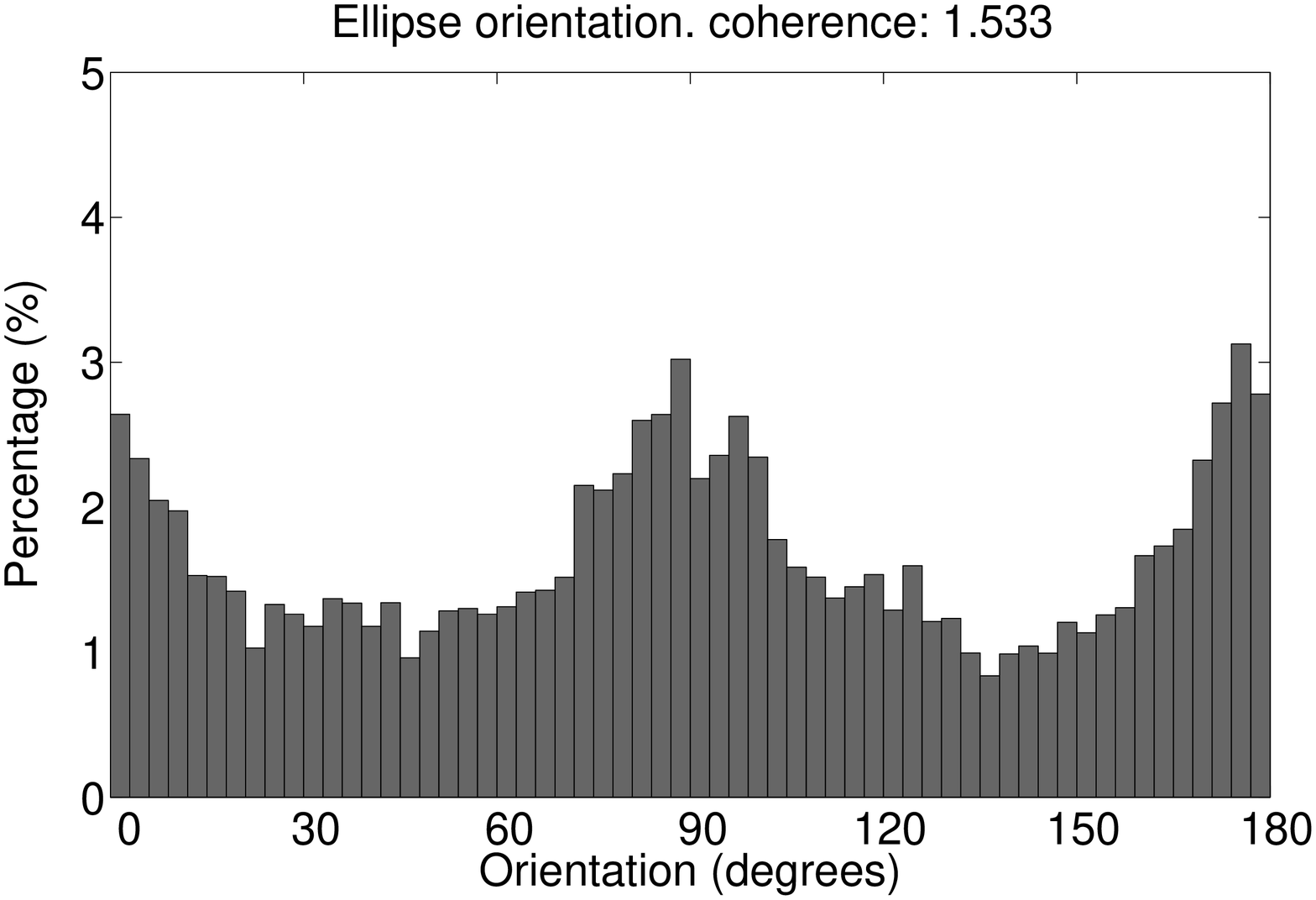,width=5.8cm}}
\rput[bl](11.6,4.35){\epsfig{file=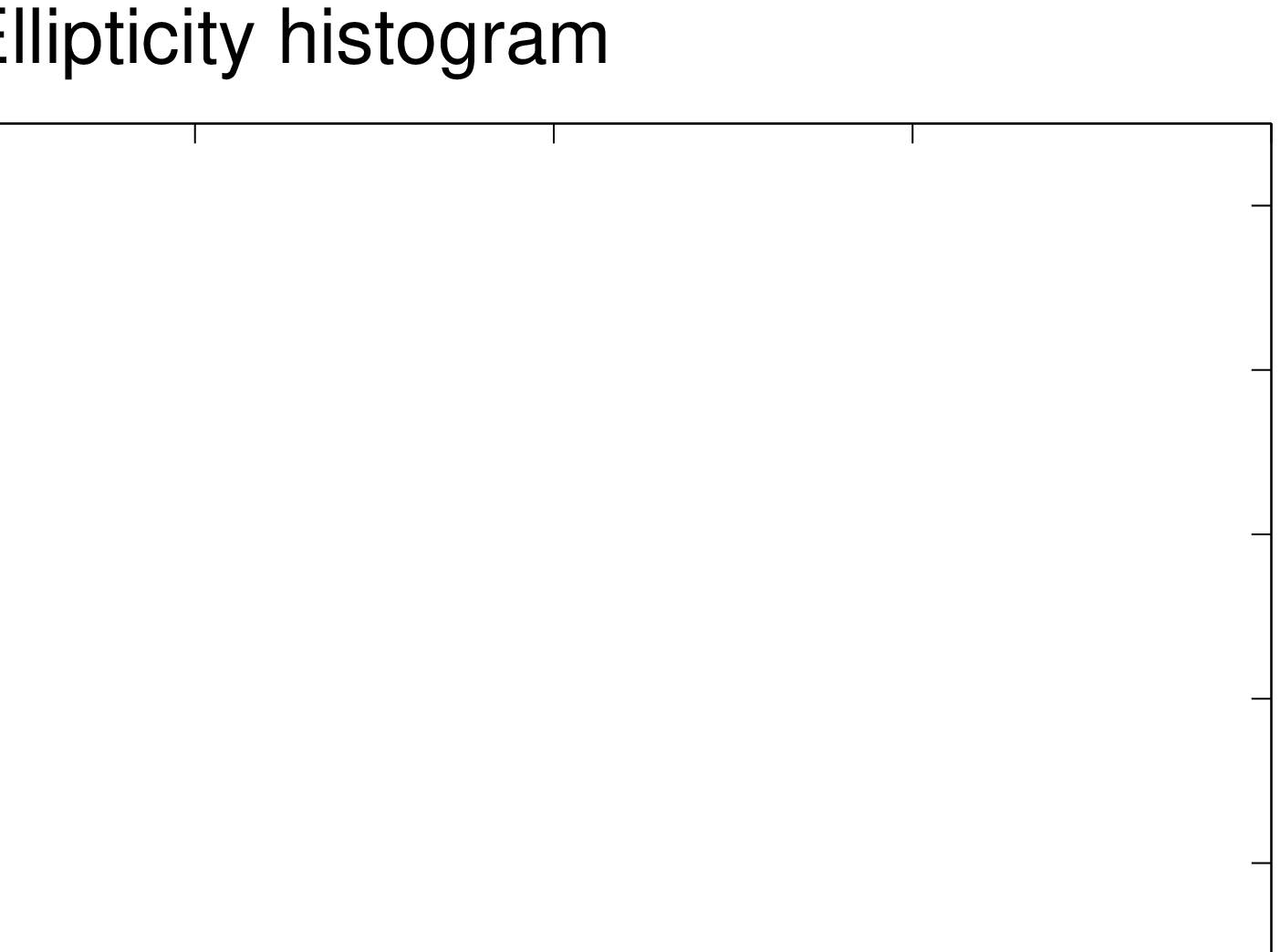,width=5.8cm}}

\rput[bl](0,0){\epsfig{file=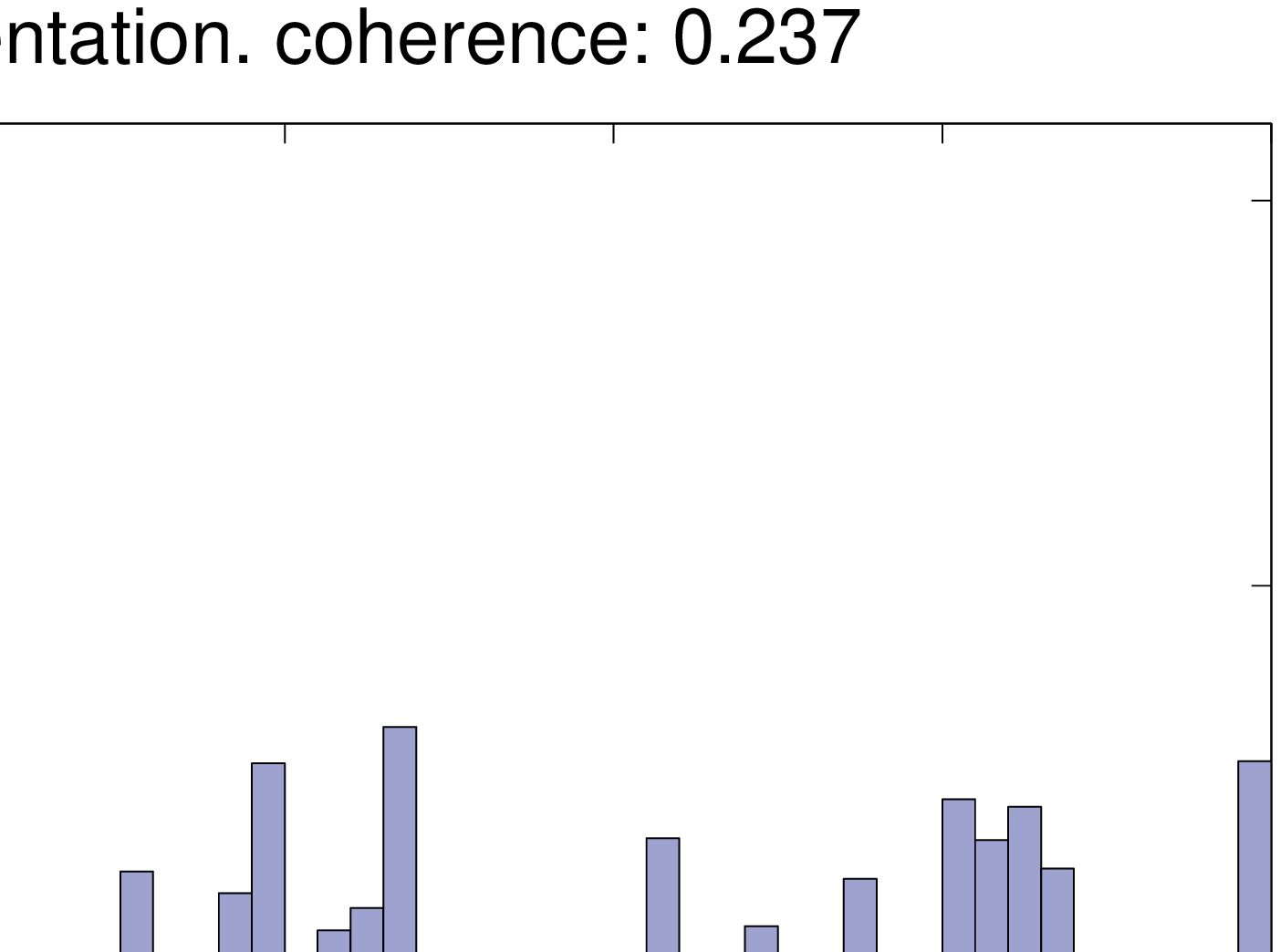,width=5.8cm}}
\rput[bl](5.8,0){\epsfig{file=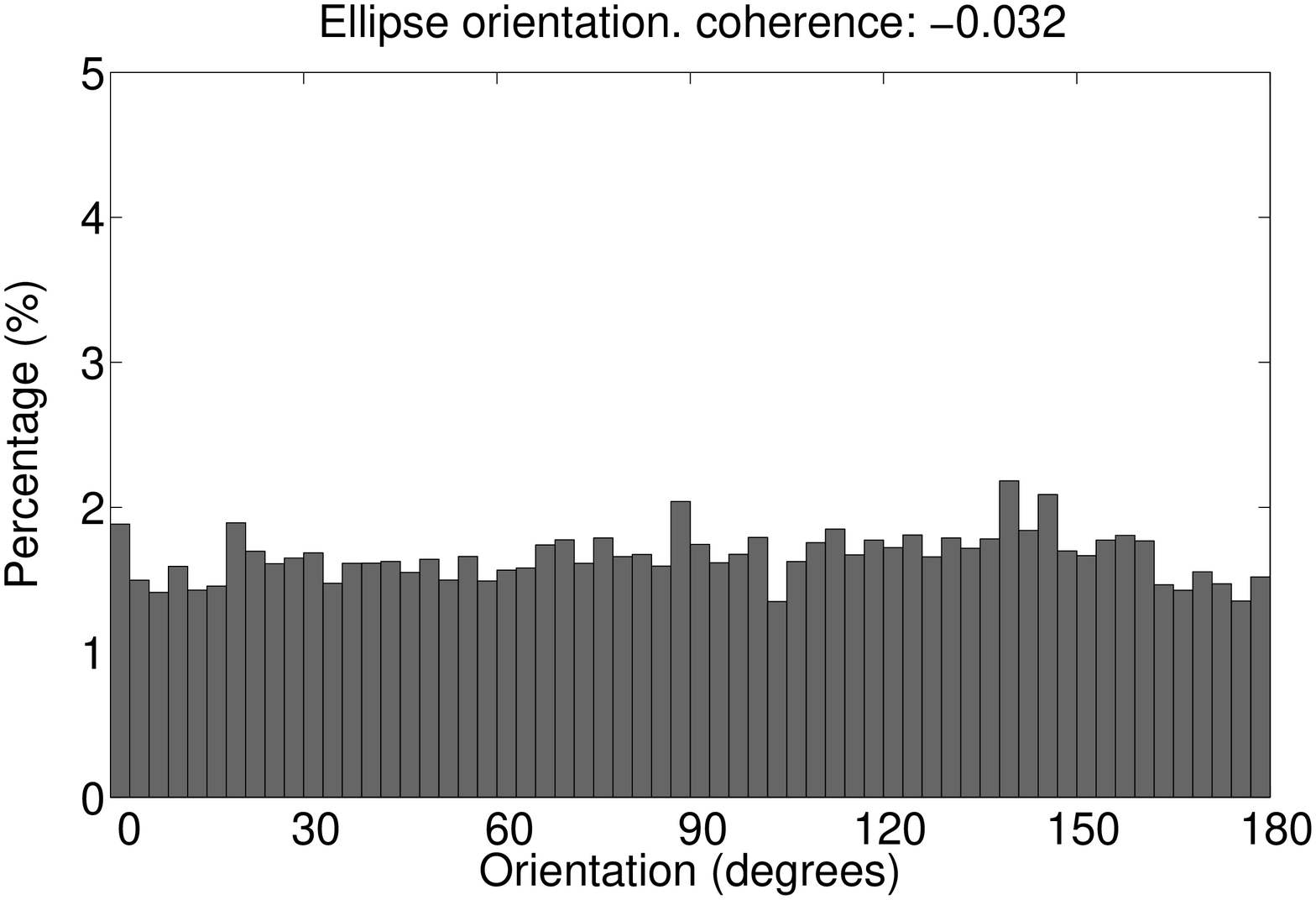,width=5.8cm}}
\rput[bl](11.6,0){\epsfig{file=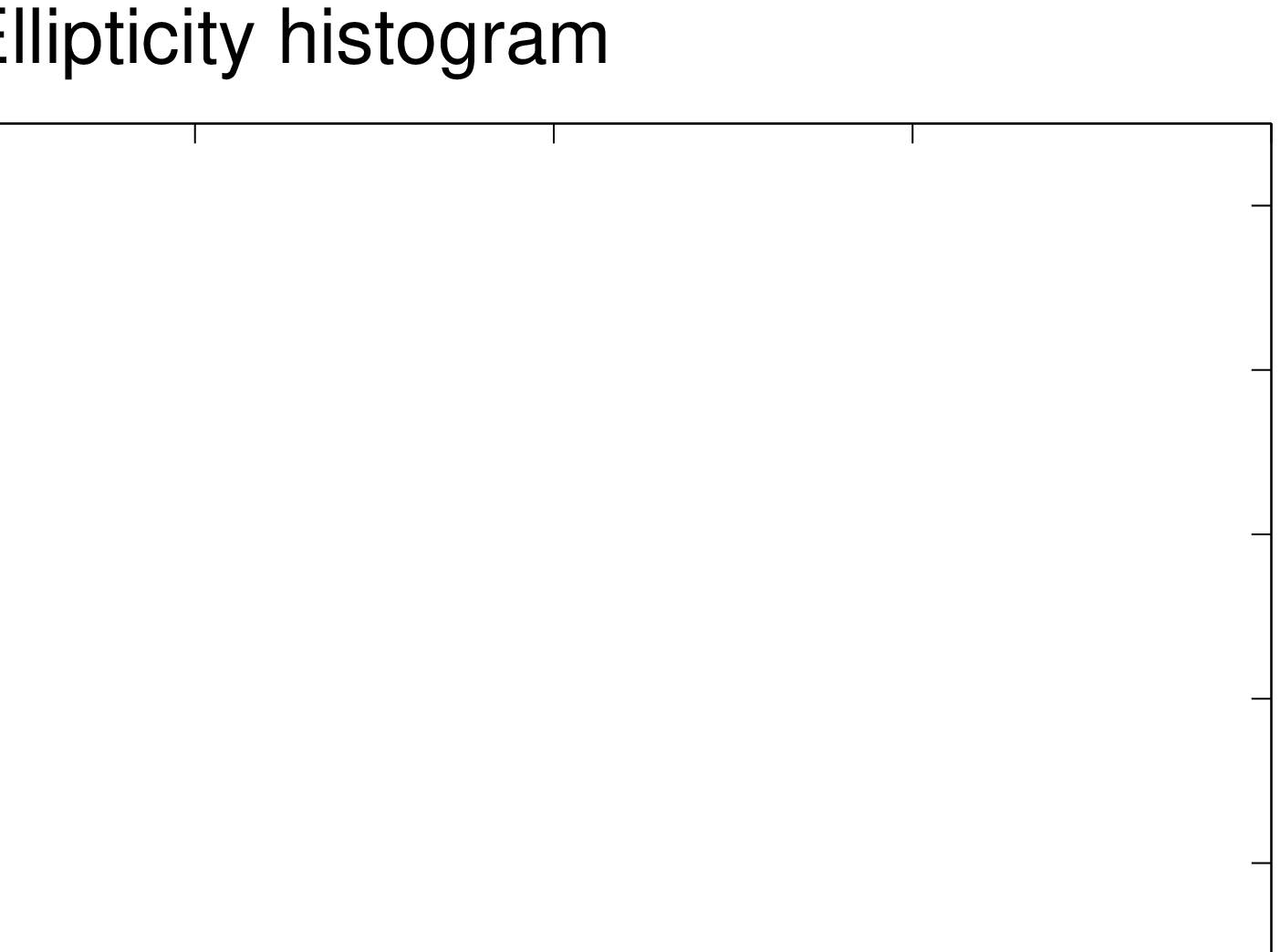,width=5.8cm}}

\psframe[linewidth=3pt,linecolor=brown](-0.1,4.3)(17.5,8.5)
\psframe[linewidth=3pt,linecolor=green](-0.1,-0.1)(17.5,4.1)

\rput[bl](0.4,8.2){\text{\small{\bf a}}}
\rput[bl](6.4,8.2){\text{\small{\bf b}}}
\rput[bl](12.2,8.2){\text{\small{\bf c}}}
\rput[bl](0.4,3.8){\text{\small{\bf d}}}
\rput[bl](6.4,3.8){\text{\small{\bf e}}}
\rput[bl](12.2,3.8){\text{\small{\bf f}}}

\endpspicture
\end{minipage}
\caption{Speed anisotropy in cylinder environments distorts the shape of the grids (top row) as compared to exploration with isotropic speed (bottom). (a,d) Average orientation distribution of long grid axes (gray) plotted in front of the average orientation distribution of the three grid axes (blue). The coherence of grid orientation is indicated at the top of each panel. The blue bars in d are the same measure as the distribution shown in Fig.~\ref{fig-grid-ori-bound}f; (b,e) Average orientation distribution of the major axes of the ellipses. At the top of each panel, the coherence in ellipse orientation is noted. It is calculated similarly as the coherence in grid orientation, but assuming 90-degree periodicity instead of 30-degree; (c,f) Average distribution of ellipticity. The distribution in c is plotted again as the red empty bars in f.}
\label{fig-diff-gridellipse-aniso}
\end{figure*}

Averaged over 70 independent simulations in cylinder environments, the orientation of the long grid axis concentrates along the preferred directions of the speed profile, i.e. 0 and 90 degrees (gray bars in Fig.~\ref{fig-diff-gridellipse-aniso}a). The other two grid axes only broadly orient to 60/120 or 30/150 degrees respectively (light blue bars in Fig.~\ref{fig-diff-gridellipse-aniso}a), with low coherence of grid orientation. In contrast, with constant speed in cylinder environments, both the average orientation distribution of grid axes and of the long grid axis are fairly uniform, reflected in even lower coherence score of grid orientation (Fig.~\ref{fig-diff-gridellipse-aniso}d).

With speed anisotropy in cylinder environments, we observe that grid maps are more elliptical than those developed from exploration with constant speed (Fig.~\ref{fig-diff-gridellipse-aniso}c vs. Fig.~\ref{fig-diff-gridellipse-aniso}f). The difference is rather subtle, however, as the fluctuations in the length of the grid axes, and our choosing always the major axis of the best fitting ellipse, whichever it is, obviously produce an ellipticity measure distributed above 1, even with constant speed. The effect of anisotropy is more salient in orienting the ellipses, Fig.~\ref{fig-diff-gridellipse-aniso}b, around the preferred directions of the speed profile, a clear difference from the uniformly distributed ellipse orientation of simulations with isotropic speed, Fig.~\ref{fig-diff-gridellipse-aniso}e. The difference in orienting ellipses can be quantitatively measured by defining {\em cross-trial ellipse orientation coherence score} for the one-dimensional average distribution of ellipse orientations. The cross-trial ellipse orientation coherence score is calculated the same way as the cross-trial grid orientation coherence score, except that 90-degree periodicity is assumed, instead of 30-degree periodicity. With speed anisotropy in cylinder environments, the cross-trial coherence in ellipse orientation is as high as 1.533. However, with constant speed, still in cylinder environments, the corresponding coherence score is close to zero, indicating a uniform ellipse orientation distribution. Speed anisotropy orients ellipses rather loosely, with a width at half-peak around 30 degrees. It is reflected also in which of the three grid axes tends to be the longest, but without apparently forcing a rigid orientation of the grids themselves, Fig.~\ref{fig-diff-gridellipse-aniso}a. 

\begin{figure*}
\centering
\pspicture(0,0)(15,8.2)

\rput[bl](0,4.2){\epsfig{file=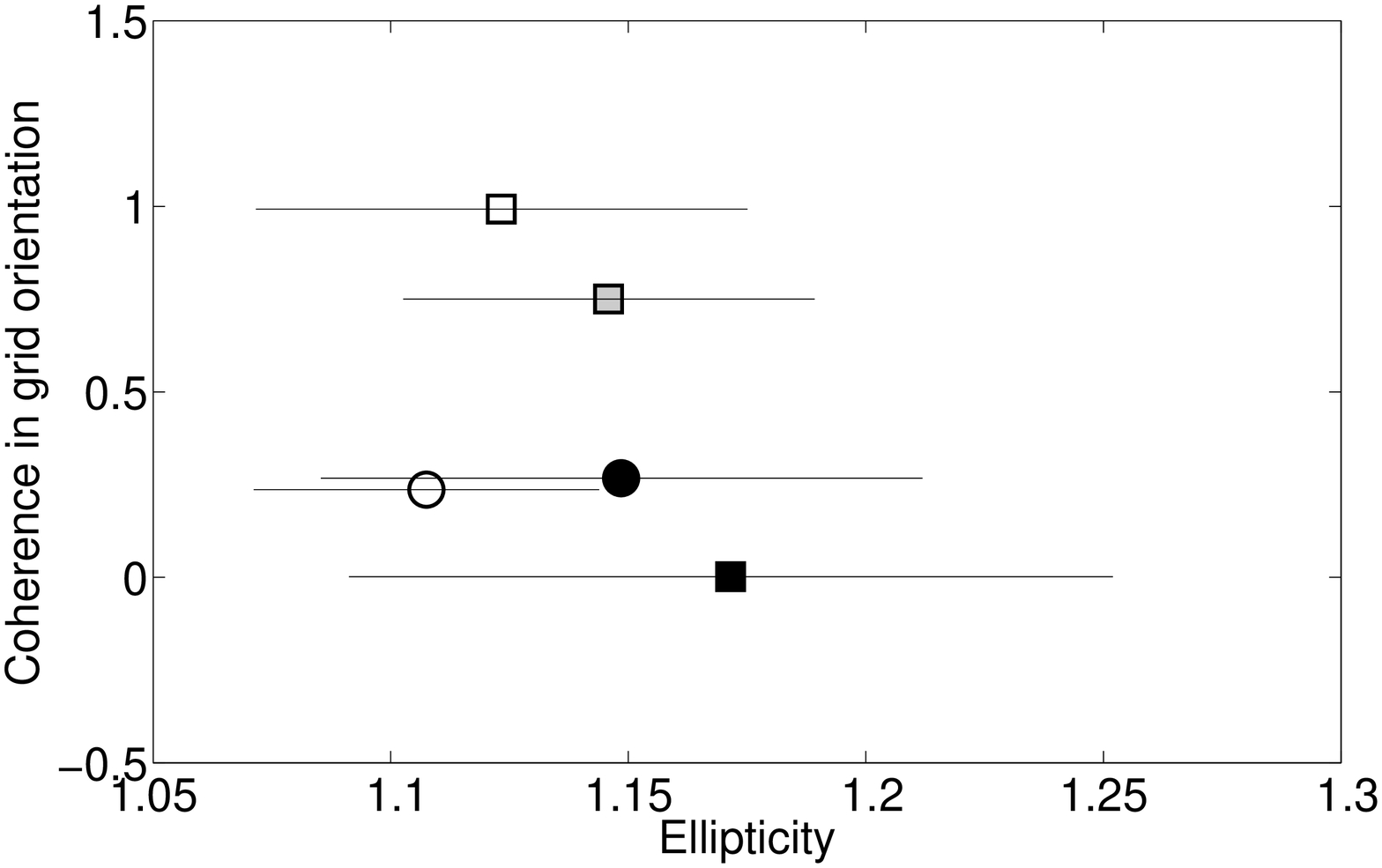,width=5.8cm}}
\rput[bl](5.9,4.2){\epsfig{file=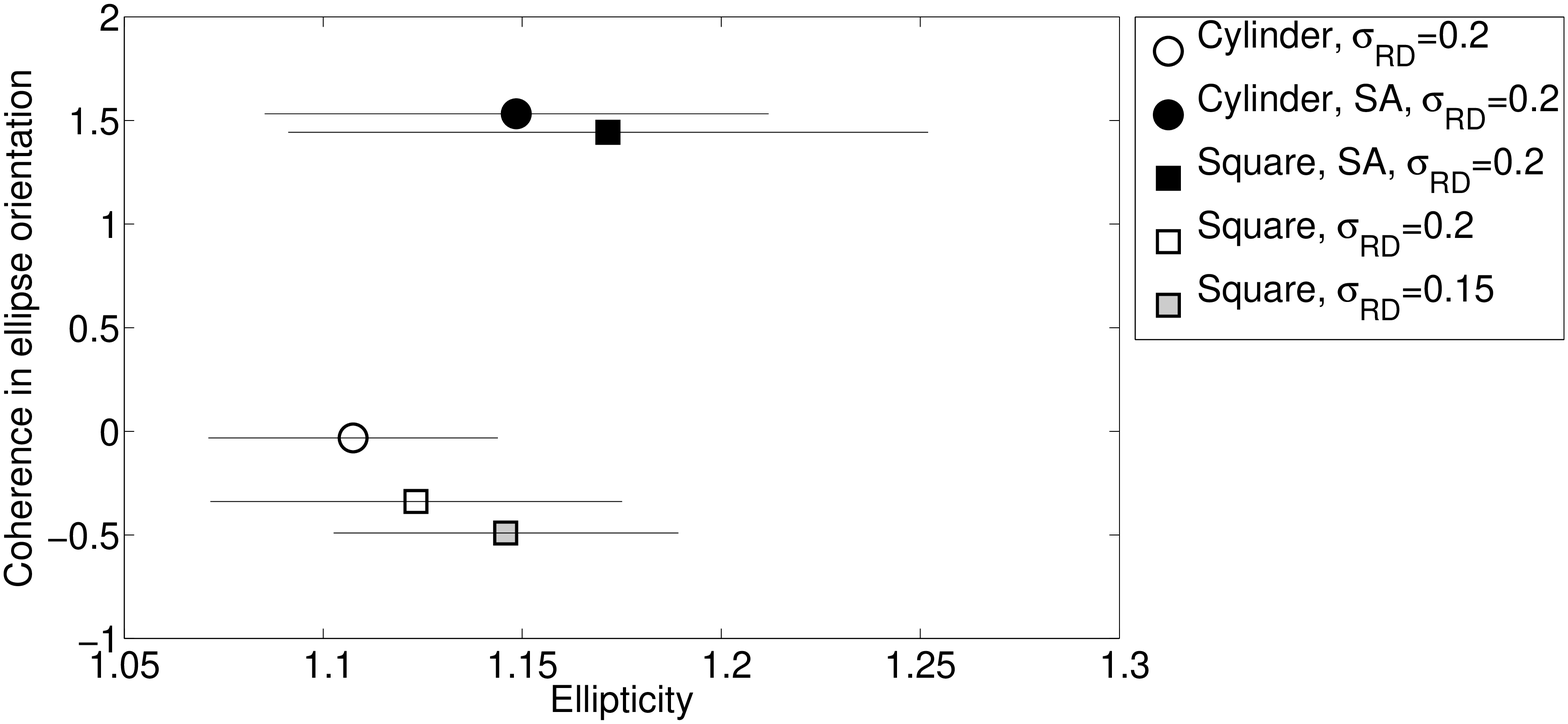,height=3.7cm}}
\rput[bl](0,0){\epsfig{file=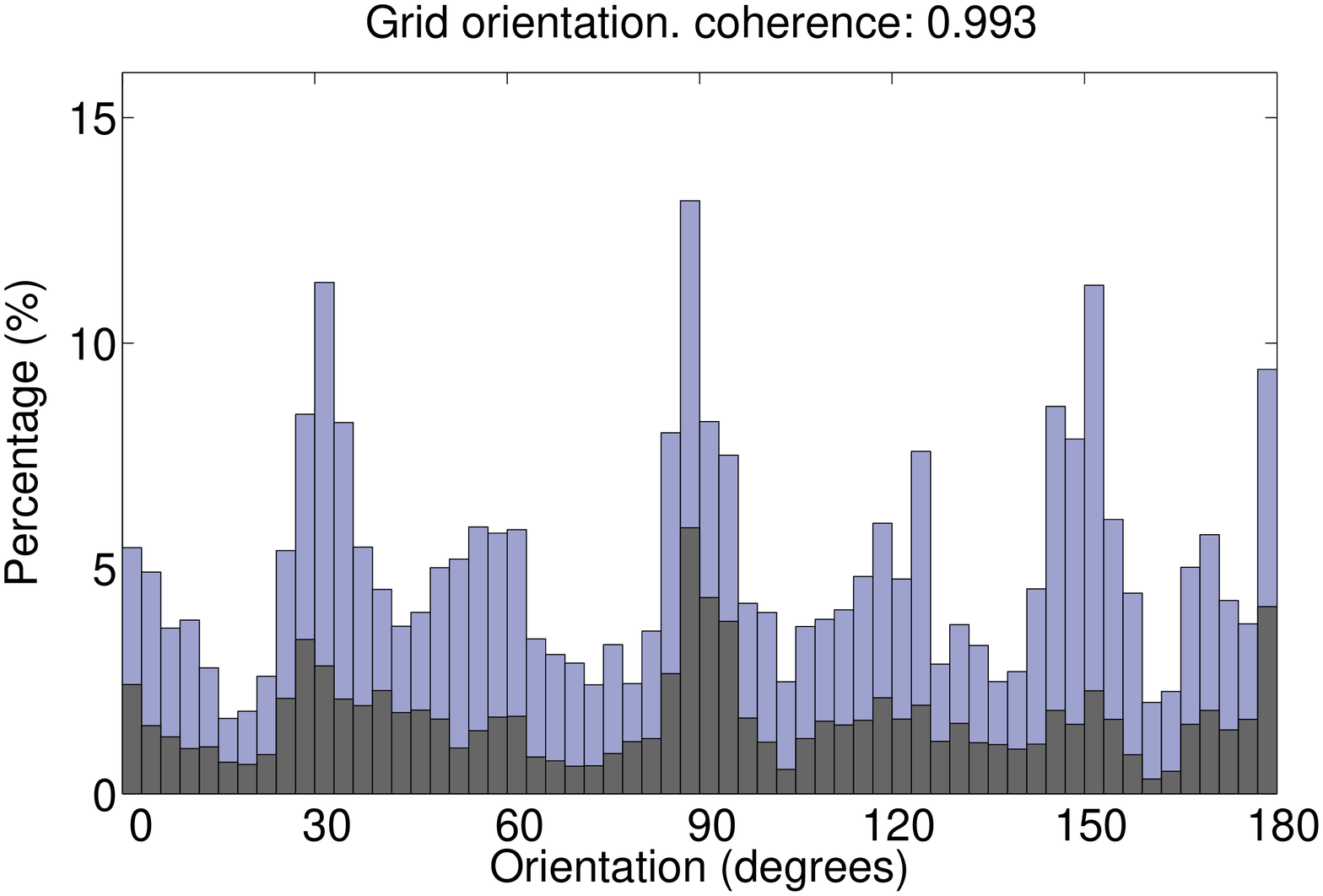,width=5.8cm}}
\rput[bl](5.9,0){\epsfig{file=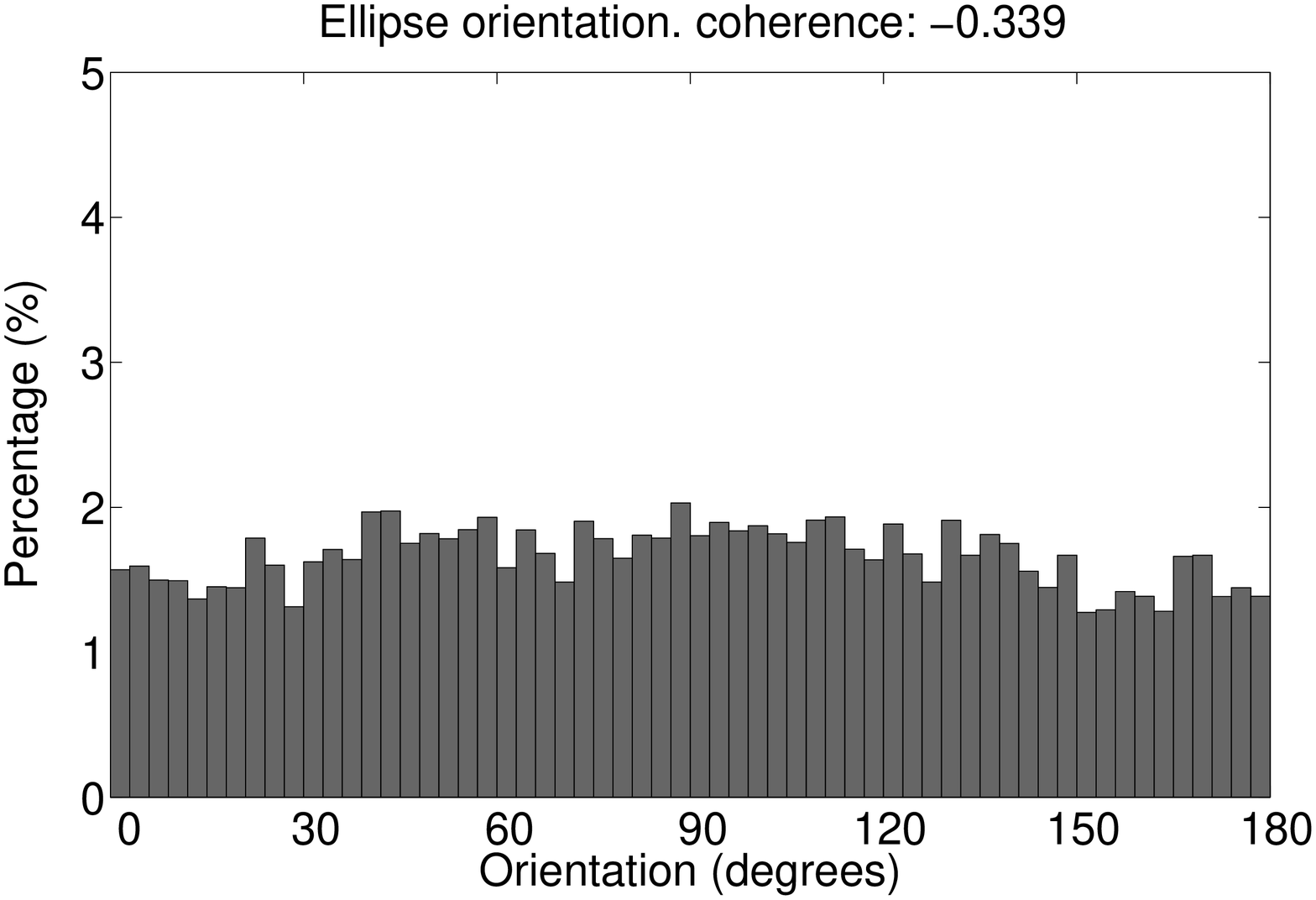,width=5.8cm}}

\psframe[linewidth=3pt,linecolor=cyan](-0.1,-0.1)(11.8,4.1)

\rput[bl](0.2,8){\text{\small{\bf a}}}
\rput[bl](6.4,8){\text{\small{\bf b}}}
\rput[bl](0.2,3.8){\text{\small{\bf c}}}
\rput[bl](6.4,3.8){\text{\small{\bf d}}}
\endpspicture
\caption{Speed anisotropy and RD anisotropy have different effects on grid shape and orientation. (a,b) Five situations are compared with respect to coherence in grid/ellipse orientation and to ellipticity, based on the average over 70 independent simulations. The horizontal lines indicate $\pm$ one standard deviation in ellipticity.  The two panels share the same legend to the right;  (c) The average orientation distribution of the long grid axes (gray, in front) and the average orientation distribution of all three grid axes (blue, in the back) for the simulations in square environments without speed anisotropy, the same data as shown also in Fig.~\ref {fig-grid-ori-bound}d. The coherence in grid orientation is noted at the top of the panel. (d) The average distribution of ellipse orientation, also for the simulations in square environments without speed anisotropy, with the corresponding coherence at the top.} 
\label{fig-comp-speed-boundary}
\end{figure*}
Fig.~\ref{fig-comp-speed-boundary} shows a summary of the differences between speed anisotropy and RD anisotropy, once coupled with RC interactions, with respect to three measures, namely, cross-trial coherence in grid orientation, cross-trial coherence in ellipse orientation, and mean ellipticity. One can see that both anisotropies slightly distort the shape of grid maps into a somewhat more pronounced elliptical arrangement of the firing fields. Speed anisotropy induces on average more distortion (with our parameters), but the average effect is overshadowed by the large variability from unit to unit, in what is an asymmetric ellipticity distribution with a long tail. The qualitative difference is in what common orientation emerges: speed anisotropy loosely orients {\em ellipses} toward the fast directions (larger cross-trial coherence in ellipse orientation), but it does not enforce a tight grid orientation, across simulations, on the aligned grids (dark circle in Fig.~\ref{fig-comp-speed-boundary}a-b). In contrast, RD anisotropy orients {\em grids} toward preferred RDs (larger cross-trial coherence in grid orientation, white and gray squares in Fig.~\ref{fig-comp-speed-boundary}a-b), but not ellipses. When combined with RD anisotropy, speed anisotropy prevails (the dark square in Fig.~\ref{fig-comp-speed-boundary}a-b). This is likely because speed aniso-tropy induces the long grid axes to lie along either 0 or 90 degrees, and it constraints the lengths (shorter) and orientation (not at 60 degrees) of the other two grid axes; with RD anisotropy, the arrangement of the fields maybe more dependent on the details of the trajectory taken in each simulation, which does not appear to restrict much the orientation of the ellipses, seemingly leaving the grids free to orient themselves. We may conclude that the shape of the environment and speed anisotropy cause apparently similar, but subtly distinct effects.

\subsection{Grid alignment without collaterals}\label{anal_quadrupole}

In the simulations above, collateral interactions among would-be conjunctive units tend to align their developing grid fields along common axes. We have considered two additional factors that influence the alignment, the shape of the environment, if different from cylindrical, and anisotropy in running speed. The latter factor, in particular, enhances the ellipticity of the resulting grids, and affects their ellipse orientation. Its effects are however secondary, in the simulations, to the primary effect produced by the collaterals. It is important to note that ellipticity and some degree of common orientation, however, can be produced also by speed anisotropy on its own, in the {\em absence} of collaterals. This can be understood by considering a simple abstract model, which extends the one earlier considered in \cite{Kro+08} to account for the development of grid fields at the single unit level. The model describes a single unit in a very large environment, which then for all practical purposes has no shape. Hence the abstract model focuses on the effect of speed anisotropy, which is modeled similarly as in the simulations, extricating it from the other two factors, the presence of collaterals, and the shape of the environment. The model, described in the Appendix, leads to a relation between the degree of anisotropy and the degree of ellipticity, and it indicates two orientations, at roughly 90 degree of each other, of the grids and of the ellipses that best match the assumed quadrupole anisotropy. Hence it predicts that grids and ellipses would have one of two common orientations even in the absence of collateral interactions. 

\begin{figure*}
\centering
\begin{minipage}{17.4cm}
\pspicture(0,0)(17.4,4.3)

\rput[bl](0,0){\epsfig{file=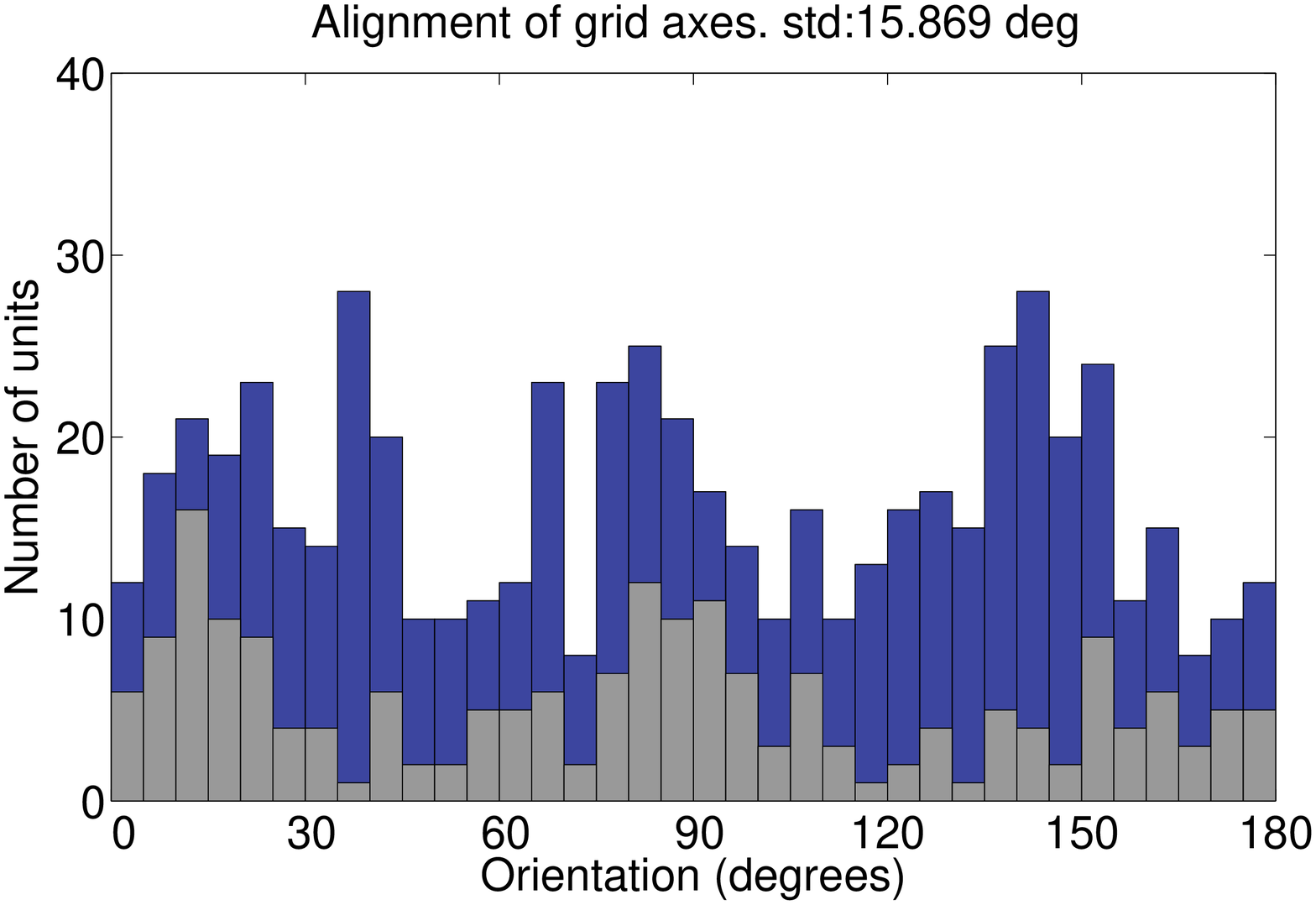,width=5.8cm}}
\rput[bl](5.8,0){\epsfig{file=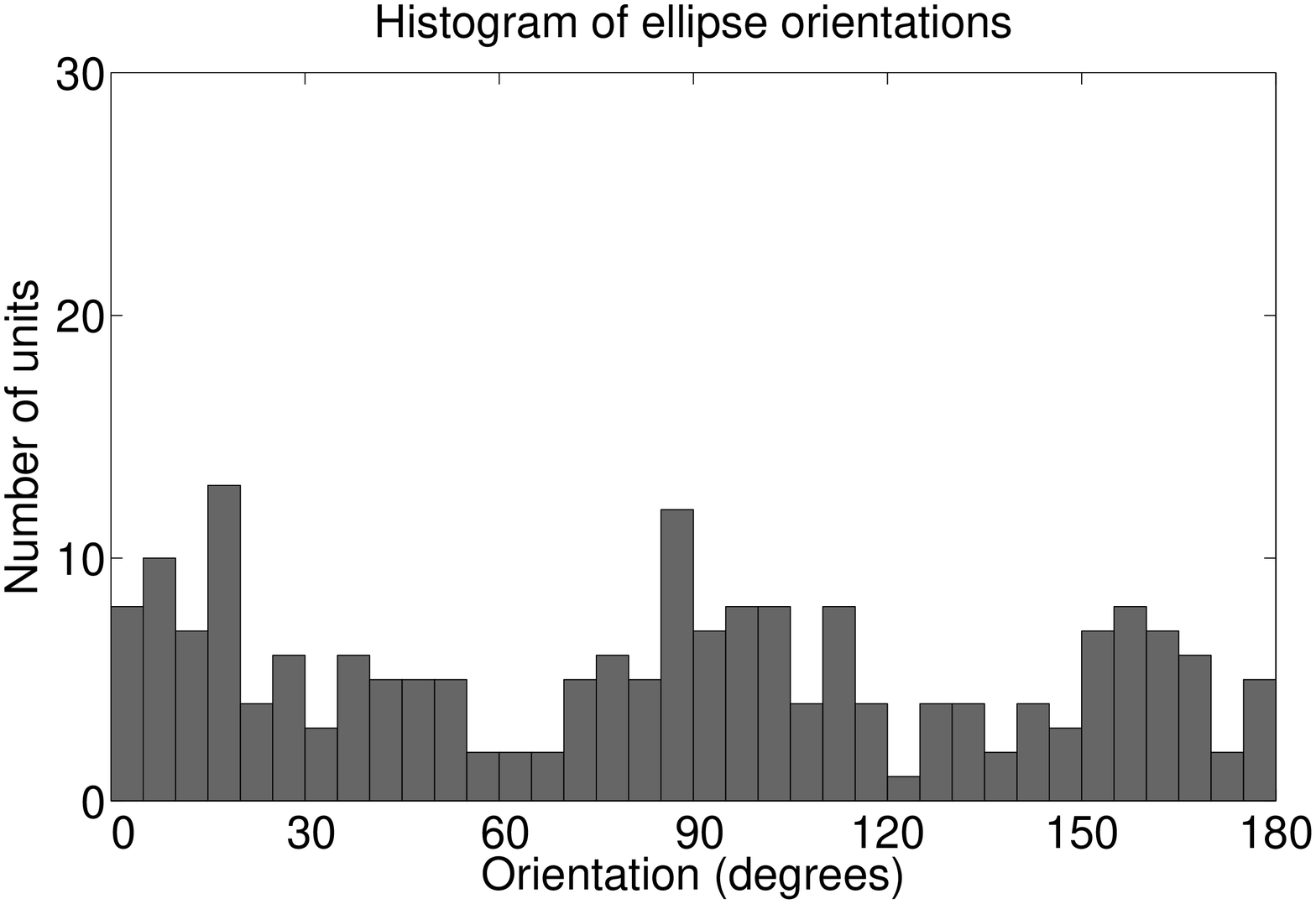,width=5.8cm}}
\rput[bl](11.6,0){\epsfig{file=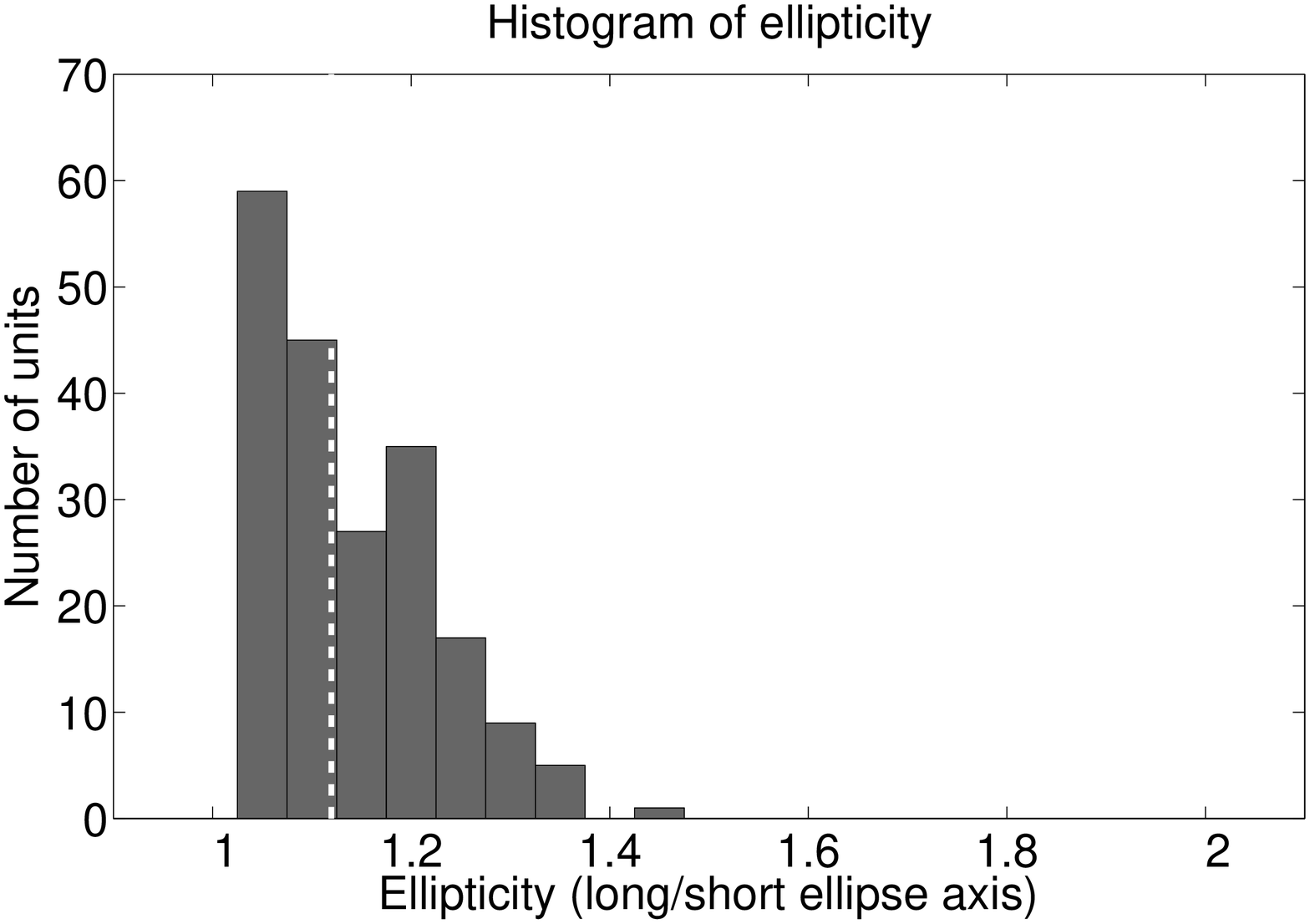,width=5.8cm}}

\rput[bl](0.4,4){\text{\small{\bf a}}}
\rput[bl](6.4,4){\text{\small{\bf b}}}
\rput[bl](12.2,4){\text{\small{\bf c}}}
\endpspicture
\end{minipage}
\caption{Grids show a weak tendency to align due to speed anisotropy alone, without collateral connections in the network ($\rho=0$). (a) Histogram of the orientations of the grid axes (blue, in the back) together with the histogram of the long grid axes (light gray, in the front). The coherence in grid alignment is low (the mean standard deviation, in degrees, of the orientation averaged over the three grid axes is shown at the top of the panel, and it is much higher than with collaterals, shown in Fig.~\ref {fig-aniso}e); (b) Orientation histogram of the major axes of the ellipses; (c) Histogram of ellipticity.}
\label{orient_no_collaterals}
\end{figure*}
The effect predicted by the analytical model is reproduced in {\em ad hoc} simulations. Fig.\ref{orient_no_collaterals} shows that in the presence of speed anisotropy only, without collateral interactions, grid units develop with an enhanced ellipticity, and with one of the grid axes preferentially aligned along one of the two orthogonal directions with higher mean speed (in the simulation of Fig.~\ref{orient_no_collaterals}, this happens to be the one at 90 degrees). The variance in ellipse orientation is considerable, larger than what is observed with speed anisotropy {\em and} collateral interactions (Fig.~\ref{orient_no_collaterals}b
vs. Fig.~\ref{fig-aniso}g). However, the degree of ellipticity is similar, irrespective of the existence of collateral interactions (Fig.~\ref{orient_no_collaterals}c
and Fig.~\ref{fig-aniso}i). 
The first element that is missing, without collateral interactions, is crucially the tight alignment of the grids with each other. This is because, in our model, collateral interactions align the grids with each other through mutual iterative convergence, whereas without collaterals the alignment only reflects single unit adaptation to what is effectively a broad shallow valley in a free energy landscape. We present the abstract model in the Appendix for clarity, but we believe it to be unlikely that, in the absence of collateral interactions, speed anisotropy alone can establish a common grid orientation with the tight alignment seen in experimental data. The second element that is missing without collateral interactions is the invariance in the relative phase of grid maps across multiple environments, which we will examine in the next section.

\section{Grid realignment in multiple environments}\label{sec-remap}

Until now, we have only considered grid alignment in a single environment. Under dramatic environmental changes, both the hippocampal and entorhinal neurons develop new maps, in a process called global remapping~\citep{Colgin2008}. While hippocampal place fields seem to shuffle randomly during global remapping, grid cells behave in a population-coherent manner. The new maps preserve the same relative phases as the old maps, and maintain a common grid orientation, not necessarily the same one as before~\citep{Fyh+07}. In this section, we address the question of how grid fields may possibly align in multiple environments. 

We simulate the network in two different cylinder environments, with identical conditions as in Section~\ref{sec-sylind}, except for the number of units. The total number of conjunctive units is 500. The total number of place units is 900. In each environment, 500 of the place units are active. The number of place units that are active in both environments is 100, i.e. 20\% of the active place units in one environment. The place fields of the place units in the first environment are completely different from those in the second environment. However the preferred HDs of the conjunctive units are the same across environments. The training is interleaved in the two environments, with 2000 epochs in each environment, 3000 steps in each epoch. This leads to $1.2\times 10^7$ training steps in total.

\begin{figure*}
\centering
\begin{minipage}{17.4cm}
\pspicture(0,0)(17.4,21.3)
\rput[bl](0.2,18.3){\epsfig{file=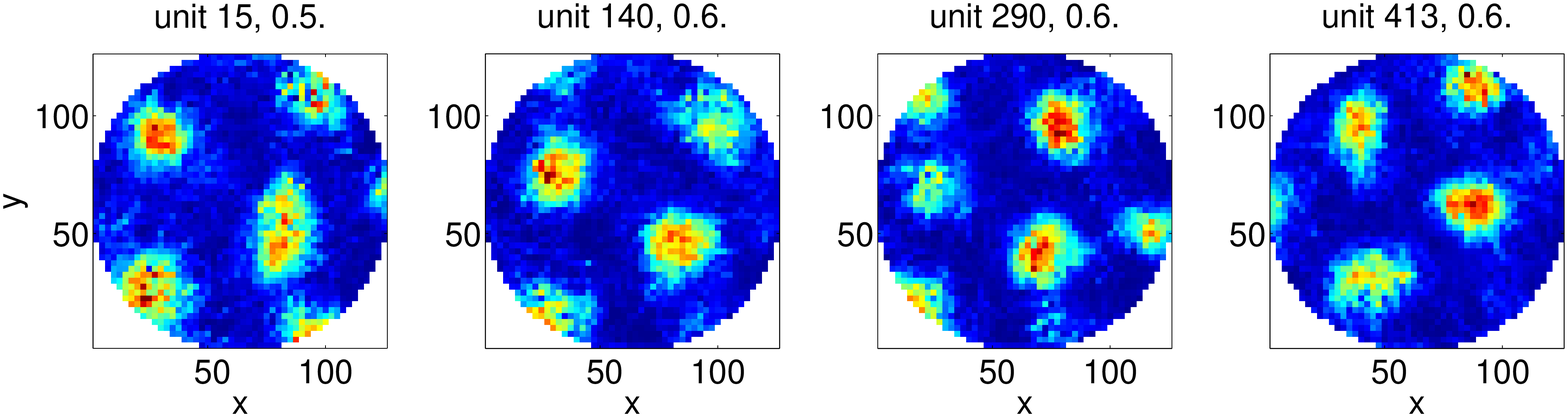,,width=10cm}}
\rput[bl](0.2,15.4){\epsfig{file=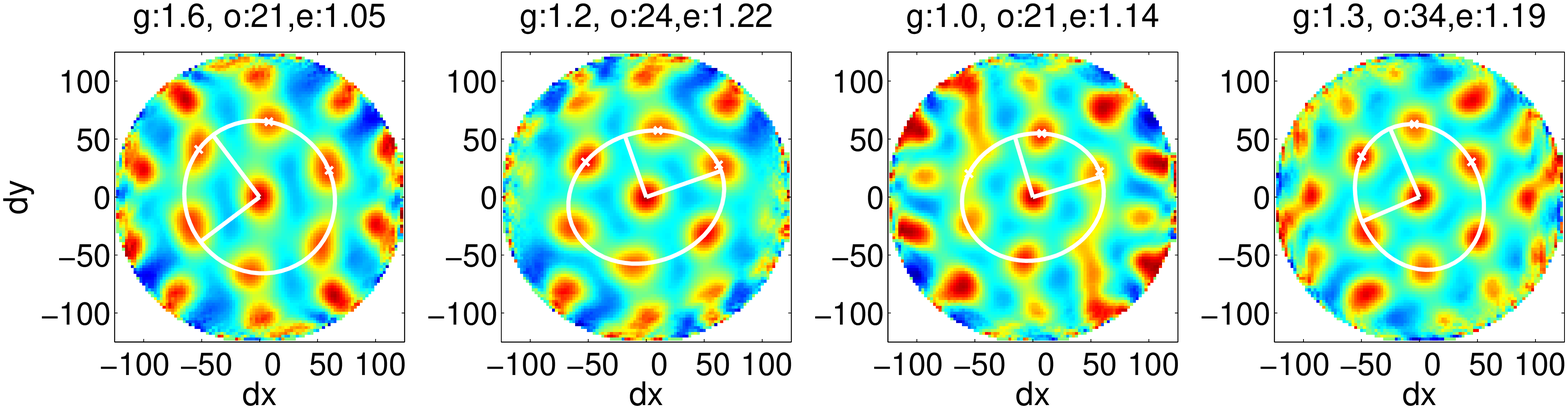,width=10cm}}
\rput[bl](10.2,15.4){\epsfig{file=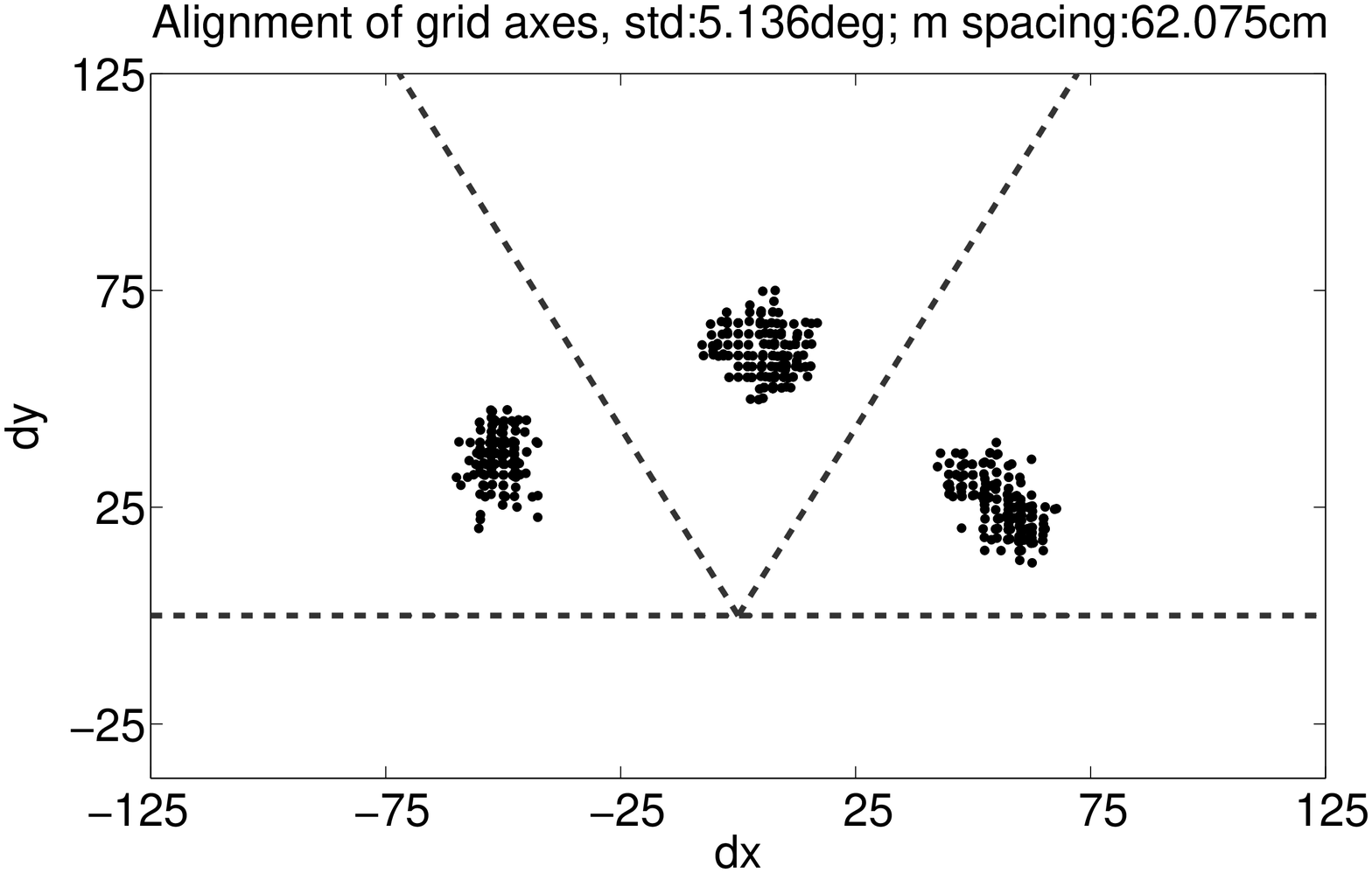,width=7cm}}

\rput[bl](0,10.8){\epsfig{file=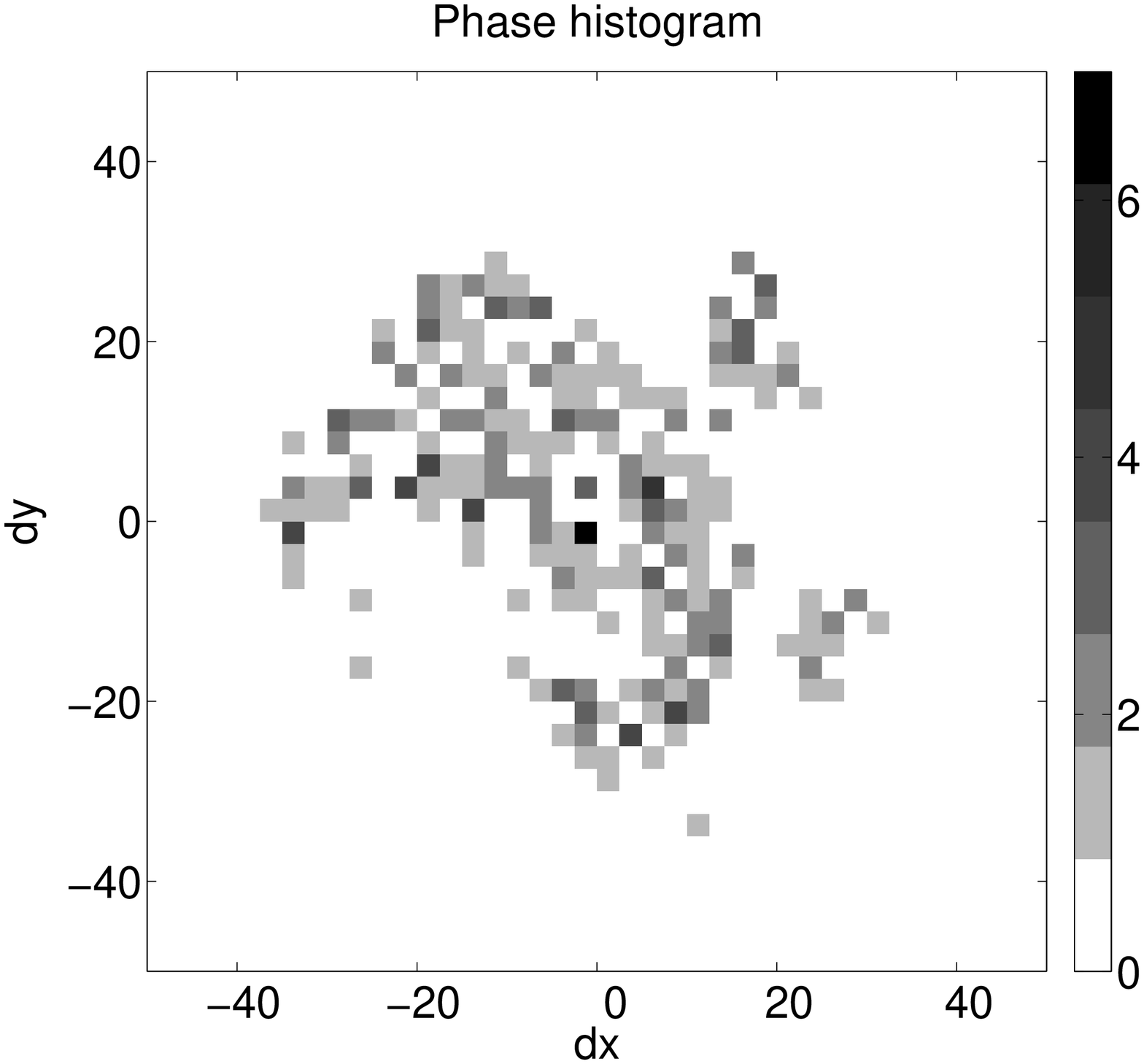,height=4.4cm}}
\rput[bl](5.2,10.8){\epsfig{file=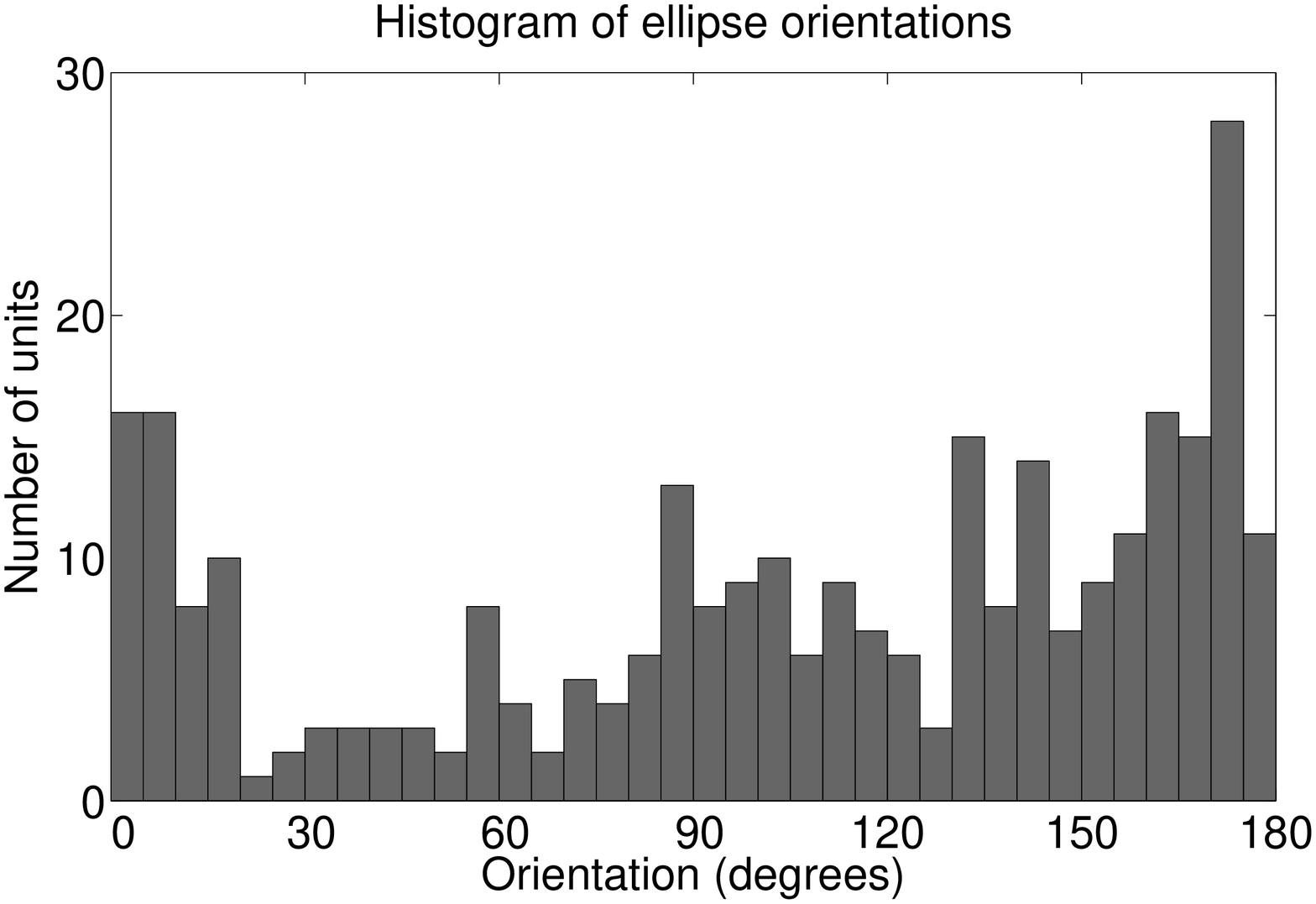,width=5.8cm}}
\rput[bl](11.1,10.8){\epsfig{file=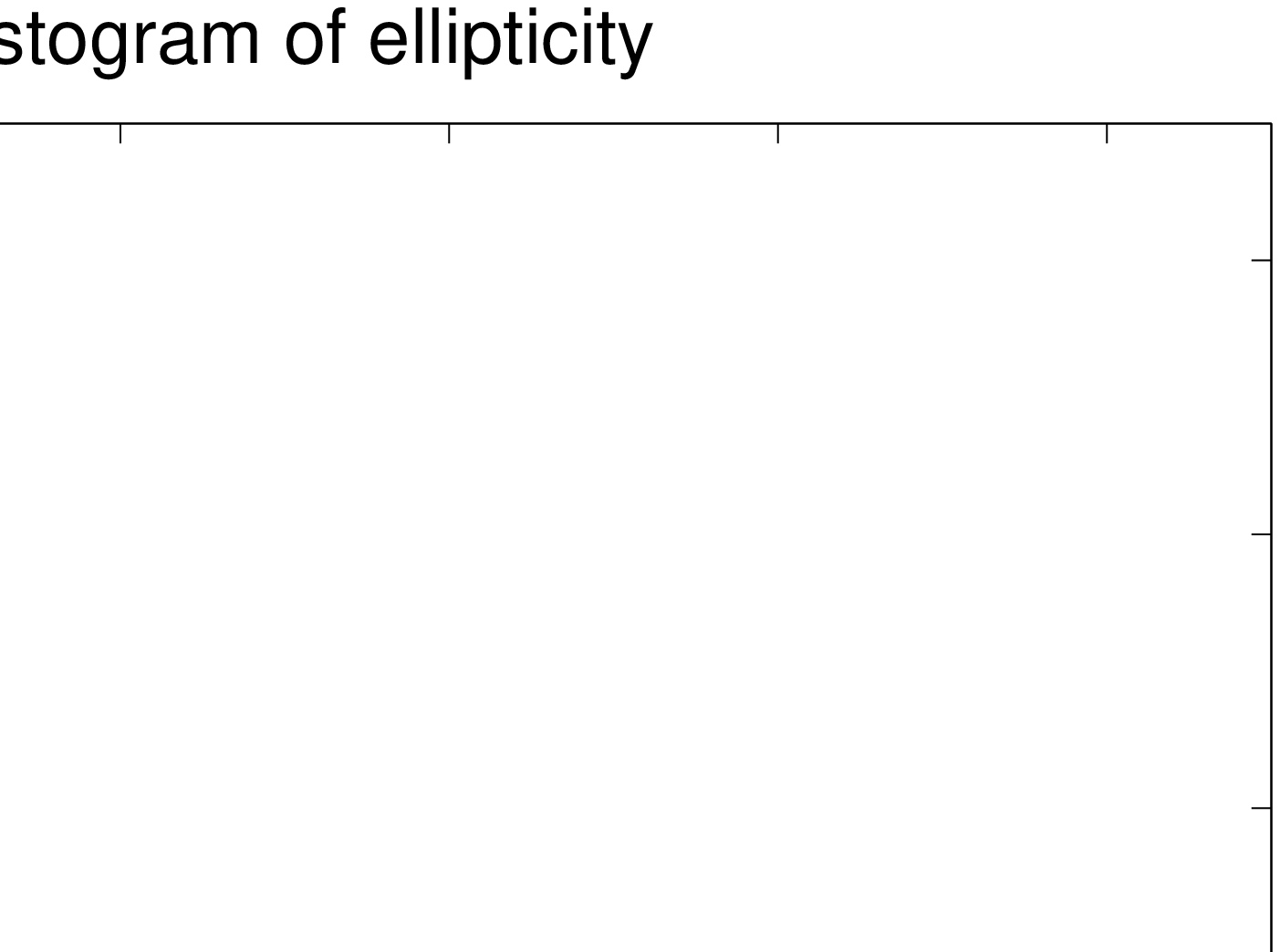,width=5.8cm}}

\rput[bl](0.2,7.5){\epsfig{file=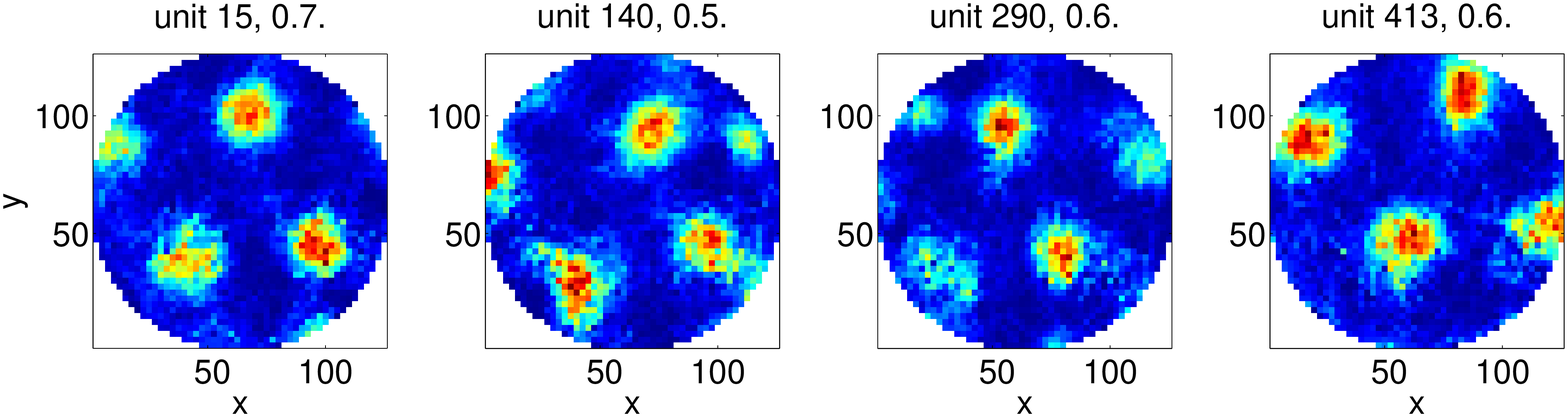,width=10cm}}
\rput[bl](0.2,4.7){\epsfig{file=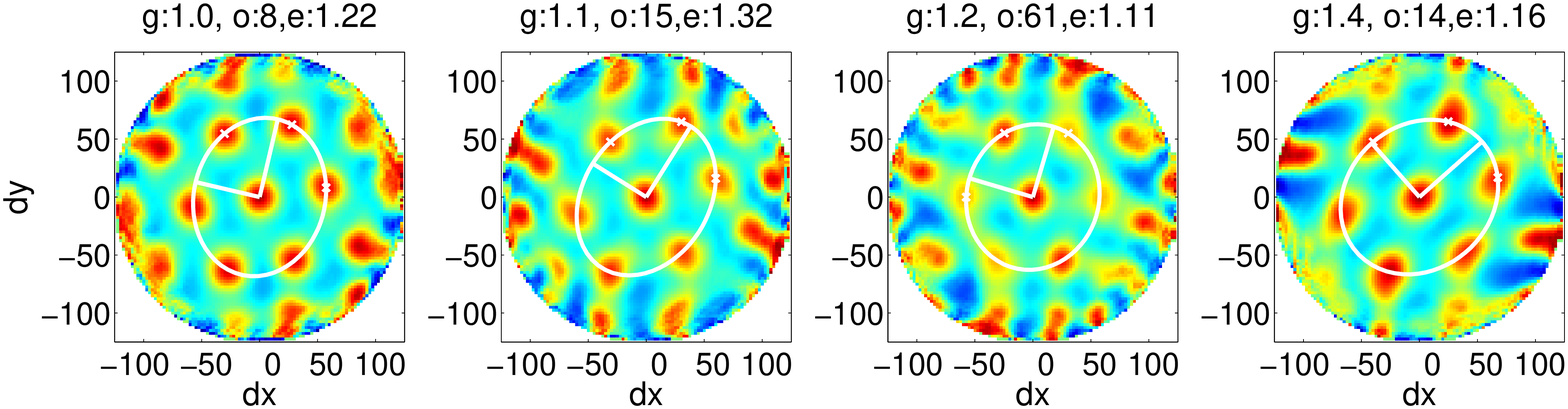,width=10cm}}
\rput[bl](10.2,4.7){\epsfig{file=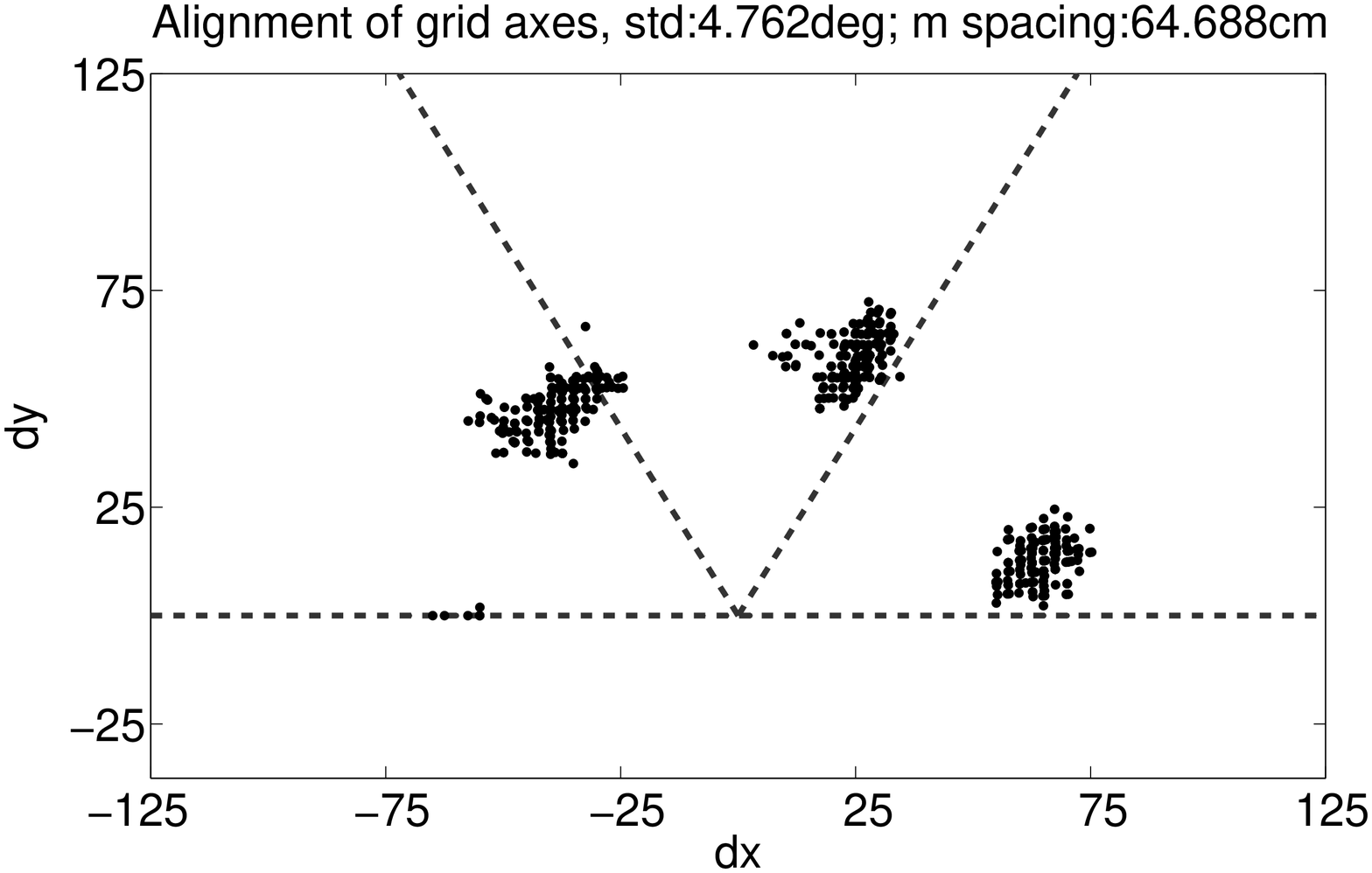,width=7cm}}
\rput[bl](0,0){\epsfig{file=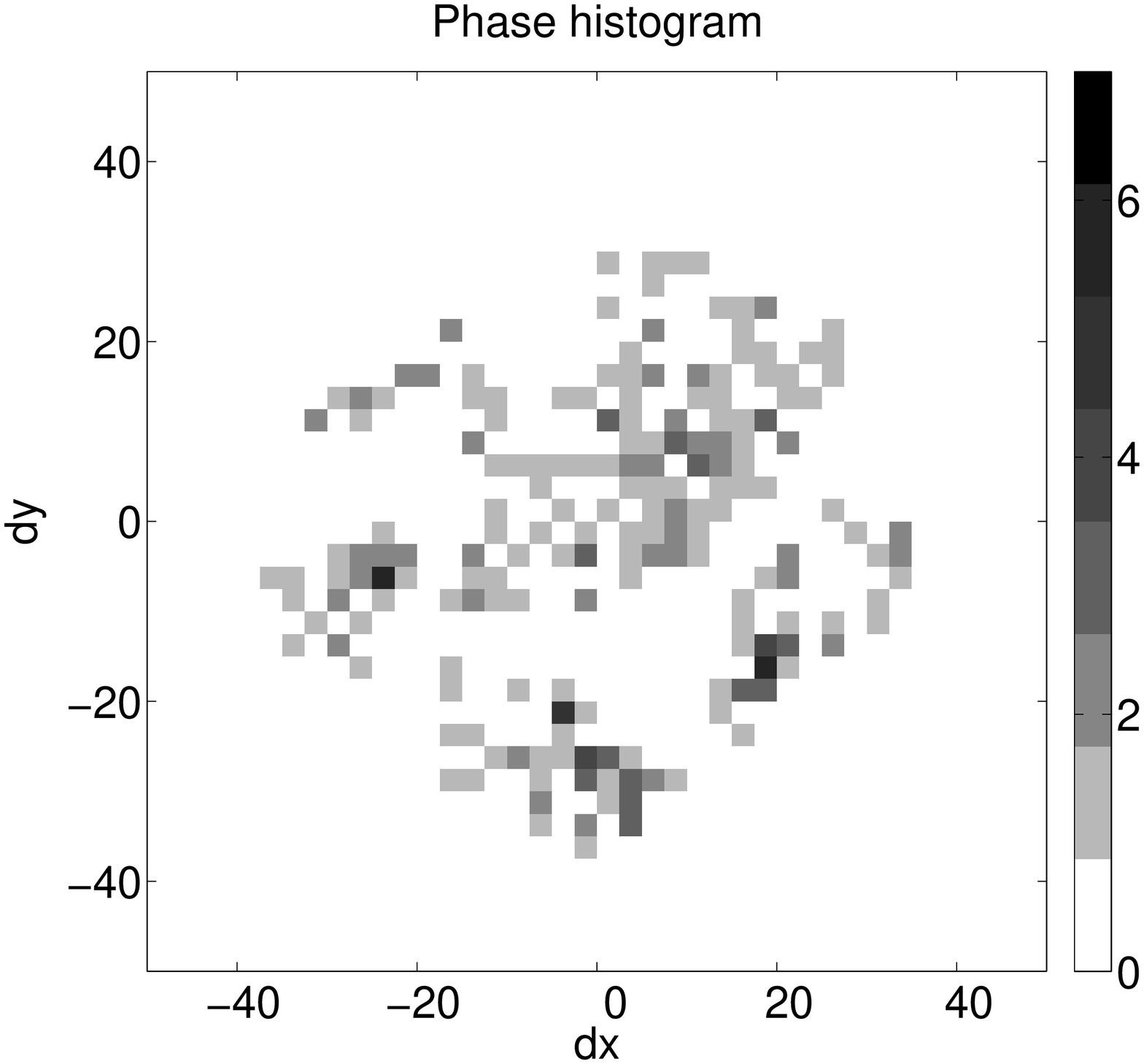,height=4.5cm}}

\rput[bl](5.2,0){\epsfig{file=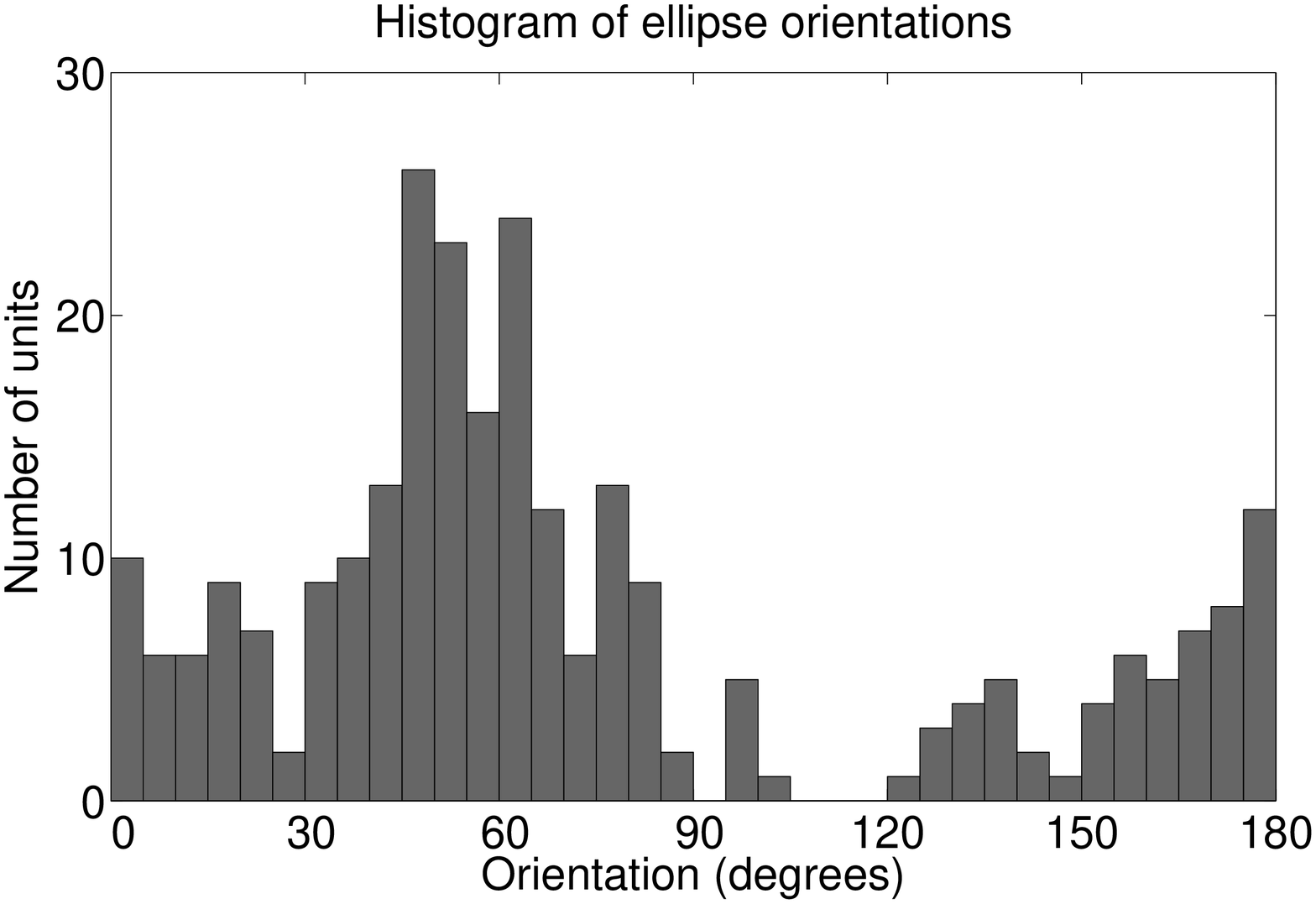,width=5.8cm}}
\rput[bl](11.1,0){\epsfig{file=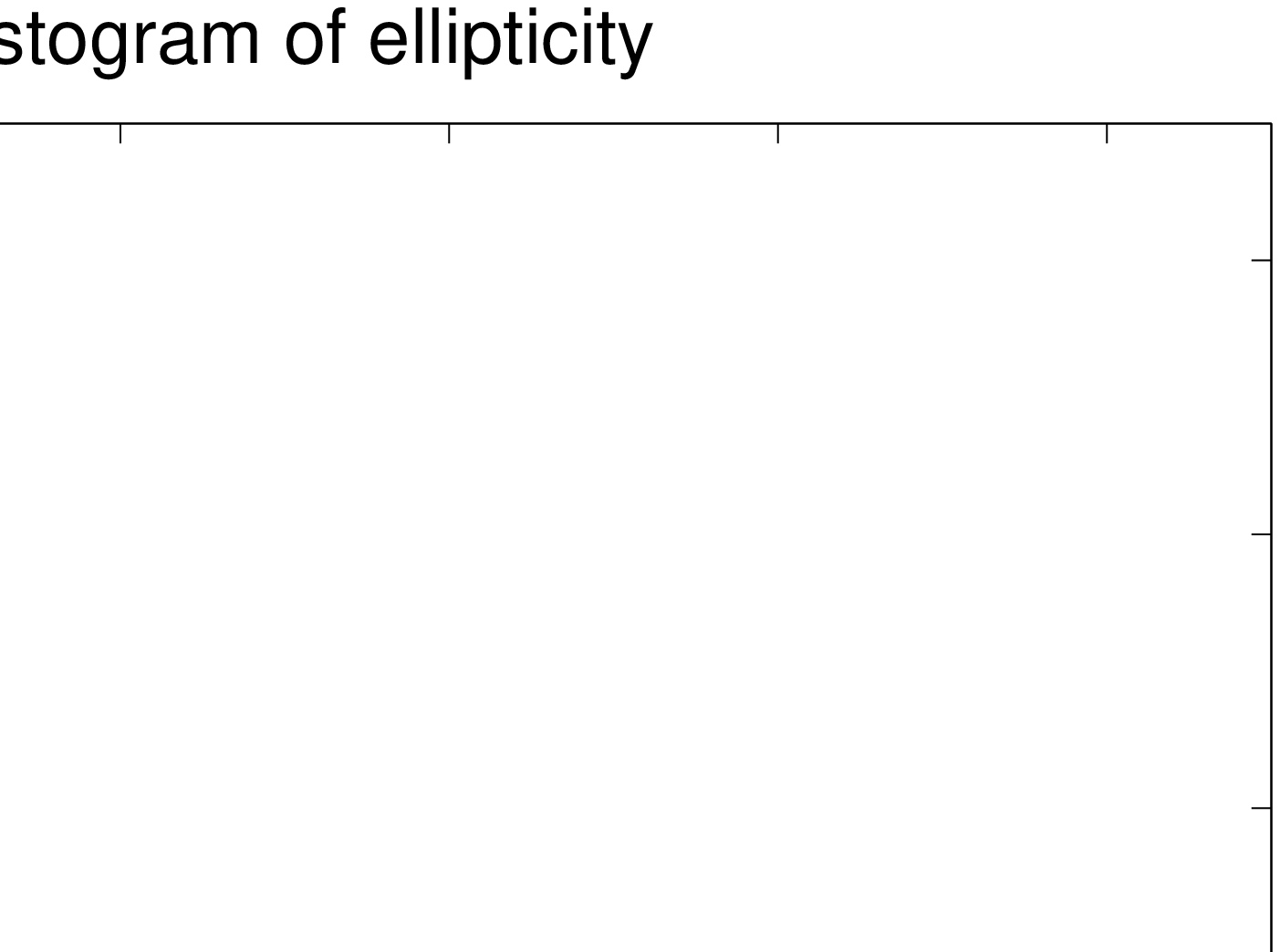,width=5.8cm}}

\psframe[linewidth=3pt,framearc=.1,linecolor=yellow](0,10.7)(17.3,21.2)
\psframe[linewidth=3pt,framearc=.1,linecolor=magenta](0,-0.1)(17.3,10.4)

\rput[bl](0.4,20.8){\text{\small{\bf a}}}
\rput[bl](11,20){\text{\small{\bf b}}}
\rput[bl](0.4,15.2){\text{\small{\bf c}}}
\rput[bl](5.8,15){\text{\small{\bf d}}}
\rput[bl](11.8,15){\text{\small{\bf e}}}

\rput[bl](0.4,10){\text{\small{\bf f}}}
\rput[bl](11,9.4){\text{\small{\bf g}}}
\rput[bl](0.4,4.3){\text{\small{\bf h}}}
\rput[bl](5.8,4.3){\text{\small{\bf i}}}
\rput[bl](11.8,4.3){\text{\small{\bf j}}}
\endpspicture
\end{minipage}
\caption{Grid realignment in environment A (top) and environment B (bottom). (a,f) Examples of the spatial fields and autocorrelograms of four different units in the two environments; The unit number and maximal firing rate (in arbitrary units) are indicated above each rate map. The gridness score, grid orientation (in degrees, between the x-axis and the nearest grid axis) and ellipticity are indicated above each autocorrelogram; (b,g) Scatter plots of the three peaks found in autocorrelograms. The angular offset of the grids in environment B relative to the grids in environment A is 9 degrees clockwise. Noted at the top of the panel are the mean spacing and coherence score in grid alignment (mean standard deviation, in degrees, of the orientations averaged over the three grid axes); (c,h) Two dimensional histograms of spatial phases relative to the best grid in each environment. Each spatial bin represents a $2.5 cm \times 2.5 cm$ area; white indicates zero units with a phase in that spatial bin, with more units indicated by progressively darker gray; (d,i) Histograms of the ellipse orientations; (e,j) Histograms of the ellipticity, i.e. the ratio between the length of the major and minor axes of an ellipse, across all units.}
\label{fig-remapping}
\end{figure*}

Fig.~\ref{fig-remapping}a,f compares the spatial fields and corresponding autocorrelograms of four example conjunctive units in two environments. In both environments, the grids are aligned, but to different orientations (Fig.~\ref{fig-remapping}b,g). The angular offset of the grids in the second environment relative to the grids in the first environment is about 9 degrees counterclockwise. The spatial phases of the units relative to the best grid in each environment do not show any clear pattern or clustering (Fig.~\ref{fig-remapping}c,h), as in the simulations with a single learned environment. 

The distribution of ellipse orientations is quite different across environments (Fig.~\ref{fig-remapping}d,i), but the ellipticity distribution is comparable, with a similar median at 1.16 (Fig.~\ref{fig-remapping}e,j). This value is similar to that of the simulations with speed anisotropy. This is because 20\% of the place units are active in both environments but in randomly different positions. Since conjunctive cells need to maintain population coherence, this randomness acts as a source of noise, imposing additional constrains on conjunctive units that move their maps away from perfect gridness. This can be seen as a third source of ellipticity. The other two sources discussed in previous sections, i.e. speed anisotropy and RD anisotropy, introduce ellipticity during grid development by breaking the symmetry of the trajectories of the simulated rat. The ellipticity caused by grid realignment in multiple environments is due to the overlap of the population codes of contextual information, and is imposed by the structure of the network. 

We then determine whether the relative spatial phases of the units are kept invariant across environments. For this we select a sub-population, by taking into account only units that have good grid maps in both environments (gridness score $> 0.75$). The firing map in environment B of each unit in the population is rotated counterclockwise at multiples of 4.5 degrees and cross-correlated with the corresponding map in environment A, as in~\citep{Fyh+07}.  The cross-environment crosscorrelograms thus obtained are averaged over the population, to get the mean cross-environment crosscorrelograms for this sub-population. 

\begin{figure*}
\centering
\begin{minipage}{16.6cm}
\pspicture(0,0)(16.6,16.5)
\rput[bl](0,0){\epsfig{file=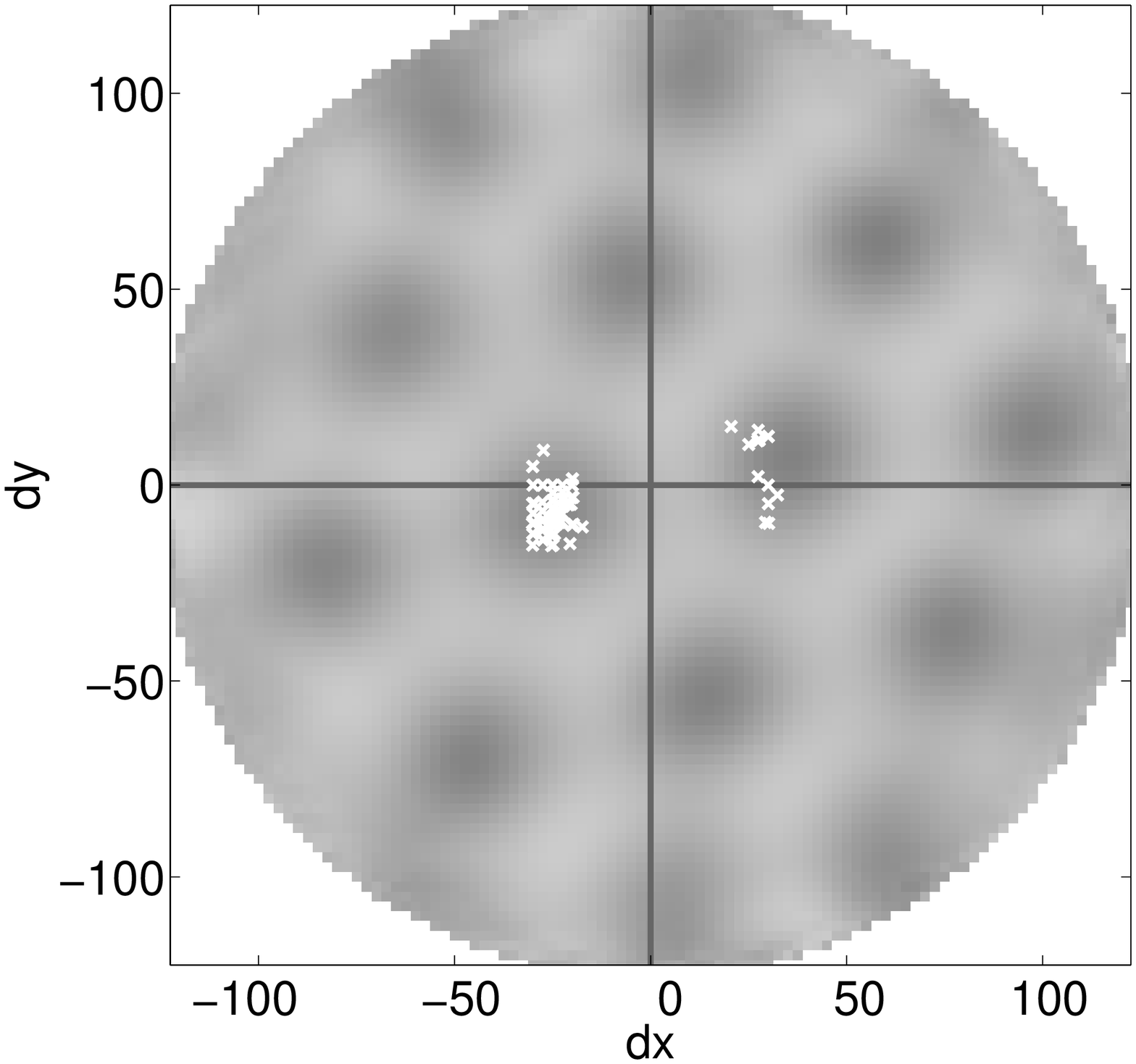,height=6.4cm}}
\rput[bl](7.3,0){\epsfig{file=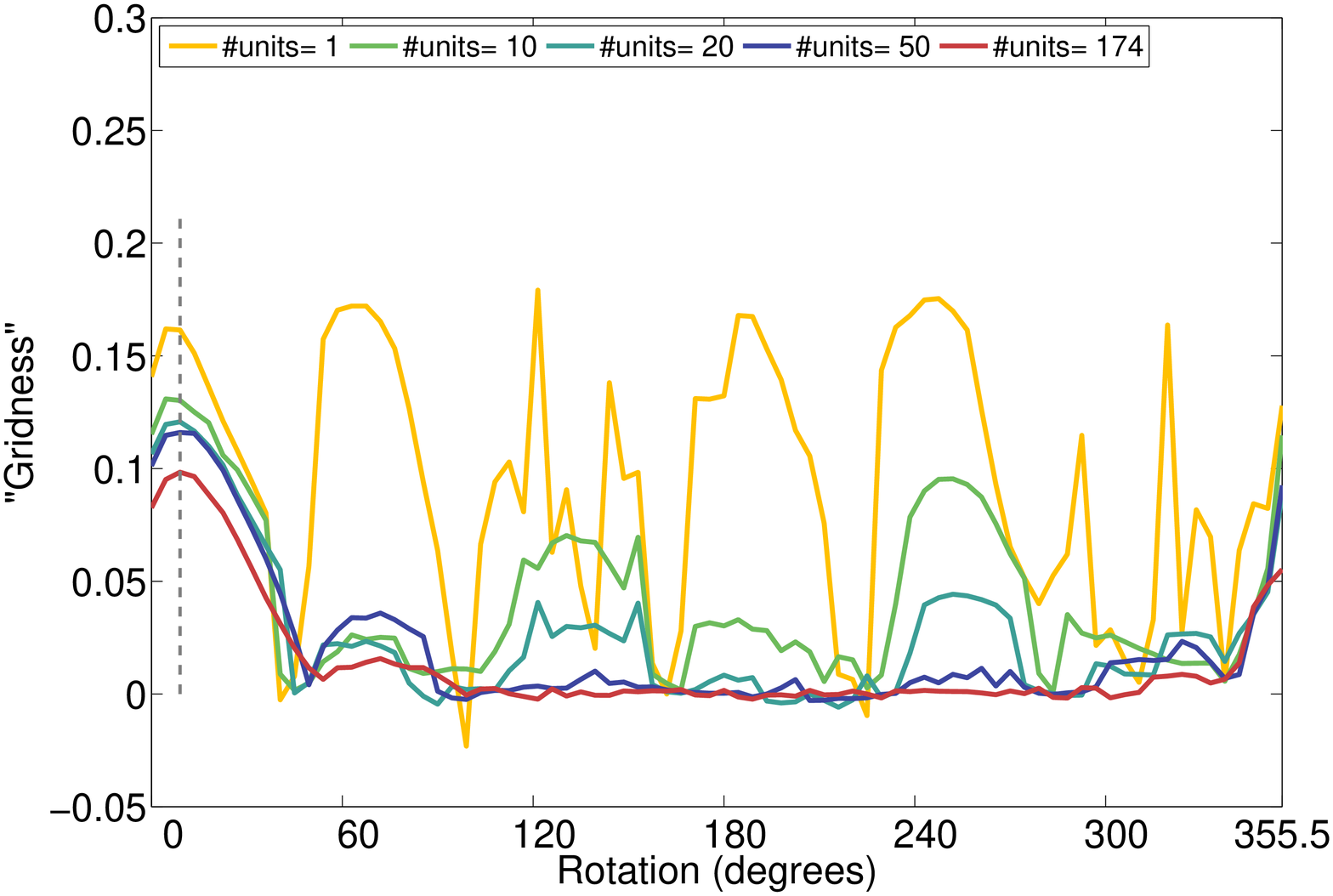,height=6.4cm}}
\rput[bl](0,6.8){\epsfig{file=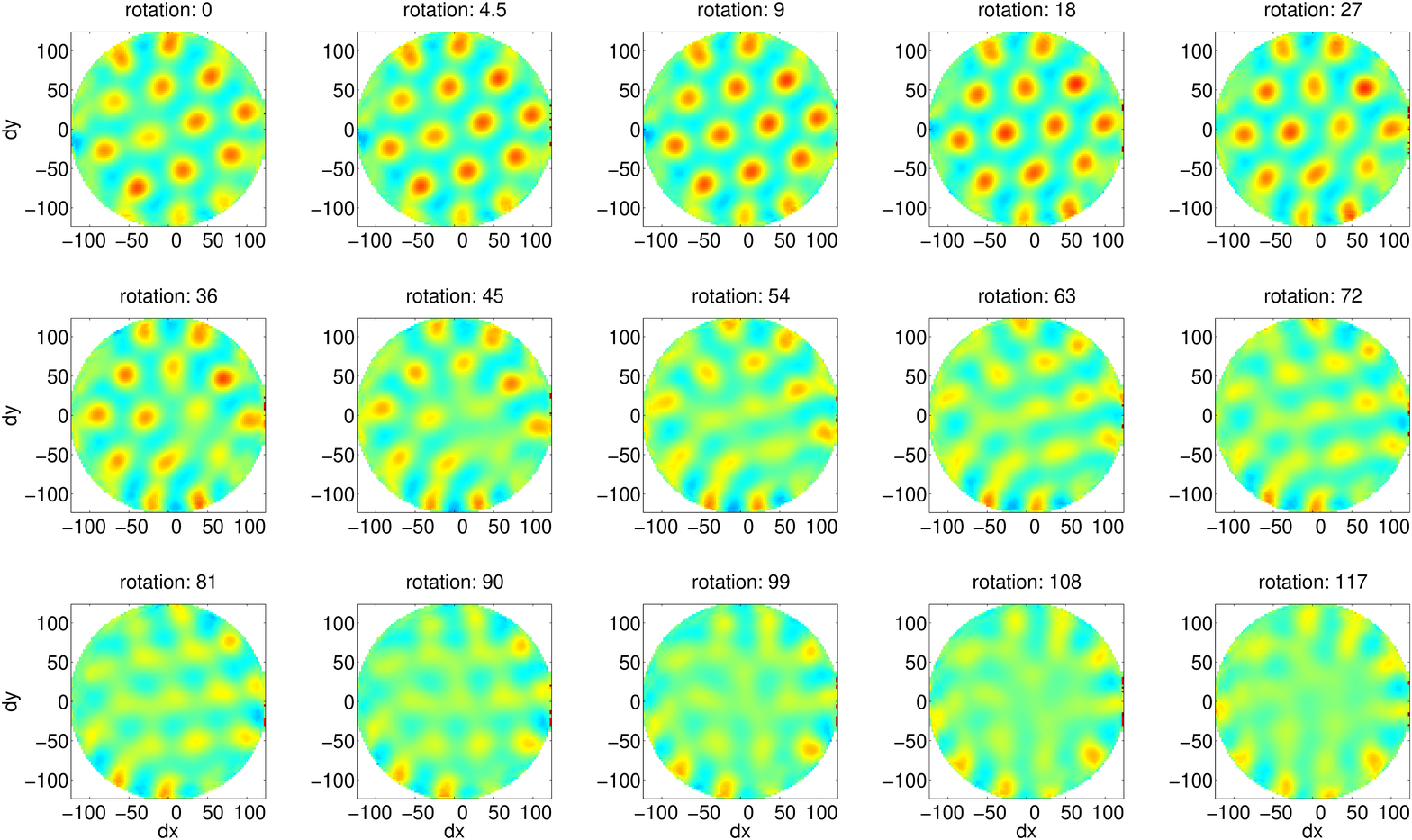,width=16.2cm}}

\rput[bl](0.4,16.4){\text{\small{\bf a}}}
\rput[bl](0.4,6.4){\text{\small{\bf b}}}
\rput[bl](7.4,6.4){\text{\small{\bf c}}}
\endpspicture
\end{minipage}
\caption{The phases of the grids in the two environments rotate and shift together. (a) The mean cross-environment crosscorrelogram of the population after the optimal counterclockwise rotation of 9 degrees is shown, together with some other rotations. The rotation angle is noted on the top of each panel. The mean cross-environment crosscorrelogram shows the best grid structure in the case of the specific rotations that (roughly) match the angular offset between the grid axes in the two environments. The Mann-Whitney test~\citep{Zar1998} shows that the mean cross-environment crosscorrelogram after 9-degree rotation has larger mean than the crosscorrelogram after zero-degree rotation (p=0.0015); (b) After the optimal rotation, the phase shifts of the same units concentrate in a small region with size similar to the field of a grid. Note that the phase shifts of a small number of units fall in a different area, due to the periodicity of the grid and to small fluctuations in the peaks of the individual crosscorrelograms. The gray crosscorrelogram shown in the background is the mean cross-environment crosscorrelogram after the optimal rotation, as shown in a; (c) The ``gridness'' of the mean cross-environment crosscorrelogram when the population set is increased from a single grid ($\#$ units 1) to the whole selected population. The gray broken line indicates the optimal rotation leading to the highest gridness of the cross-environment crosscorrelogram.}
\label{fig-grid-displacement}
\end{figure*}

\begin{figure}
\centering
\begin{pspicture}(-2.5,-1)(2.5,3)
\newrgbcolor{lightblue}{0 0.6 0.8}
\newrgbcolor{lightgreen}{0.1961 0.8039 0.1961}
\newrgbcolor{orangered}{0.9490 0 0.3373}

\psline[linecolor=lightgray,linewidth=0.5pt,linestyle=dashed]{-}(1,-1)(1,3)
\psline[linecolor=lightgray,linewidth=0.5pt,linestyle=dashed]{-}(2,-1)(2,3)
\psline[linecolor=lightgray,linewidth=0.5pt,linestyle=dashed]{-}(0,-1)(0,3)
\psline[linecolor=lightgray,linewidth=0.5pt,linestyle=dashed]{-}(-1,-1)(-1,3)
\psline[linecolor=lightgray,linewidth=0.5pt,linestyle=dashed]{-}(-2,-1)(-2,3)

\psplot[linecolor=lightgray,linewidth=0.5pt,linestyle=dashed]{-2}{2}{x 0.5774 mul}
\psplot[linecolor=lightgray,linewidth=0.5pt,linestyle=dashed]{-2}{2}{x 0.5774 mul 1.1547 add}
\psplot[linecolor=lightgray,linewidth=0.5pt,linestyle=dashed]{-2}{1.2}{x 0.5774 mul 2.3094 add}

\psplot[linecolor=lightgray,linewidth=0.5pt,linestyle=dashed]{-2}{2}{x -0.5774 mul}
\psplot[linecolor=lightgray,linewidth=0.5pt,linestyle=dashed]{-2}{2}{x -0.5774 mul 1.1547 add}
\psplot[linecolor=lightgray,linewidth=0.5pt,linestyle=dashed]{-1.2}{2}{x -0.5774 mul 2.3094 add}

\pspolygon[linewidth=0,linecolor=white,fillstyle=solid,fillcolor=lightgray](0,0)(1.1547;30)(2;60)(1.1547;90)

\psline[linecolor=lightblue,linewidth=2pt]{->}(2.5;180)(2.5;0)
\psline[linecolor=lightgreen,linewidth=2pt]{->}(1.5;300)(3;120)
\psline[linecolor=orangered,linewidth=2pt]{->}(3;60)(1.5;240)
\put(2.2,0.2){$e_1$}%
\put(-1.4,2.6){$e_2$}
\put(-0.9,-1){$e_3$}

\psarc[linewidth=1pt,linecolor=gray,linestyle=dotted,dotsep=1pt]{->}{1.3}{50}{110}

\psline[linecolor=lightgreen,linewidth=1pt,linestyle=dashed]{<-}(0.5554,0.6443)(0.69554,0.40157)
\psline[linecolor=lightgreen,linewidth=1pt,linestyle=dashed]{<-}(-0.4446,1.2216)(-0.304482,0.978908)

\psline[linecolor=orangered,linewidth=1pt,linestyle=dashed]{<-}(0.5554,0.6443)(0.13756,-0.0794203)
\psline[linecolor=orangered,linewidth=1pt,linestyle=dashed]{<-}(-0.4446,1.2216)(-0.862418,0.497917)

\psline[linecolor=lightblue,linewidth=1pt,linestyle=dashed]{<-}(0.5554,0.6443)(0,0.6443)
\psline[linecolor=lightblue,linewidth=1pt,linestyle=dashed]{<-}(-0.4446,1.2216)(-1,1.2216)

\readdata{\igriddata}{initgrid.data}
\listplot[plotstyle=dots,linecolor=brown,dotsize=5pt]{\igriddata}

\readdata{\rgriddata}{rotatedgrid.data}
\listplot[plotstyle=dots,linecolor=gray,dotsize=5pt]{\rgriddata}

\psdot[dotstyle=o,dotsize=0.15,linecolor=gray,fillcolor=white](0.5554,0.6443)

\put(-0.3,-0.4){$(0, 0)$}

\end{pspicture}
\caption{Rotating a grid map by 60 degrees results in a non-linear spatial phase transform. The spatial periodicity of a grid can be decomposed into the periodicity along the projection axes $e_i$, defined perpendicularly to the three grid axes. The orientation of and the period/wavelength along the third projection axis is constrained by the other two projection axes. The spatial phases of grids with the same spacing and orientation can be uniquely represented by points in a rhombus (area in gray). The sides of the rhombus are parallel to two of the grid axes, and the length of the side is equal to grid spacing. The spatial phases are only distinguishable in a rhombus area, because the rhombus is exactly one period length if projected onto two of the projection axes, $e_1$ and $e_2$. Points outside the rhombus are equivalent to those in the rhombus, modulo the wavelength along each projection axis. A grid with non-zero phase (shown in brown) is rotated by 60 degrees (indicated by the dotted arc) around the origin. The phase of the rotated grid (in gray) is equivalent to the white dot in the gray rhombus, since the coordinates (dashed arrows) of the two points along the projection axes are the same, modulo the wavelength. The phase change caused by the rotation depends on the initial phase of the grid.}
\label{fig-phase-change-rotate}
\end{figure}
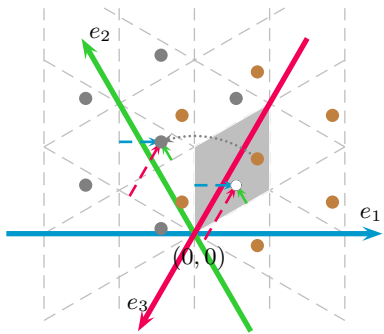

The mean crosscorrelogram of two groups of grid maps shows whether the grid maps are related by the same shift transformation. When two grids have the same orientation and spacing, the crosscorrelogram between them shows grid structure. In addition, the relative spatial shift between them determines the direction and distance that the corresponding crosscorrelogram is displaced away from the center. Cross-correlating one group of grids, which have the same orientation and spacing but distributed spatial phases, with a second group of grids, which are a shifted version of the first group for the same amount, will result in a set of crosscorrelograms with grid pattern and identical offset away from the center. The offset is just the common shift between the two groups. Adding the crosscorrelograms between the grids in the two groups preserves a clear grid structure. If a common rotation of multiples of 60 degrees is applied, instead of a common shift, the second group of grids still have the same orientation as the original grids, and each individual crosscorrelogram still keeps a grid structure, but the phases are changed differently for each grid (Fig.\ref{fig-phase-change-rotate}), producing crosscorrelograms with peaks appearing in distributed locations. Thus, the mean crosscorrelogram of the two groups does not maintain a grid pattern. For a common rotation other than multiples of 60 degrees, the rotated grids do not align to the same orientation as the original grids any more, making it impossible for the mean crosscorrelogram to be a grid. Therefore, for two groups of grids, the mean crosscorrelogram appears to be a grid only when there is a coherent shift between the two groups.

Fig.\ref{fig-grid-displacement}a shows that only with a counterclockwise rotation close to 9 degrees does the mean cross-environment crosscorrelogram show a grid structure. This is the angle that aligns the grid axes of both environments in Fig.\ref{fig-remapping}b,g. Fig.~\ref{fig-grid-displacement}a thus indicates that after the angular offset is counter-balanced, the spatial shifts between the grid fields in the two environments are the same (Fig.~\ref{fig-grid-displacement}b). For other rotations, in contrast, the spatial shifts are different across units, giving rise to an increasingly flat mean cross-environment crosscorrelogram. The absolute angular offset between the grid orientations in two environments is not important. Note that our simulations have fixed HD selectivity across environments, whereas in rats the preferred HDs of both conjunctive cells and HD cells appear to rotate coherently by the same amount in different environments~\citep{Sargolini2006}. Therefore the angular offset between the two sets of grid maps in our simulations can be considered as the angular shift after correcting for the HD selectivity shift across environments.   

Fig.\ref{fig-grid-displacement}c shows the ``gridness'' scores of the mean cross-environment crosscorrelogram with respect to the amount of rotation when a different number of units from the network is included in the analysis. The ``gridness'' score defined here is similar to the one in the method described in Section~\ref{sec-sylind}. The difference is that the ring circling the six maxima is centered on the closest maxima from the center of the crosscorrelogram, and that the difference in covariance instead of the correlation coefficient is used to define gridness. This is because most of the mean cross-environment crosscorrelograms are quite flat, giving unstable correlation coefficients when normalizing with small variance. As can be seen in Fig.\ref{fig-grid-displacement}c, with 20 units as a population, it is already possible to infer the angular mismatch between the fields in two environments. When only one unit is considered, the gridness of the cross-environment crosscorrelogram is roughly periodic with period 60 degrees. This indicates that with the population code expressed by grid cells it is possible to keep a unique orientation correspondence across environments in spite of the orientation symmetry of the grids, and to differentiate locations in an environment by the relative spatial phases of the grid fields. To represent places on a larger scale, multiple grid spacings are needed~\citep{Sol+06,Rol+06,Fiete2008,Si2009a}.

\section{Discussion}\label{sec-discuss}
Several conclusions can be extracted from these simulations. In the first place, we have described how alignment can be achieved in the model through recurrent collaterals, as long as they are associated with HD information. Secondly, we have shown that running-direction anisotropy in behavior, induced by the shape of the environment, is enough to distort the grid maps into ellipses, but it has the side-effect, of orientating one of the grid axes with one of the walls. At the same time, whether because too weak or localized, this kind of anisotropy fails to align the distortion, measured by the long grid axis or ellipse direction, into a common orientation. We have then explored the effect of running speed anisotropy, which has a milder influence on grid orientation but a stronger one on the orientation of the ellipses. In this scenario, recurrent collaterals have the effect of amplifying through population coherence what could be seen as a distortion of the idealized grid map already at a single cell level. This, however, may be regarded as a side effect of the function of recurrent collaterals in aligning the grid maps. Such a function may be useful, we hypothesize, in order to achieve global remapping with an invariant structure of relative phases, which, as we show in the last part of the paper, our model reproduces when the virtual rat explores two different rooms.

Whether grid alignment results from network interactions or as a consequence of a single unit process is a very interesting and complex question. In this respect, the literature presents opposing views. 

In models of interference between multiple oscillators, the grid maps are the result of information processing within a single neuron~\citep{Bur+07,Giocomo2007,Burgess2008,Hasselmo2008}. Each grid cell is then independent, and the common alignment comes from the fact that they all share the same path integration inputs. The different spatial phases are achieved by anchoring each grid cell to a different set of spatial inputs. This selective anchoring is in a way analogous to what happens with our place inputs, but it plays in the oscillator-interference models the minor role of correcting the errors of path integration, which is the main source of information used to construct the grid maps.

A completely opposite view to the single unit perspective comes from the attractor models~\citep{McN+06,Fuh+06,Guanella2007,burak2009,Navratilova2011}. In these, grid maps do not emerge individually, but as an emergent property of the population, so that grid maps and grid alignment are intrinsically inseparable. In our model, as shown by Kropff and Treves (2008), grid cells with no recurrent collaterals can develop grid maps with independent random orientations if they are left to interact weakly through inhibition, a situation that reminds us of the anatomy of layer II of medial entorhinal cortex. The introduction of collateral connections with the necessary head direction information to disambiguate firing sequences, as suggested earlier and proved here, can achieve grid alignment, as observed in experiments, thus offering a possible explanation of why these elements are found together in layers III to VI and are both absent in layer II. In our model, the common grid orientation lies on the attractor manifold created by the collateral interactions. This attracting property is a feature shared by the attractor models. In the latter, however, the collateral interactions have the full job of creating both grid maps and grid alignment. 

The alignment of grid cells in layer II is perhaps inherited from the deeper layers, which send strong projections to superficial layers~\citep{Haeften2003}. This hypothesis, not developed in the present work, represents an interesting direction of modeling research, which could focus on how conjunctive cells and grid cells may develop together in different layers, and perhaps even self-organize into laminar structures~\citep{Kro+08,Treves2003}.     

From the point of view of our model, the best way to address the question of a single unit versus a population origin of grid alignment is to interfere with the head direction source of information, assuming with some plausibility that this signal is not directly generated in the entorhinal cortex. Without head direction information, sequences in the grid cell network would be ambiguous. It would be interesting to study experimentally the effects that this kind of manipulation would produce in layers III to VI, where we assume that collateral connections work hand by hand with head direction information.

The functional difference between feed-forward connections and collateral connections can be appreciated in global remapping. In each environment, only a subset of place units is active. Feed-forward connections therefore relay environment-specific inputs to the conjunctive units. The collateral connections, however, signal the interactions between conjunctive units by the environment-independent geometrical structures they encode. With the same feed-forward and collateral connections, the network is able to provide an efficient metric encoding of space for all environments, since the emerging grid maps are essentially the same set of grids, almost identical after a unique rotation and translation. A natural question to ask is how the collateral weights can be learned. Addressing this important issue will be very instructive for the attractor models as well, and therefore requires a separate article to be elaborated in depth.

As we show in the present work, a weaker form of alignment of grid cells can be produced with independent single units, if the behavior of the rat includes some form of anisotropy. This alignment, however, would imply no population coherence in relative phases with global remapping. Still, anisotropy might be the reason why grid maps present distortion or ellipticity. This is not the first time that behavior is proposed to influence the firing of grid cells. In the hairpin maze experiment~\citep{Derdikman2009}, rats covered a two dimensional space but walking through a series of corridors in such a way that the behavioral constraints were those typical of exploration in one dimension. As a result, the grid cells showed a fragmented map with several one-dimensional patches instead of the canonical triangular structure. In our model, behavior alone creates distortion in the symmetry of grid maps, which is amplified and made coherent by recurrent collaterals. In a way, the distortion observed by Stensland and colleagues (2010), coherent at the population level, might be the price that the system has to pay in order to get a metric system of space that is consistent across environments.

It is poorly understood how much the geometry of a map in a new environment depends on the previous experience of the rat. It has been shown that in some over-trained animals grid maps emerge very fast in a novel environment~\citep{Hafting2005}, but we are far from being able to generalize this observation to all rats, experienced and \naive~\citep{Bar+09}. It is possible that the perfect grid pattern observed in some rats is a result of the successful generalization of an isotropic distribution of trajectories in 2 dimensions, related to a vast experience in open field environments at the right age of development, possibly starting around P20. Such extensive exposure to the open field is far from being an exclusive natural condition for all rats. It would be interesting to study the adult grid maps of rats raised in a different environment, for example in a system of tunnels. If there is a critical time window when the geometry of the grid system is developed, these animals might have, as adults, grid cells with peculiar properties even after prolonged training in the usual 2 dimensional square boxes. Another interesting aspect of the problem is what happens during this training period. Perhaps extensive learning can compensate for part of the distortion observed in grid maps. The main problem of such a study is that untrained animals exhibit a very poor coverage of 2 dimensional environments, making it hard to compare the geometry of their spatial maps with trained controls.

The network model produces qualitatively similar results over a large portion of parameter space. The delay parameter $\tau$ is critical to avoid the collapse of the sub-population with similar HD into a regime of synchronous dynamics. One wonders whether this delay in synaptic transmission might be realized by NMDA synapses, or other complex form of coupling among conjunctive cells.

It is possible to extend the network model to incorporate path integration cues into the inputs. Grids in such a network would still be stable even with week spatial inputs, and can serve as spatial codes useful in robot exploration tasks, where the robot has to rely on its own location estimation~\citep{Si_2007b,Fra+07b,Samu2009,Milford2010}.

\section*{Acknowledgements}
This work is supported by the European 7th Framework Program SpaceBrain and the Norwegian NOTUR project. We are grateful for enlightening discussions to all colleagues in the Kavli Institute and in the Spacebrain EU collaboration.

\appendix

\section{Abstract ellipse model}\label{Appendix}

\begin{figure}
\begin{center}
\begin{pspicture}(-3,-3)(3,2.8)
\newrgbcolor{lightblue}{0 0.6 0.8}
\newrgbcolor{lightgreen}{0.1961 0.8039 0.1961}
\newrgbcolor{orangered}{0.9490 0 0.3373}

\psline[linewidth=1pt]{->}(3;180)(3;0)
\psline[linewidth=1pt]{->}(3;315)(3.5;135)
\psline[linewidth=1pt]{->}(3;67.5)(3;247.5)

\psline[linecolor=lightblue,linewidth=0.5pt,linestyle=dashed]{-}(1,-2)(1,2)
\psline[linecolor=lightblue,linewidth=0.5pt,linestyle=dashed]{-}(2,-2)(2,2)
\psline[linecolor=lightblue,linewidth=0.5pt,linestyle=dashed]{-}(0,-2)(0,2)
\psline[linecolor=lightblue,linewidth=0.5pt,linestyle=dashed]{-}(-1,-2)(-1,2)
\psline[linecolor=lightblue,linewidth=0.5pt,linestyle=dashed]{-}(-2,-2)(-2,2)

\psplot[linecolor=lightgreen,linewidth=0.5pt,linestyle=dashed]{-2}{2}{x}
\psplot[linecolor=lightgreen,linewidth=0.5pt,linestyle=dashed]{-1}{2}{x 1.4142 sub}
\psplot[linecolor=lightgreen,linewidth=0.5pt,linestyle=dashed]{-2}{1}{x 1.4142 add}
\psplot[linecolor=lightgreen,linewidth=0.5pt,linestyle=dashed]{-2}{-0.5}{x 1.4142 2 mul add}

\psplot[linecolor=orangered,linewidth=0.5pt,linestyle=dashed]{-2}{2}{x -0.4142 mul}
\psplot[linecolor=orangered,linewidth=0.5pt,linestyle=dashed]{-2}{2}{x -0.4142 mul -1.4142 add}
\psplot[linecolor=orangered,linewidth=0.5pt,linestyle=dashed]{-2}{2}{x -0.4142 mul 1.4142 add}

\psdot[dotstyle=o,dotsize=0.2,linecolor=orangered,fillcolor=orangered](1,1)
\psdot[dotstyle=o,dotsize=0.2,linecolor=orangered,fillcolor=orangered](0,1.4142)
\psdot[dotstyle=o,dotsize=0.2,linecolor=orangered,fillcolor=orangered](-1,0.4142)
\pscircle[linecolor=orangered](-1,-1){0.08}
\pscircle[linecolor=orangered](0,-1.4142){0.08}
\pscircle[linecolor=orangered](1,-0.4142){0.08}

\psline[linewidth=2pt,linecolor=lightblue]{-}(0;0)(1;0)
\psline[linewidth=2pt,linecolor=lightgreen]{-}(0;0)(1;135)
\psline[linewidth=2pt,linecolor=orangered]{-}(0;0)(1.3066;247.5)

\put(2.6,0.2){$e_1$}
\put(0.9,0.2){$\lambda_1$}
\put(-2.2,2.4){$e_2$}
\put(-0.8,0.8){$\lambda_2$}
\put(-1.5,-2.7){$e_3$}
\put(-0.9,-1.4){$\lambda_3$}
\end{pspicture}
\end{center}
\caption{Non-evenly separated grid axes, i.e. not at $\pi/3$ of each other, imply non-even lengths of the axes themselves, to preserve periodicity, resulting in an elliptical arrangement of the vertices. In this example, the grid axes (indicated by the broken lines) are oriented towards at 90, 45 and 157.5 degrees. Orthogonal to them are the three projection axes entering the 2D Fourier transform, at angles $e_i$ of 0, 135, and 247.5 degrees, respectively. The vertices of the grid are arranged on lines at 3 different distances $\lambda_i$ from each other (and from the origin).}
\label{fig-ellip-grid}
\end{figure}
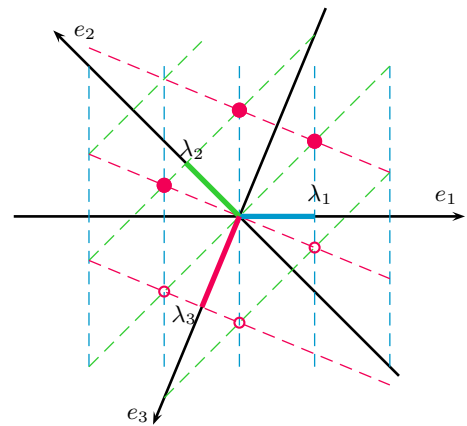

What does a non-circular, but elliptical arrangement of the grid map imply, geometrically? 

If any two of the three grid axes are not separated by 60 degrees, the lengths of the grid axes cannot be equal any more, for the grid to retain its canonical periodicity. As Fig.~\ref{fig-ellip-grid} illustrates, the grid vertices occur periodically along lines (broken in the Figure; those passing through the origin are called grid axes) orthogonal to the three projection axes (the black axes $e_i$ in Fig.~\ref{fig-ellip-grid}), provided such lines share 3-way intersections. If the grid is modeled as a simple sum of 3 cosines, the distance between the broken lines of each orientation is the wavelength $\lambda_i\equiv 2\pi/k_i$ of the corresponding cosine, which has a $\vec{k}$-vector of length $k_i$ oriented at $e_i$ from the $x$-axis. To preserve grid structure, the inverse wavelength $k_3$ along the third projection axis $e_3$ is constrained so that the peaks (red broken lines in Fig.~\ref{fig-ellip-grid}) coincide with those along each of the other two projection axes (blue and green broken lines). This leads to two equations relating $\lambda_3$ to $\lambda_1$ and $\lambda_2$ and to the angles between them. In the Figure, it is assumed that $e_1=0$, but it is easy to write the following equations considering the general case of nonzero $e_1$

\begin{align*}
\lambda_3 &= \frac{\lambda_1}{\cos(e_2-\pi/2-e_1)}\cos(e_3-e_2-\pi/2),\\
\lambda_3 &= \frac{\lambda_2}{\cos(e_2-\pi/2-e_1)}\cos(e_3-e_1+\pi/2).
\end{align*}
They lead to a solution for $e_3$ and $\lambda_3$ which may be expressed for example as
\begin{align*}
e_3 &= \arccos\left[\frac{\lambda_1\cos(e_2-e_1)+\lambda_2}{\sqrt{\lambda_1^2+\lambda_2^2+2\lambda_1 \lambda_2\cos(e_2-e_1)}}\right]+\pi+e_1,\\
\lambda_3 &= \frac{\lambda_1 \lambda_2}{\sqrt{\lambda_1^2+\lambda_2^2+2\lambda_1 \lambda_2\cos(e_2-e_1)}}.
\end{align*}
It is straightforward to see that these relations translate into an equivalent, but simpler relation between the $\vec{k}$-vectors
\begin{equation}
\vec{k}_1+\vec{k}_2+\vec{k}_3=0.
\end{equation}

What determines the exact position of the $\vec{k}$-vectors?

Considering the abstract model discussed in detail in \cite{Kro+08}, of the development of an individual grid, the model does not distinguish between a would-be pure grid and a conjunctive unit. It assumes that at the end of the developmental process the firing map $\Psi_i(\vec{x})$ of the unit minimizes a functional $L$, which in its basic version takes the form
\begin{equation}
L={1\over A}\int_Ad\vec{x}[\nabla\Psi(\vec{x})]^2+{\gamma\over A}\int_Ad\vec{x}\Psi(\vec{x})\int_Ad\vec{x'}\Psi(\vec{x}')K(|\vec{x}'-\vec{x}|),
\end{equation}
where $K(|\Delta\vec{x}|)$ is a kernel expressing neuronal fatigue. The first term of the cost function expresses a penalty for maps that vary too much across space, and a preference for smooth maps. This can be understood from the point of view of the smoothness of the spatial inputs and the smoothness of the neuronal transfer function. The second term of the cost function expresses a penalty for maps in which a neuron has to fire for very long periods of time, subject to neural fatigue. In \cite{Kro+08} the grid pattern emerges as a compromise solution between these opposite requirements. There we give two examples of the kernel, which can be treated analytically. The analysis involves going into Fourier space, where the firing map is decomposed into 2D Fourier modes 
\begin{equation}
\Psi_i(\vec{x})=a_0+\sum_i a_i \cos(\vec{k}_i\cdot\vec{x}+\phi_i)
\end{equation}
and the functional becomes
\begin{equation}
L= \frac{1}{2}\sum_i a_i^2k_i^2+\gamma_0^2\tilde{K}(0)+{\gamma\over 2}\sum_i a^2_i\tilde{K}(k_i)
\end{equation}
where $\tilde{K}$ simply denotes the 2D Fourier transform of the kernel, and $k$, again, is the length of the 2D vector $\vec{k}$.

As discussed in \cite{Kro+08}, among 2D periodic solutions the favored ones are those with a superposition of 3 cosines
\begin{equation}
\Psi(\vec{x})= 1 + (2/3)\sum_{i=1}^3 \cos(\vec{k}_i\cdot\vec{x})
\end{equation}
where for simplicity we have omitted the phases by suitably defining the origin.

To model a degree of speed anisotropy, with faster runs e.g. along the axes or along the diagonals of a square environment, one can simply add a quadrupole term to the functional $L$, the precise form of which is not essential, like the precise form of the kernel $K$ is not essential either. One simple choice is
\begin{eqnarray}
L^{\rm anis}&=& {1\over 2}\sum_i a_i^2k_i^2+\gamma_0^2\tilde{K}(0)+{\gamma\over 2}\sum_i a^2_i\tilde{K}(k_i)\nonumber\\ &+&\zeta \sum_i \left[(k_{xi}^3+k_{yi}^3)/k_i^3 - 1/\sqrt{2}\right],
\end{eqnarray}
where the quadrupole term, with a strength $\zeta$, adds a cost $\zeta (\sqrt{2}-1)/\sqrt{2}$ for each $\vec{k}$ aligned with the $x$- or $y$-axis, and no cost for those oriented at $\pm \pi/4$, hence favoring $\vec{k}$ vectors along the diagonals (rotating the quadrupole by  $\pi/4$ one can favor instead those aligned with the axes).

\begin{figure*}
\centering
\begin{pspicture}(0,0)(17.4,5)
\rput[bl](0,0){\epsfig{file=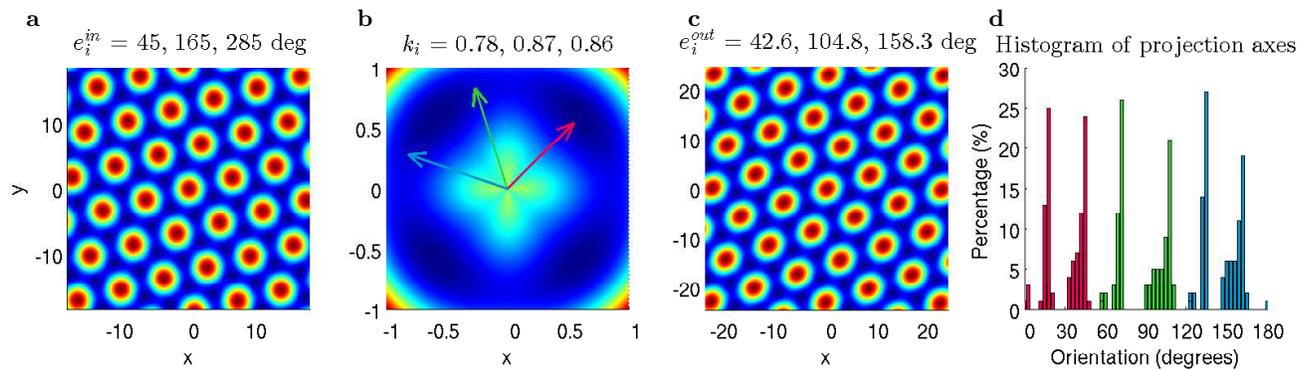,width=17.4cm}}

\rput[bl](0.4,4.6){\text{\small{\bf a}}}
\rput[bl](4.8,4.6){\text{\small{\bf b}}}
\rput[bl](9.2,4.6){\text{\small{\bf c}}}
\rput[bl](13.2,4.6){\text{\small{\bf d}}}

\end{pspicture}
\caption{The input (a) and output (c) produced by the gradient descent algorithm that adjusts the $\vec{k}_i$ vectors to minimize a cost function $L^{\rm anis}$ that includes a quadrupole term (panel b, in Fourier space). Depending on the exact initial condition, the algorithm converges towards the theoretically optimal solution (which has angles $e_i= -110, 20, -45$ deg and the two longer $\vec{k}_i$ vectors of identical length) but it stops before reaching it, as solutions with a jitter of $\pm 2$ deg have the same cost, to the fourth significant digit. Panel c shows in different colors the orientation distribution of the three $e_i$s, clustering into two modes.  However, the shallow landscape of the cost function (for any reasonable form of the quadrupole anisotropy) indicates that the tight alignment seen experimentally cannot result from single-unit mechanisms alone.}
\label{fig-abstract-mod}
\end{figure*}

The addition of the extra term makes it hard to get analytical solutions, but numerical minima of the cost function can be found with a gradient descent algorithm that takes into account the additional constraint on  the triplet of $\vec{k}_i$ vectors and their amplitudes $a_i$, $i=1,2,3$. We performed several numerical minimizations of $L^{\rm anis}$ using as initial conditions perfect grids oriented in all possible directions , with the 3 $\vec{k}_i$ at $2\pi/3$ of each other, their length such as to minimize $L$, and $a_i\equiv 2/3$. The numerical algorithm rapidly converges to a solution with one somewhat shorter $\vec{k}$ vector aligned with one of the two orthogonal directions of faster speed (in Fig.\ref{fig-abstract-mod}, $3\pi/4$) and the other two vectors somewhat longer and more than $2\pi/3$ of each other. Whatever the initial orientation of the original symmetric grid, the algorithm finds a solution very close to either of the two ``exact'' solutions, the one shown in Fig.\ref{fig-abstract-mod} and the equivalent reflection along the vertical or horizontal axis. In this way, when only the single cell level effect is considered, the shorter vector, indicating roughly the direction of the ellipse, stands at  $\pm \pi/4$ (or at $0, \pi/2$ if the faster speed are along the axes). 

\bibliographystyle{apsrev}
\bibliography{ref}

\end{document}